\documentclass[twocolumns,structabstract]{aa}
\usepackage{graphicx}
\usepackage{amssymb}
\usepackage{txfonts}
\usepackage{hyperref}
\usepackage{ulem}
\usepackage[table]{xcolor}
\usepackage{hhline}

\renewcommand{\l}{\ell}
\newcommand{\ie}{\textit{i.e.}}
\newcommand{\eg}{\textit{e.g.}}
\newcommand{\cf}{\textit{cf.}}

\newcommand{\numax}{\nu_{\mathrm{max}}}
\newcommand{\Teff}{T_{\mathrm{eff}}}
\newcommand{\erel}{\varepsilon_{\mathrm{rel.}}}
\newcommand{\enorm}{\varepsilon_{\mathrm{norm.}}}
\newcommand{\brel}{b_{\mathrm{rel.}}}
\newcommand{\bnorm}{b_{\mathrm{norm.}}}
\newcommand{\HeI}{He\,{\small I}}
\newcommand{\HeII}{He\,{\small II}}
\renewcommand{\emph}[1]{\textit{#1}}
\newcommand{\Msun}{\mathrm{M}_{\odot}}
\newcommand{\Rsun}{\mathrm{R}_{\odot}}
\newcommand{\Lsun}{\mathrm{L}_{\odot}}
\newcommand{\SolutionCell}[1]{\multicolumn{1}{c}{\cellcolor{lightgray}\textcolor{black}{\textit{#1}}}}
\newcommand{\GridCell}[1]{\multicolumn{1}{c}{\cellcolor{white}\textcolor{black}{#1}}}
\newcommand{\GlitchCell}[1]{\multicolumn{1}{c}{\cellcolor{gray}\textcolor{white}{#1}}}
\newcommand{\InversionCell}[1]{\multicolumn{1}{c}{\cellcolor{black}\textcolor{white}{#1}}}

\begin{document}

\title{SpaceInn hare-and-hounds exercise:
       Estimation of stellar properties using space-based asteroseismic data}
\titlerunning{SpaceInn hare-and-hounds exercise}

\author{D.~R. Reese\inst{1,2,3}, W.~J. Chaplin\inst{1,2}, G.~R. Davies\inst{1,2},
A. Miglio\inst{1,2}, H. M. Antia\inst{4}, W.~H. Ball\inst{5, 6}, S. Basu\inst{7}, G.
Buldgen\inst{8}, J. Christensen-Dalsgaard\inst{2}, H.~R. Coelho\inst{1,2},
S. Hekker\inst{6,2}, G. Houdek\inst{2}, Y. Lebreton\inst{9,10},
A. Mazumdar\inst{11}, T.~S. Metcalfe\inst{12}, V. Silva Aguirre\inst{2}, D.
Stello\inst{13,2}, K. Verma\inst{4}}
\authorrunning{Reese et al.}

\institute{School of Physics and Astronomy, University of Birmingham,
           Edgbaston, Birmingham, B15 2TT, UK
           \and
           Stellar Astrophysics Centre (SAC),
           Department of Physics and Astronomy, Aarhus University,
           Ny Munkegade 120, DK-8000 Aarhus C, Denmark
           \and
           LESIA, Observatoire de Paris, PSL Research University, CNRS, Sorbonne
           Universit{\'e}s, UPMC Univ. Paris 06, Univ. Paris Diderot, Sorbonne Paris
           Cit{\'e}, 92195 Meudon, France \\
           \email{daniel.reese@obspm.fr}
           \and
           Tata Institute of Fundamental Research, Homi Bhabha Road, Mumbai
           400005, India
           \and
           Institut f{\"u}r Astrophysik, Georg-August-Universit{\"a}t
           G{\"o}ttingen, Friedrich-Hund-Platz 1, 37077 G{\"o}ttingen, Germany
           \and
           Max-Planck-Institut f{\"u}r Sonnensystemforschung,
           Justus-von-Liebig-Weg 3, 37077 G{\"o}ttingen, Germany
           \and
           Department of Astronomy, Yale University, P. O. Box 208101, New
           Haven, CT 065208101, USA
           \and
           Institut d'Astrophysique et G{\'e}ophysique de l'Univert{\'e} de
           Li{\`e}ge, All{\'e}e du 6 ao{\^u}t 17, 4000 Li{\`e}ge, Belgium
           \and
           Observatoire de Paris, GEPI, CNRS UMR 8111, F-92195 Meudon, France
           \and
           Institut de Physique de Rennes, Universit{\'e} de Rennes 1, CNRS UMR
           6251, F-35042 Rennes, France
           \and
           Homi Bhabha Centre for Science Education, TIFR, V. N. Purav Marg,
           Mankhurd, Mumbai 400088, India
           \and
           Space Science Institute, Boulder, CO 80301, USA
           \and
           Sydney Institute for Astronomy (SIfA), School of Physics, University
           of Sydney, NSW 2006, Australia
          }

\date{}

\abstract
{
Detailed oscillation spectra comprising individual frequencies for numerous
solar-type stars and red giants are either currently available, \eg\ courtesy of
the CoRoT, \emph{Kepler}, and K2 missions, or will become available with the
upcoming NASA TESS and ESA PLATO 2.0 missions.  These data can lead to a precise
characterisation of these stars thereby improving our understanding of stellar
evolution, exoplanetary systems, and the history of our galaxy.
}
{
Our goal is to test and compare different methods for obtaining stellar
properties from oscillation frequencies and spectroscopic constraints. 
Specifically, we would like to evaluate the accuracy of the results and
reliability of the associated error bars, and see where there is
room for improvement.
}
{
In the context of the SpaceInn network, we carried out a hare-and-hounds
exercise in which one group, the hares, produced ``observed'' oscillation
spectra for a set of $10$ artificial solar-type stars, and a number of hounds
applied various methods for characterising these stars based on the data
produced by the hares.  Most of the hounds fell into two main groups.  The first
group used forward modelling (\ie\ applied various search/optimisation
algorithms in a stellar parameter space) whereas the second group relied on
acoustic glitch signatures.
}
{
Results based on the forward modelling approach were accurate to
$1.5\,\%$ (radius), $3.9\,\%$ (mass), $23\,\%$ (age), $1.5\,\%$
(surface gravity), and $1.8\,\%$ (mean density), as based on the root-mean
square difference.  Individual hounds reached different degrees of
accuracy, some of which were substantially better than the above average
values. For the two $1$ M$_{\odot}$ stellar targets, the accuracy on the age is
better than $10\,\%$ thereby satisfying the requirements for the PLATO 2.0
mission.  High stellar masses and atomic diffusion (which
in our models does not include the effects of
radiative accelerations) proved to be sources
of difficulty.  The average accuracies for the acoustic radii of the base of the
convection zone, the \HeII\ ionisation, and the $\Gamma_1$ peak located between
the two He ionisation zones were $17\,\%$, $2.4\,\%$, and $1.9\,\%$,
respectively.  The results from the forward modelling were on average more
accurate than those from the glitch fitting analysis as the latter seemed to be
affected by aliasing problems for some of the targets.
}
{
Our study indicates that forward modelling is the most accurate way of
interpreting the pulsation spectra of solar-type stars.  However, given its
model-dependent nature, such methods need to be complemented by
model-independent results from, \eg, glitch analysis. Furthermore, our results
indicate that global rather than local optimisation algorithms should be used in
order to obtain robust error bars.
}


\maketitle

\section{Introduction}

Determining accurate stellar properties through asteroseismology is fundamental
for various aspects of astrophysics.  Indeed, accurate stellar properties help
us to place tighter constraints on stellar evolution models. 
Furthermore, the accuracy with which the properties of exoplanets are
determined depends critically on the accuracy of the properties of their host
stars \citep[\eg][and references therein]{Guillot2011}.  Last but not least,
obtaining accurate stellar properties is an integral part of characterising
stellar populations in the Milky Way and reconstructing its history
\citep[\eg][]{Miglio2013, Casagrande2014}.

With the advent of past and current high precision space photometry missions,
namely CoRoT \citep{Baglin2009}, \emph{Kepler} \citep{Borucki2009}, and its
re-purposed version K2 \citep{Howell2014}, detailed asteroseismic spectra
comprising individual frequencies of solar-like oscillations have become
available for hundreds of solar-type stars \citep[\eg][]{Chaplin2014},
including planet-hosting stars \citep[\eg][]{Davies2016}, as well
as thousands of red giants \citep[\eg][]{Mosser2010, Stello2013}, including
recent detections in data collected by K2 \citep{Chaplin2015, Stello2015}. 
Upcoming space missions, such as TESS \citep{Ricker2014} and PLATO 2.0
\citep{Rauer2014} will increase this number even more.  Furthermore,
combining the data from these missions with highly accurate
parallaxes obtained via the \emph{Gaia} mission \citep{Perryman2001}
will lead to tighter constraints on stellar properties.

Obtaining stellar properties from pulsation spectra is a non-linear inverse
problem, which may have multiple local minima in the relevant parameter space
\citep[\eg][]{Aerts2010}. Accordingly, this has led to the development of a
variety of techniques, both in the context of helio- and asteroseismology, for
finding stellar properties and associated error bars as well as best-fitting
models.  Indeed, as described in \citet{Gough1985}, there are various ways of
interpreting helioseismic data, namely the forward modelling approach or
``repeated execution of the forward problem'' as Gough puts it -- where one
seeks to find models whose oscillation frequencies provide a good match
to the observations --, analytical approaches which include asymptotic methods
and glitch fitting, and formal inversion techniques (these typically rely on
linearising the relation between frequencies and stellar structure, and
inverting it subject to regularity constraints).  The same techniques also apply
to asteroseismology, although the number of available pulsation frequencies is
considerably smaller than in the solar case given that observations are
disk-averaged, and the ``classical'' parameters (\eg\ $\Teff$, [Fe/H],
luminosity) are determined with larger uncertainties.  It therefore becomes
crucial to compare these methods in terms of accuracy (\ie\ how close the result
is to the actual value and how realistic the error bars are) and computational
cost (given the large number of targets which have been or will be observed by
space missions).

An ideal approach for carrying out such a comparison would be to test these
methods on stars for which independent estimates of stellar properties are
available.  This has been done in various works \citep[\eg][]{Bruntt2010,
Miglio2005, Bazot2012, Huber2012, SilvaAguirre2012}, where stellar masses
deduced from orbital parameters in binary systems, and/or radii from a
combination of astrometry and interferometry in nearby systems were used either
as a test of seismic results or as supplementary constraints. An alternate
approach is to carry out a hare-and-hounds exercise.  In such an exercise, one
group, the ``hares'' produces a set of ``observations'' from theoretical stellar
models.  These are then sent to other groups, the ``hounds'', who try to deduce
the general properties of these models based on the simulated observations.  An
obvious limitation of hare-and-hounds exercises is that they are unable to test
the effects of physical phenomena which are present in real stars but not in our
models due to current limitations in our theory.  Various hare-and-hounds
exercises have been carried out in the past or are ongoing, to test various
stages of seismic inferences, namely mode parameter extraction from light
curves, seismic interpretation of pulsation spectra, or the two combined.  For
instance, \citet{Stello2009} investigated retrieving general stellar properties
from seismic indices and classical parameters in the framework of the asteroFLAG
consortium \citep{Chaplin2008}.  However, with the large amount of high-quality
seismic data currently available from space missions CoRoT, \emph{Kepler}, and
K2, and the specifications for the upcoming PLATO 2.0 mission \citep{Rauer2014},
it is necessary to push the analysis further by testing the accuracy with which
stellar properties can be retrieved from pulsation spectra composed of
individual frequencies along with classical parameters including luminosities
based on \emph{Gaia}-quality parallaxes.  Indeed, the availability of
large numbers of individual pulsation frequencies as opposed to average seismic
parameters allows us to apply detailed stellar modelling techniques, thereby
leading to an improved characterisation of the observed stars, especially of
their ages.  Accordingly, the main objective of this SpaceInn
hare-and-hounds exercise has been to test how accurately one can retrieve
general stellar properties from such data.  Furthermore, we wanted to compare
convection zone depths obtained from best-fitting models with those obtained
from an independent analysis of so-called acoustic glitches.  In this paper, we
describe the exercise and its results.  The following section focuses on the
theoretical models upon which the observations are based.  This is then followed
by a description of the ``hounds'' and their different techniques for retrieving
stellar properties.  Section~\ref{sect:results} gives the results and is
subdivided into four parts, the first one dealing with general stellar
properties, the second one with properties related to the base of the convection
zone, the third with properties of the \HeII\ ionisation zone, and the last with
comparisons between the acoustic structure of the target stars and that of some
of the best-fitting solutions.  A discussion concludes the paper.

\section{The ``observational'' data}

\subsection{The models and their pulsation modes}

\begin{figure}[htbp]
\begin{center}
\includegraphics[width=\columnwidth]{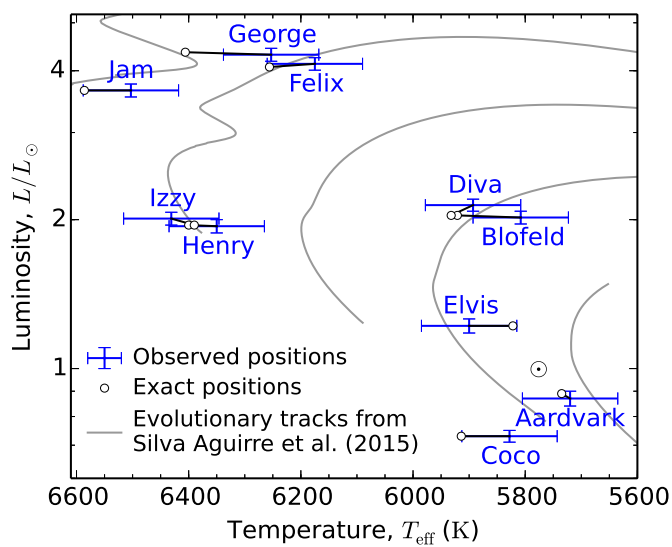}
\caption{HR diagram showing the exact and ``observed'' positions of the 10
stellar targets.  Evolutionary tracks from \citet{SilvaAguirre2015} and the Sun's
position have also been included for the sake of comparison.  \label{fig:HR}}
\end{center}
\end{figure}

A set of $10$ solar-type stellar models were selected as target models in this
hare-and-hounds exercise.  These models were chosen from a broad range of
stellar masses and temperatures in the cool part of the HR diagram where
oscillations have been routinely detected in main-sequence and subgiant stars,
deliberately including difficult cases, in order to test the limits of the
various fitting procedures used by the hounds.  Hence, the models went from
$5735$ to $6586$ K in effective temperature, $0.73$ to $4.36$ $\Lsun$ in
luminosity, and $0.78$ to $1.33$ $\Msun$ in mass.   Their ``observational''
properties, \ie\ the ones communicated to the hounds, as well as the exact
values are given in Table~\ref{tab:observational_properties}. 
Table~\ref{tab:composition} gives the compositions of the models.
Figure~\ref{fig:HR} shows their positions (both exact and ``observed'') in an HR
diagram.

\begin{table*}[htbp]
\caption{``Observational'' parameters of the stellar targets.
\label{tab:observational_properties}}
\centering
\begin{tabular}{lcccccccccc}
\hline
\hline
\textbf{Name} &
$\Teff^{\mathrm{obs}}$  & 
$\Teff^{\mathrm{exact}}$  & 
$\left(L/\Lsun\right)_{\mathrm{obs}}$ &
$\left(L/\Lsun\right)_{\mathrm{exact}}$ &
$\Delta \nu_{\mathrm{obs}}$ &
$\Delta \nu_{\mathrm{exact}}$ &
$\nu_{\mathrm{max}}^{\mathrm{obs}}$ &
$\nu_{\mathrm{max}}^{\mathrm{exact}}$ &
[Fe/H]$_{\mathrm{obs}}$ &
[Fe/H]$_{\mathrm{exact}}$ \\
\hline
Aardvark & $5720 \pm 85$ & $5735$  &  $0.87 \pm 0.03$  &  $0.89$ &  $149.6 \pm 2.9$  &  $144.7$  &  $3503 \pm 165$  &  $3372$ & $ 0.02 \pm 0.09$ &  $ 0.00$ \\
Blofeld  & $5808 \pm 85$ & $5921$  &  $2.02 \pm 0.06$  &  $2.04$ &  $ 97.4 \pm 1.9$  &  $ 94.3$  &  $1750 \pm 100$  &  $2015$ & $ 0.04 \pm 0.09$ &  $ 0.09$ \\
Coco     & $5828 \pm 85$ & $5914$  &  $0.73 \pm 0.02$  &  $0.73$ &  $160.9 \pm 3.3$  &  $162.5$  &  $3634 \pm 179$  &  $3587$ & $-0.74 \pm 0.09$ &  $-0.70$ \\
Diva     & $5893 \pm 85$ & $5932$  &  $2.14 \pm 0.06$  &  $2.04$ &  $100.0 \pm 1.9$  &  $ 96.0$  &  $2059 \pm 101$  &  $2031$ & $ 0.03 \pm 0.09$ &  $ 0.16$ \\
Elvis    & $5900 \pm 85$ & $5822$  &  $1.22 \pm 0.04$  &  $1.22$ &  $118.3 \pm 2.4$  &  $120.2$  &  $2493 \pm 127$  &  $2606$ & $ 0.04 \pm 0.09$ &  $ 0.00$ \\
Felix    & $6175 \pm 85$ & $6256$  &  $4.13 \pm 0.12$  &  $4.07$ &  $ 70.0 \pm 1.4$  &  $ 69.6$  &  $1290 \pm 66 $  &  $1337$ & $ 0.06 \pm 0.09$ &  $ 0.00$ \\
George   & $6253 \pm 85$ & $6406$  &  $4.31 \pm 0.13$  &  $4.36$ &  $ 68.8 \pm 1.4$  &  $ 70.5$  &  $1311 \pm 67 $  &  $1356$ & $-0.03 \pm 0.09$ &  $ 0.00$ \\
Henry    & $6350 \pm 85$ & $6400$  &  $1.94 \pm 0.06$  &  $1.95$ &  $117.6 \pm 2.3$  &  $116.7$  &  $2510 \pm 124$  &  $2493$ & $-0.35 \pm 0.09$ &  $-0.36$ \\
Izzy     & $6431 \pm 85$ & $6390$  &  $2.01 \pm 0.06$  &  $1.95$ &  $114.6 \pm 2.3$  &  $116.1$  &  $2319 \pm 124$  &  $2481$ & $-0.34 \pm 0.09$ &  $-0.25$ \\
Jam      & $6503 \pm 85$ & $6586$  &  $3.65 \pm 0.11$  &  $3.65$ &  $ 86.4 \pm 1.7$  &  $ 86.6$  &  $1758 \pm 89 $  &  $1785$ & $ 0.09 \pm 0.09$ &  $ 0.00$ \\
\hline
\end{tabular}
\tablefoot{The ``exact'' large separations are a least-squares fit to all of the
modes, using the ``observational'' error bars to decide the weights on each
mode.  The ``exact'' $\nu_{\mathrm{max}}$ values were obtained by applying the
$\nu_{\mathrm{max}}$ scaling relation using the reference values given in
Table~\ref{tab:reference_values}.}
\end{table*}

\begin{table*}[htbp]
\caption{Chemical composition of the stellar targets. \label{tab:composition}}
\centering
\begin{tabular}{lccccccc}
\hline
\hline
\textbf{Name} &
$X_0$ &
$Z_0$ &
$X_{\mathrm{surf}}$ &
$Z_{\mathrm{surf}}$ &
$(Z/X)_{\mathrm{surf}}$ &
$(Z/X)_{\mathrm{surf},\odot}$ \\
\hline
Aardvark & 0.71550 & 0.01755 & 0.71543 & 0.01755 & 0.02453 & 0.0245 \\
Blofeld  & 0.71400 & 0.02000 & 0.78280 & 0.01579 & 0.02018 & 0.0165 \\
Coco     & 0.74140 & 0.00360 & 0.74132 & 0.00360 & 0.00486 & 0.0245 \\
Diva     & 0.72600 & 0.02600 & 0.72593 & 0.02600 & 0.03582 & 0.0245 \\
Elvis    & 0.71550 & 0.01755 & 0.71543 & 0.01755 & 0.02453 & 0.0245 \\
Felix    & 0.71550 & 0.01755 & 0.71543 & 0.01755 & 0.02453 & 0.0245 \\
George   & 0.71550 & 0.01755 & 0.71543 & 0.01755 & 0.02453 & 0.0245 \\
Henry    & 0.72600 & 0.01000 & 0.78010 & 0.00825 & 0.01058 & 0.0245 \\
Izzy     & 0.72600 & 0.01000 & 0.72593 & 0.01000 & 0.01378 & 0.0245 \\
Jam      & 0.71550 & 0.01755 & 0.71543 & 0.01755 & 0.02453 & 0.0245 \\
\hline
\end{tabular}
\tablefoot{The last column specifies the reference solar value of
$(Z/X)$ used to obtain the value of [Fe/H].  Blofeld and Henry include
atomic diffusion
(see Table~\ref{tab:fundamental_properties}), thereby leading to different
surface abundances.  Even in the other models, the value of $X$ decreases
slightly due to Deuterium burning.}
\end{table*}

The models were calculated using the CLES stellar evolution code
\citep{Scuflaire2008}. The equation of state was based on OPAL 2001
\citep{Rogers2002} using the tabulated $\Gamma_3-1$ values.  OPAL opacities
\citep{Iglesias1996}, complemented with \citet{Ferguson2005} opacities at low
temperatures, were used.  The nuclear reaction rates came from the NACRE
compilation \citep{Angulo1999} and included the revised
$^{14}$N(p,$\gamma$)$^{15}$O reaction rate from \citet{Formicola2004}.  
Convection was implemented through standard mixing-length theory using
solar-calibrated values of the mixing length, $\alpha_{\textrm{MLT}}$
\citep{Boehm-Vitense1958}.  Atomic diffusion based on the prescription
given in \citet{Thoul1994} was included in specific cases. 
This approach includes gravitational settling, as well as the effects of
temperature and composition gradients, but neglects radiative accelerations
\citep[see also][for a review of this and other approaches]{Thoul2007}. A
radiative grey atmosphere using the Eddington approximation
\citep[\eg][]{Unno1966} was included in most models and extended from the
photosphere, $T=\Teff$, to an optical depth of $\tau = 10^{-3}$.  The
fundamental properties of the targets are given in
Table~\ref{tab:fundamental_properties}, where the models have been sorted
according to mass (for reasons which will become apparent later on).

In most cases, the pulsation modes were calculated with InversionKit
2.1\footnote{This code is currently available at:
\href{http://bison.ph.bham.ac.uk/\~dreese/InversionKit/} {\tt
http://bison.ph.bham.ac.uk/$\sim$dreese/InversionKit/}}, using 4th order
calculations, various sets of equations involving either Lagrangian and
Eulerian pressure perturbations, and the mechanical boundary condition $\delta
P = 0$.  For two of the models, the ADIPLS code
\citep{Christensen-Dalsgaard2008b} was used instead, so as to apply an
isothermal boundary condition, since this boundary condition is not currently
implemented in InversionKit.  This was used as a way of simulating surface
effects, \ie\ offsets between observed and modelled frequencies which occur as a
result of our poor modelling of the near-surface layers of the star
\citep[\eg][]{Kjeldsen2008}, and attenuating the fact that the atmosphere was
truncated in one of the models.  The frequencies are given in
Appendix~\ref{sect:pulsation_frequencies}.

\begin{table*}[htbp]
\caption{Fundamental properties of the stellar
targets.\label{tab:fundamental_properties}}
\centering
\begin{tabular}{lcccccccccccccc}
\hline
\hline
\textbf{Name} &
$M$  &
$R$  &
$\bar{\rho}$ &
$\log(g)$ &
$t$ &
$r_{\mathrm{BCZ}}$ &
$\tau_{\mathrm{Tot.}}$ &
$\tau_{\mathrm{BCZ}}$ &
$\alpha_{\mathrm{MLT}}$ &
$\alpha_{\mathrm{ov}}$ &
\textbf{Diff.} &
\textbf{Mix.} &
\textbf{Atm.} &
\textbf{B.C.} \\
& (M$_{\odot}$) & (R$_{\odot}$) & (g.cm$^{-3}$) & (dex)& (Gyrs) & ($R$) & (s) & (s) & & & & & \\
\hline
Coco     & 0.78 &  0.815 &  2.029 &  4.508 &  9.616 &  0.746 & 2999 & 1293 & 1.6708 &  --  & No  & GN93  & Edd.  & $\delta P = 0$  \\
Aardvark & 1.00 &  0.959 &  1.596 &  4.474 &  3.058 &  0.731 & 3371 & 1405 & 1.6708 &  --  & No  & GN93  & Edd.  & $\delta P = 0$  \\
Elvis    & 1.00 &  1.087 &  1.097 &  4.365 &  6.841 &  0.727 & 4045 & 1656 & 1.6708 &  --  & No  & GN93  & Edd.  & Isoth.      \\
Henry    & 1.10 &  1.138 &  1.051 &  4.367 &  2.055 &  0.839 & 4199 & 2280 & 1.8045 &  --  & Yes & GN93  & Edd.  & $\delta P = 0$  \\
Izzy     & 1.10 &  1.141 &  1.041 &  4.364 &  2.113 &  0.849 & 4219 & 2353 & 1.6708 &  --  & No  & GN93  & Edd.  & $\delta P = 0$  \\
Blofeld  & 1.22 &  1.359 &  0.684 &  4.257 &  2.595 &  0.838 & 5107 & 2811 & 1.6738 & 0.10 & Yes & AGS05 & Trun. & Isoth.      \\
Diva     & 1.22 &  1.353 &  0.693 &  4.261 &  4.622 &  0.767 & 5082 & 2288 & 1.6708 & 0.10 & No  & GN93  & Edd.  & $\delta P = 0$  \\
Felix    & 1.33 &  1.719 &  0.369 &  4.091 &  2.921 &  0.842 & 7014 & 3830 & 1.6708 & 0.05 & No  & GN93  & Edd.  & $\delta P = 0$  \\
George   & 1.33 &  1.697 &  0.383 &  4.102 &  2.944 &  0.875 & 6930 & 4156 & 1.6708 & 0.25 & No  & GN93  & Edd.  & $\delta P = 0$  \\ 
Jam      & 1.33 &  1.468 &  0.592 &  4.228 &  1.681 &  0.905 & 5666 & 3718 & 1.6708 & 0.05 & No  & GN93  & Edd.  & $\delta P = 0$  \\
\hline
\end{tabular}
\tablefoot{$M$ is the mass, $R$ the radius, $\bar{\rho}$ the mean density, $g$ the surface
gravity, $t$ the stellar age, $r_{\mathrm{BCZ}}$ the radius at the base of the
convection zone, $\tau_{\mathrm{Tot.}}$ the acoustic radius,
$\tau_{\mathrm{BCZ}}$ the acoustic radius at the base of the convection zone,
and $\alpha_{\mathrm{ov}}$ the overshooting parameter.  ``Diff.'' represents
atomic diffusion, ``Mix.'' the abundances mixture, ``Atm.'' the atmosphere,  ``Edd.''
a radiative grey Eddington atmosphere, ``Trun.'' a truncated atmosphere, and ``B.C.'' boundary
condition on the pulsation modes.  The abundances mixtures were GN93
\citep{Grevesse1993} and AGS05 \citep{Asplund2005}. Reference values are given
in Table~\ref{tab:reference_values}.}
\end{table*}

\begin{table}
\caption{Reference values. \label{tab:reference_values}}
\begin{tabular}{lcc}
\hline
\hline
\textbf{Quantity} &
\textbf{Value} &
\textbf{Reference} \\
\hline
$G$ (in $\mathrm{cm}^3\mathrm{g}^{-1}\mathrm{s}^{-2}$) & $6.6742 \times 10^{-8}$ & 2002 CODATA \\
$M_{\odot}$ (in g)    & $1.9884 \times 10^{33}$ & \citep{Cox2000}$^a$\\
$R_{\odot}$ (in cm)   & $6.9599 \times 10^{10}$ & \citep{Allen1973} \\
$L_{\odot}$ (in erg.s$^{-1}$) & $3.8422 \times 10^{33}$ & \\
$T_{\mathrm{eff}\,\odot}$ (in K)         & 5777  & \\
$\Delta \nu_{\odot}$ (in $\mu$Hz)        & 135.1 & \citep{Huber2009, Huber2011} \\
$\nu_{\mathrm{max}\,\odot}$ (in $\mu$Hz) & 3090  & \citep{Huber2009, Huber2011} \\
\hline
\end{tabular}
\tablefoot{Some of the above values are in fact outdated and do not represent the
most accurate value available.  They merely play the role of reference values in
this article. \\
$^a$ based on the ratio $GM_{\odot}/G$, using the above value of $G$ and the
value $GM_{\odot} = 1.32712440 \times 10^{26}$ cm$^3\,$s$^{-2}$ from
\citet{Cox2000} and references therein.}
\end{table}

In order to test the effects of different physical assumptions, a number of
models came in pairs and a triplet in which at least one of the properties was
modified.  These model groupings can easily be recognised in
Table~\ref{tab:fundamental_properties} since their members have the same masses.
Hence, Aardvark and Elvis differ according to age and boundary condition on the
pulsation modes.  The main difference between Henry and Izzy is that the latter
was calculated with atomic diffusion as prescribed in \citet{Thoul1994}
and has a slightly different metallicity.  Blofeld and Diva have different
abundance mixtures, slightly different metallicities, a different treatment of
the atmosphere, and different boundary conditions for the pulsation modes. 
Furthermore, Blofeld includes atomic diffusion whereas Diva does not.
In the last group of stars, Jam is significantly younger, and George has a
different overshoot parameter.

\subsection{Generating ``observational'' data}

From the above theoretical values, a set of ``observational'' data was produced
by incorporating noise.  These data included classical parameters, namely $\Teff$,
$L/\Lsun$ and [Fe/H], as well as seismic constraints, which included
individual frequencies as well as $\Delta\nu$, the average large frequency
separation, and $\numax$, the frequency of maximum oscillation power. These
observational data and associated error bars were then made available to the
hounds through a dedicated website\footnote{
\href{http://bison.ph.bham.ac.uk/hare\_and\_hounds1\_spaceinn/index.php/Main\_Page}
     {\tt http://bison.ph.bham.ac.uk/hare\_and\_hounds1\_spaceinn/}
\href{http://bison.ph.bham.ac.uk/hare\_and\_hounds1\_spaceinn/index.php/Main\_Page}
     {\tt index.php/Main\_Page}}
and are given in Table~\ref{tab:observational_properties} for the global
properties, and Tables~\ref{tab:freq_hares1} to~\ref{tab:freq_hares3} for the
individual frequencies.  A detailed description of how the error bars were
chosen and noise added is given in the sections that follow.

\subsubsection{Global parameters}

Estimates of the global or average seismic parameters $\numax$ and $\Delta\nu$
were provided as guideline data. We took the pristine values
($\numax$ was obtained using Eq.~(\ref{eq:numax}), and $\Delta\nu$ was 
calculated via a least squares fit to the model frequencies) and added
random Gaussian noise commensurate with the typical precision in those
quantities expected from the analysis of a shorter 1-month dataset
(\ie\ 5\,\% in $\numax$; 2\,\% in $\Delta\nu$).  We note that detailed
modelling was performed by the hounds using the individual frequencies, which
have a much higher information content than the above average/global seismic
parameters. The pristine effective temperatures and metallicities [Fe/H] were
perturbed by adding Gaussian deviates having standard deviations of 85\,K and
0.09\,dex (again, as per the assumed formal uncertainties). Finally, we assumed
a 3\,\% uncertainty on luminosities from \emph{Gaia} parallaxes, with most of
that due to uncertainty in the bolometric correction.

\subsubsection{Pulsation frequencies}

Each artificial star's fundamental properties ($T_{\rm eff}$, $M$ and $R$) were
used as input to scaling relations from which basic parameters of the
oscillation spectrum were calculated, and from there, the expected precision in
the frequencies.

The dominant frequency spacing of the oscillation spectra is the large
separation $\Delta\nu$.  In main-sequence stars, each $\Delta\nu$-wide segment
of the spectrum will contain significant power due to the visible $\l=0$, 1 and 2
modes, and, in the highest S/N observations, small contributions from modes of
$\l=3$. The integrated power in each segment will therefore correspond to the
power due to the radial mode, multiplied by the sum of the visibilities (in
power) over $\l$.

Let us define $A_{\rm max}$ to be the equivalent radial-mode amplitude at the
centre of the p-mode envelope, i.e., at $\nu_{\rm max}$, the frequency of
maximum oscillation power.  This frequency was calculated according to
\citep[\eg][]{Kjeldsen1995, Belkacem2011}:
  \begin{equation} 
  \nu_{\rm max} \simeq \left(\frac{M}{\Msun}\right)
  \left(\frac{R}{\Rsun}\right)^{-2} \left(\frac{T_{\rm
      eff}}{\rm T_{\rm eff\,\odot}}\right)^{-1/2}\, \nu_{\rm
    max\,\odot},
  \label{eq:numax}
  \end{equation}
with the commonly adopted solar values given in
Table~\ref{tab:reference_values}.

We also define the factor $\zeta$ to be the sum of the normalised mode
visibilities (in power), \ie,
 \begin{equation}
 \zeta = \sum_{\l} V_{\l}^2,
 \label{eq:vvis}
 \end{equation}
where $V_0=1$ (by definition). The visibilities for the non-radial modes are
largely determined by geometry. Here, we adopt values of $V_1=1.5$, $V_2=0.5$
and $V_3=0.03$ \citep[see][]{Ballot2011}.

If we re-bin the power spectrum into $\Delta\nu$-wide segments, the maximum
power spectral density in the segment at the centre of the spectrum will be
 \begin{equation}
 H_{\rm max} = \left(\frac{A_{\rm max}^2}{\Delta\nu}\right) \zeta.
 \label{eq:henv}
 \end{equation}
We used the scaling relations in \citet{Chaplin2011} to calculate $A_{\rm
max}$ and hence $H_{\rm max}$ for each artificial star using the fundamental
properties of the models as input and assuming observations were made with the
\emph{Kepler} instrumental response, which affects the mode amplitudes.
We note that the responses of CoRoT and PLATO are similar, while that of TESS is
redder, implying lower observed amplitudes.

A Gaussian in frequency provides a reasonable description of the shape of the
power envelope $H_{\rm env}(\nu)$ defined by the binned spectrum. We used the
scaling relations in \citet{Mosser2012} to calculate the full width at half
maximum (\textsc{fwhm}) of the Gaussian power envelope of each star, and hence
the power $H_{\rm env}(\nu)$ in the envelope as a function of frequency (that
power being normalised at the maximum of the envelope by $H_{\rm max}$).

Estimates of the frequency-dependent heights $H(\nu)$ shown by individual radial
modes were calculated according to \citep[\eg][]{Campante2014}:
 \begin{equation}
 H(\nu) = \left( \frac{2A(\nu)^2}{\pi \Gamma(\nu)} \right) 
   \equiv \left( \frac{2H_{\rm env}(\nu)\Delta\nu}{\pi \Gamma(\nu)\zeta} \right)
 \label{eq:h}
 \end{equation}
Here, $\Gamma(\nu)$ are the \textsc{fwhm} linewidths of the individual
oscillation peaks. Frequency dependent linewidth functions were fixed for each
star using Eq.~1 in \citet{Appourchaux2014} and Eq.~2 in
\citet{Appourchaux2012}.

Next, the background power spectral density, $B(\nu)$, across the frequency
range occupied by the modes is dominated by contributions from granulation and
shot noise. We used the scaling relations in \citet{Chaplin2011} to estimate the
frequency dependent background of the stars, assuming each was observed as a
bright \emph{Kepler} target having an apparent magnitude in the \emph{Kepler}
bandpass of $K_{\mathrm{p}} = 8$.  We note that PLATO 2.0 will also show similar noise
levels.  From there we could calculate a frequency dependent
background-to-height (radial-mode equivalent) ratio for each star, \ie,
 \begin{equation}
 \beta(\nu) = \frac{B(\nu)}{H(\nu)}.
 \label{eq:btoh}
 \end{equation}
The expected frequency precision in the radial modes is then given by
\citep{Libbrecht1992, Toutain1994, Chaplin2007}:
 \begin{equation}
 \sigma(\nu) = \left(\frac{\mathcal{F}[\beta(\nu)]\Gamma(\nu)}{4\pi T}\right)^{1/2},
 \label{eq:sig}
 \end{equation}
where we assumed continuous observations spanning $T=1\,\rm yr$, and the
function in $\beta(\nu)$ is defined according to:
 \begin{equation}
 \mathcal{F}[\beta(\nu)] = \sqrt{1+\beta(\nu)} 
             \left(\sqrt{1+\beta(\nu)}+\sqrt{\beta(\nu)}\right)^3
 \label{eq:fsig}
 \end{equation}

Estimation of the frequency precision in the non-radial modes depends not only
on the non-radial mode visibility -- which changes $\beta(\nu)$ relative to the
radial-mode case -- but also on the number of observed non-radial components and
their observed heights (which depends on the angle of inclination of the
star) and on how well resolved the individual components are (which depends on
the ratio between the frequency splitting and the peak linewidth). While
accounting for the change in $\beta(\nu)$ is trivial, correcting for the other
factors is somewhat more complicated \citep[\eg][]{Toutain1994, Chaplin2007}. We
could of course have simulated the actual observations, and applied our usual
analysis techniques to extract frequencies and uncertainties to pass on for
modelling. However, for this exercise we deliberately sought to avoid conflating
error or bias from the frequency extraction with error or bias from the
modelling. We therefore adopted an empirical correction factor $e_{\l}$ for each
angular degree, $\l$, based on results from fits to the oscillation spectra of
several tens of high-quality \emph{Kepler} targets. These factors may be
regarded as being representative; values for individual stars will vary,
depending on the specific combination of individual stellar and seismic
parameters and the inclination angle of the star.

If the model computed eigenfrequencies of the star are $\nu_{n\l}$, we therefore
estimated formal uncertainties on each frequency using:
 \begin{equation}
 \sigma_{n\l} = e_{\l}\,\sigma(\nu_{n\l}),
 \label{eq:sigl}
 \end{equation}
where $e_0=1.0$, $e_1=0.85$, $e_2=1.60$ and $e_3=6.25$.  This takes no account
of whether a mode would in principle be detectable in the observed spectrum.
Having first applied a coarse cut to the frequency list of a given star, by
removing frequencies having $\sigma_{n\l} > 5\,\rm \mu Hz$, we then ran mode
detection tests \citep{Chaplin2002, Chaplin2011} on the remaining frequencies
using their predicted background-to-height ratios
 \begin{equation}
 \beta_{n\l} = \beta(\nu_{n\l})/V_{\l}^2,
 \label{eq:betal}
 \end{equation}
and linewidths $\Gamma_{n\l} = \Gamma(\nu_{n\l})$ as input, each time assuming
$T=1\,\rm yr$. Only those modes that passed our tests were retained (a 1\,\%
false-alarm threshold), to give a final list of frequencies $\nu_{n\l}$ and
formal uncertainties $\sigma_{n\l}$ for each star. These selected frequencies
were perturbed by adding random Gaussian deviates having a standard deviation
equal to $\sigma_{n\l}$, to yield the ``observational'' frequencies that were
passed to the modellers.

\section{The hounds}

The ``hounds'' were subdivided into two main groups.  The first group applied a
forward modelling approach to find optimal models, stellar properties, and
associated error bars.  The second group fitted the acoustic glitch signatures
to characterise the base of the convection zone.  One of the hounds, KV and
HMA, applied both strategies and therefore appears in both groups.  Finally,
some other hounds applied inverse techniques as will be described below.

\subsection{Group 1: forward modelling approach}

The members in this group used various forward modelling strategies to find
optimal models which reproduce the observational data, including the detailed
seismic information provided.  From these models, they found various properties
of the star, namely mass, radius, density, $\log(g)$, and age.  Some of the
members of this group also provided the acoustic and physical radii and/or
depths of the base of the convection zone.  The main differences between the
strategies applied by the different hounds concern: 1) the search algorithm (or
optimisation procedure), 2) the stellar evolution codes along with the choice of
physics, and 3) the exact choice of observational constraints used.  This is
summarised in Table~\ref{tab:hounds_group1}.  A slightly more detailed
description is given in the following paragraphs.  We also note that a number of
the methods applied here have also been used in the KAGES project and are
consequently described in greater detail in \citet{SilvaAguirre2015}.

Several of the hounds (typically those who used frequencies as opposed to ratios)
included surface corrections on the model frequencies.  Typically, such corrections
are negative, \ie\ the model frequencies need to be decreased to match
the observations.  However, in a number of cases, the hounds had to increase
their model frequencies to match the provided frequencies.  This is perhaps
not entirely surprising since the ``observed'' frequencies come from models
and not from stars with true surface effects.

\begin{table*}[htbp]
\caption{Description of hounds in group 1 (forward modelling).
\label{tab:hounds_group1}}
\begin{tabular}{llllll}
\hline
\hline
\textbf{Method} &
\textbf{Participant(s)} &
\textbf{Optimisation procedure} &
\textbf{Evol. code} &
\textbf{Constraints} &
\textbf{References} \\
\hline
GOE        & WHB       & Grid search + Nelder-Mead    & MESA    & freq.           & \citet{Appourchaux2015} \\
YMCM       & SB        & Monte Carlo analysis         & YREC    & freq. \& ratios & \citet{SilvaAguirre2015} \\
ASTFIT     & JCD       & Scan evolutionary sequences  & ASTEC   & freq.           & \citet{SilvaAguirre2015} \\
YL         & YL        & Levenberg-Marquardt          & CESAM2k & freq.           & \citet{Lebreton2014} \\
AMP        & TM        & Genetic algorithm            & ASTEC   & freq. \& ratios & \citet{Metcalfe2009, Metcalfe2014} \\
BASTA      & VSA       & Bayesian grid scan           & GARSTEC & ratios          & \citet{SilvaAguirre2015} \\
MESAastero & DS        & Grid search + Nelder-Mead    & MESA    & freq.           & \citet{Paxton2013, Paxton2015} \\
V\&A, grid & KV \& HMA & Scan evolutionary sequences  & MESA    & freq.           & \cite{Verma2014} \\
\hline
\end{tabular}
\tablefoot{YMCM = Yale-Monte Carlo Method; ASTFIT = ASTEC Fitting method; AMP =
Asteroseismic Modeling Portal; BASTA = BAyesian STellar Algorithm; YREC = Yale
Rotating stellar Evolution Code \citep{Demarque2008}; ASTEC = Aarhus STellar
Evolution Code \citep{Christensen-Dalsgaard2008a}; CESAM = Code d'Evolution
Stellaire Adaptatif et Modulaire \citep{Morel2008}; MESA = Modules for
Experiments in Stellar Astrophysics \citep{Paxton2011, Paxton2013, Paxton2015};
GARSTEC = GARching STellar Evolution Code \citep{Weiss2008}.}
\end{table*}

\paragraph{GOE}

This approach involves several steps:
\begin{enumerate}
\item obtaining initial estimates of the model parameters and uncertainties
  using a pre-computed grid of models.  The classical and global seismic
  parameters are used when performing this step.
\item generating 10 initial guesses that populate these parameters within
  their uncertainties and with mixing-lengths sampled uniformly between 1.2 and
  2.4.  The best-fitting parameters are also included as an eleventh guess.
\item using MESA's built-in Nelder-Mead method (also known as a downhill simplex
  method or an amoeba method) to optimise the above 11 choices.  All of the
  classical constraints, namely $L$, $\Teff$, and [Fe/H], as well as individual
  frequencies including the 1-term version of the surface corrections from
  \citet{Ball2014} were used in this process.
\item gathering the above samples into one big sample.  The true global optimum is
  used as the best-fit, and uncertainties on the model parameters are derived
  from surfaces of constant $\chi^2$.  To boost numbers, points beyond the
  $\chi^2_{\mathrm{min}}+1$ surfaces were rescaled by the square root of their
  $\chi^2$ distance.
\item finding uncertainties on derived parameters by linearising about the
  best-fitting model, using $\exp(-\chi^2/2)$ as weights.
\end{enumerate}
The above models included overshoot based on \citet{Herwig2000} and
atomic diffusion as prescribed in \citet{Thoul1994}, even for massive
stars.

\paragraph{YMCM}

In the Yale-Monte Carlo Method (YMCM), a set of relevant models was calculated
for each of the stellar targets based on a Monte Carlo analysis.  Individual
frequencies including a scaled solar surface correction term were fitted to
observations, as were the classical constraints, including $L$.
Frequency ratios (based on uncorrected frequencies) were
subsequently used to check the results.  No diffusion or overshoot was used in
the models (which is expected to affect the base of the convection zone). 
Little evidence for a surface term was found.

\paragraph{ASTFIT}

In the ASTEC Fitting method (ASTFIT), grids of evolutionary tracks are used in
interpreting the data.  None of the models take atomic diffusion into
account, and convective-core overshoot is not included.  A fixed enrichment law
with $\Delta Y/\Delta Z = 1.4$ was used when constructing the grids.  Best
fitting models along the relevant tracks are found using homologous
transformations, and obtained by interpolation.  These models are selected
according to individual frequencies which are corrected for surface effects
using a scaled version of the solar surface term
(\citealt{Christensen-Dalsgaard2012} -- we note that the implementation of this
terms is slightly different to what is applied in the YMCM pipeline).  The
luminosity was not used when finding optimal models. Average results are
obtained from individual results weighted according to $\exp(-\chi^2/2)$, where
$\chi^2 = \chi^2_{\nu} + \chi^2_{\mathrm{spec}}$, the quantity $\chi_\nu^2$
being a \emph{reduced} $\chi^2$ based on the frequencies, and
$\chi^2_{\mathrm{spec}}$  a $\chi^2$ value on the classic observables (excluding
the luminosity).

\paragraph{YL}

A Levenberg-Marquardt approach was used to fit the seismic and classic
constraints, namely individual frequencies with surface corrections based on
\citet{Kjeldsen2008}, and $L$, $\Teff$, and [Fe/H].  Stellar models were
calculated on the fly and included atomic diffusion as prescribed in
\citet{Michaud1993}.  The effects of overshoot were tested in some of the more
``problematic'' stars, but the final list of results is based on models without
overshoot.  Such an approach provides both the best-fitting properties relevant
to the grid (namely mass, age, metallicity $Z/X$, helium abundance $Y$, and
mixing length) and the uncertainties on the properties. Other properties, such
as $R$, $\bar{\rho}$, $\log(g)$, $r_{\mathrm{BCZ}}/R$, are derived from the best
fitting models and consequently do not have error bars.

\paragraph{AMP}

The Asteroseismic Modeling Portal (AMP) searches the stellar parameter space
using a parallel genetic algorithm.  Stellar models and associated frequencies
are calculated on the fly in this approach.  These models include the effects of
helium diffusion \citep{Michaud1993}, but not overshoot.  The same
AMP configuration as was used in \citet{Metcalfe2014} was also applied here
-- the updated physics and fitting methods described in
\citet{Metcalfe2015} were not employed.  In
particular, individual frequencies \citep[including surface corrections based
on][]{Kjeldsen2008} and frequency ratios were simultaneously used when searching
for best-fitting models.  Likelihood-weighted mean values and associated
standard deviations are then obtained from the calculated models.  Such
properties are consistent with the properties of individual models identified by
the genetic algorithm.  It was noted that Felix and Diva were the least well
fitted.

\paragraph{BASTA}

The BAyesian STellar Algorithm (BASTA) consists in mapping out the posterior
probability distribution function by scanning a pre-computed grid of stellar
models. It uses $\Teff$, and [Fe/H] as constraints but not the luminosity. 
Given that the method relies on frequency ratios, frequency corrections for
surface effects were not used.  Some of the grids included atomic
diffusion as based on \citet{Thoul1994}, and some of them took overshooting into
account using an exponential decay on the convective velocities in the
overshooting region \citep[\cf][]{Weiss2008, SilvaAguirre2011}.  A fixed
enrichment law with $\Delta Y/\Delta Z = 1.4$ was used in the grids.  A number
of properties and associated, non-symmetric error bars are provided, but the
acoustic radii of the base of the convection zones had to be extracted from
best-fitting models.

\paragraph{MESAastero}

The MESAastero procedure is a two step process: the first step involves
generating grids of models in the vicinity of the different stellar targets. 
This provides good starting points for the second step, which is an automated
search based on a MESA's built-in Nelder-Mead method, where models are
calculated on the fly. The uncertainties were calculated as the average distance
of the set of points with $\chi^2 = \chi^2_{\mathrm{min}} + 1$, where
$\chi^2_{\mathrm{min}}$ is the $\chi^2$ value of the best-fitting
model. If the resulting $\chi^2$ landscape from the first simplex run did not
look reasonably well sampled (\eg\  it was single sided with respect to the
minimum), additional simplex runs were carried out using different starting
values.  This ensured a global minimum was found and robust uncertainties could
be derived.  The observational constraints used to find best fitting models were
$\Teff$, [Fe/H], and individual frequencies.  The surface correction recipe from
\citet{Kjeldsen2008} was included for Aardvark, Elvis, Henry, and Izzy.  The
models were constructed using [Fe/H], $M$, and $\alpha_{\mathrm{MLT}}$ as free
parameters, and used a fixed enrichment law with $\Delta Y/\Delta Z = 1.4$. 
They included an exponential prescription for overshoot based on
\citet{Herwig2000} but not diffusion.

\paragraph{V\&A, grid}

In the approach used here, 1000 evolutionary tracks with randomly selected model
properties in appropriate ranges were computed for each stellar target.  
One best model for each evolutionary track was obtained by fitting the
uncorrected model frequencies to the given frequencies. In this way, we get an
ensemble of models with different masses, initial compositions, mixing-lengths
and ages.  From this ensemble of models, a $\chi^2$ map was
calculated, thereby yielding best-fitting properties and associated error bars. 
The $\chi^2$ values were based on $L$, $\Teff$, surface metallicity, and the
average large and small frequency separations (as opposed to individual
frequencies).  The models included diffusion of He and heavy elements
\citep{Thoul1994}, except for higher-mass targets (namely Diva, Felix, George,
and Jam).  However, they did not include overshoot.

\subsection{Group 2: glitch fitting analysis}
\label{sect:glitch_group}

The second group of hounds fitted the acoustic glitch signatures in order to
obtain the acoustic depths of the base of the convection zone as well as that of
the \HeII\ ionisation zone or the $\Gamma_1$ peak nearby.  This type of method
relies on the fact that sharp features in the acoustic structure of the star,
such as the transition from a radiative to a convective zone or the presence of
an ionisation zone, lead to an oscillatory pattern in the frequency spectrum. 
The period of this pattern gives the acoustic depth of the feature whereas the
amplitude and rate of decrease with frequency is related to the
amplitude of the feature as well as to its ``sharpness'', \ie\  to whether the
feature corresponds to a discontinuity on the first, second or a higher
derivative of the acoustic profile \citep[\eg][]{Monteiro1994}.

Fitting acoustic glitch signatures differs from the forward modelling approach
in the sense that it focuses on very specific information contained within the
pulsation spectrum, rather than trying to fit the spectrum as a whole. 
Furthermore, it does so directly without making comparisons with theoretical
predictions from models (except for interpreting the amplitude of the feature in
terms of He abundance -- see, \eg, \citealt{Verma2014}), thereby making the
results model-independent.  In contrast, the forward modelling approach is
indirect and model-dependent since it ends up implicitly comparing glitch
signatures present in the observations to those obtained in theoretical models.
Furthermore, these features may be drowned out by other features
present in the pulsation spectra.

The acoustic depths obtained by these hounds were subsequently converted to
acoustic radii using the total acoustic radii.  The associated error bars were
calculated as the sum of the error bars on the acoustic depths and those coming
from the total acoustic radii.  In keeping with the approach taken in
the glitch fitting analysis, the total acoustic radii were deduced from the
large frequency separations.  However, rather than using the coarse values
provided, the latter were recalculated via a least-squares fit to all of the
observed frequencies, using the provided uncertainties to find
appropriate weights.  These values, $\Delta\nu_{\mathrm{recalc.}}$, along with
the uncertainties deduced from the least-squares fit (which only keeps
track of how the uncertainties on the frequencies propagate to the
final result but does not take into account how well the frequencies fit a
linear trend), and the resultant acoustic radii are listed in
Table~\ref{tab:tau}.  A comparison between
Table~\ref{tab:observational_properties} and Table~\ref{tab:tau} confirms the
improved accuracy of $\Delta\nu_{\mathrm{recalc.}}$.

\begin{table*}[htbp]
\caption{Comparison between different evaluations of the acoustic radius.
\label{tab:tau}}
\begin{tabular}{lrrrc}
\hline
\hline
\textbf{Target} & $\Delta\nu_{\mathrm{recalc.}}$ &
$\frac{1}{2\Delta\nu}$ &
$\tau_{\mathrm{Tot.}}=\int_0^{R_{\star}} \frac{\mathrm{d}r}{c}$ & $R_{\star}/R$ \\
& (in $\mu$Hz) & (in s) & (in s) & \\
\hline
Aardvark & $144.698 \pm 0.009$ & $3455.5 \pm 0.2$ & 3370.5 & 1.0006128 \\
Blofeld  &  $94.271 \pm 0.008$ & $5303.9 \pm 0.5$ & 5106.8 & 1.0000000 \\
Coco     & $162.474 \pm 0.013$ & $3077.4 \pm 0.2$ & 2998.7 & 1.0007149 \\
Diva     &  $96.023 \pm 0.009$ & $5207.1 \pm 0.5$ & 5082.4 & 1.0007380 \\
Elvis    & $120.231 \pm 0.009$ & $4158.6 \pm 0.3$ & 4045.3 & 1.0007033 \\
Felix    &  $69.559 \pm 0.015$ & $7188.2 \pm 1.5$ & 7014.1 & 1.0009181 \\
George   &  $70.466 \pm 0.022$ & $7095.6 \pm 2.2$ & 6930.2 & 1.0009409 \\
Henry    & $116.698 \pm 0.022$ & $4284.6 \pm 0.8$ & 4199.0 & 1.0008149 \\
Izzy     & $116.095 \pm 0.022$ & $4306.8 \pm 0.8$ & 4219.4 & 1.0007725 \\
Jam      &  $86.605 \pm 0.034$ & $5773.3 \pm 2.2$ & 5666.2 & 1.0008475 \\
\hline
\end{tabular}
\tablefoot{The smallest difference between $\tau_{\mathrm{Tot.}}$ and
$\frac{1}{2\Delta\nu}$ is 78.6 s for Coco, whereas the largest difference is
197.1 s for Blofeld.  The last column gives $R_{\star}$, the upper
integration bound used when calculating $\tau_{\mathrm{Tot.}}$.  This radius
corresponded an optical depth of $\tau_{\mathrm{opt.}}=10^{-3}$ (the last mesh
point in the stellar targets), except for Blofeld where the atmosphere was
truncated at the photospheric radius.}
\end{table*}

It is important to bear in mind that significant discrepancies can appear when
calculating the acoustic radii of stars.  This is illustrated by the differences
between the third and fourth columns in Table~\ref{tab:tau}, which
contain two different calculations of the acoustic radius.  These differences
likely stem from the fact that $\tau_{\mathrm{Tot.}}$ represents an asymptotic
value whereas $1/2\Delta\nu$ is based on modes of finite radial order. 
Another source of error includes differences in the exact definition of the
radius used as an upper integration bound in the definition of
$\tau_{\mathrm{Tot.}}$ \citep[\eg][]{Hekker2013}.  Such differences will also
affect the values given for the acoustic depth of the base of the convection
zone.  In order to derive the acoustic radius in a physically sound way, one
would need to linearly extrapolate the squared sound speed, $c^2$ from the
outer regions of the adiabatically-stratified portion of the convection zone to
the place where $c^2$ would vanish, and integrate $\mathrm{d}r/c$ to this point
\citep{Houdek2007}. In what follows, we take a more pragmatic approach which
consists in comparing the acoustic radii rather than the depths of the base of
the convection zone.  Indeed, as pointed out in \citet{Ballot2004}, this
approach mostly cancels out any differences in the precise definition of the
stellar radius used in calculating the acoustic depth.

Table~\ref{tab:hounds_group2} lists the specific frequency combinations which
were used in finding glitch signatures.  Relevant references are also provided. 
The following paragraphs then give a few more details on the methodologies of
the various hounds from this group.

\begin{table*}[htbp]
\caption{Description of hounds in group 2 (glitch fitting analysis).
\label{tab:hounds_group2}}
\begin{tabular}{llll}
\hline
\hline
\textbf{Participant(s)} &
\textbf{Seismic signature} &
\textbf{Error bars} &
\textbf{References} \\
\hline
V\&A, glitch & individual frequencies & Monte Carlo analysis & \cite{Verma2014} \\
HRC & second frequency differences & MCMC & Coelho et al. (in prep.) \\
GH & second frequency differences & Monte Carlo analysis & \citet{Houdek2007, Houdek2011} \\
AM & second frequency differences & Monte Carlo analysis & \citet{Mazumdar2012, Mazumdar2014} \\
\hline
\end{tabular}
\end{table*}

\paragraph{V\&A, glitch}
The approach taken here is method C of \citet{Verma2014}.  In this method, both
the smooth and oscillatory glitch-related components are fitted simultaneously
directly to the frequencies.  A Monte Carlo approach was used  for obtaining the
uncertainties on the glitch parameters.  This involved constructing histograms
of the parameters deduced from multiple realisations of the data.

\paragraph{HRC}
applied an MCMC approach to fitting the second frequency differences, thereby
obtaining the posterior probability distribution function of the glitch
parameters.  This allowed him to obtain optimal values along with their
associated error bars.

\paragraph{GH}
fitted glitch signatures to second frequency differences.  Specifically, this
analysis includes both ionisation stages of helium (unlike the methods
from the other glitch-fitting hounds which only include a single signature for
both ionisation stages), and adopts Airy functions and a polytropic
representation of the acoustic potential in the stellar surface layers to
account more realistically for the contribution from the \HeI\ glitch which, for
stars with surface temperatures similar to the Sun, lies partially in the
evanescent region of acoustic modes.  This leads to deeper (larger) acoustic
depths, since these are measured relative to the acoustic radius determined from
linearly extrapolating $c^2$ to the place where it vanishes (as describe above)
as opposed to the location indicated by $1/2\Delta\nu$.  Accordingly, the
acoustic radii reported throughout the article for this method will be
underestimated due to the use of $1/2\Delta\nu$ in the conversion from acoustic
depths to acoustic radii.  For the parameters, which relate the fitting
coefficients of the \HeI\ glitch to those of the \HeII\ glitch, the constant
solar values of Houdek \& Gough (2007) were adopted for all model fits.  Finally,
the error bars were deduced from a Monte Carlo analysis, much like what was done
by V\&A, glitch.

\paragraph{AM}\footnote{This stands for A. Mazumdar and not A. Miglio.}
also fitted the second frequency differences.  Once more, the error bars were
deduced from a Monte Carlo analysis.

\subsection{Others: inversion techniques}

Besides these two main groups of hounds, GB and DRR applied the mean
density inversions described in \citet{Reese2012}, but had different strategies
for selecting the reference models.  For each target, GB selected a reference
model via the Levenberg-Marquardt algorithm, using the average large and small
frequency separations, $r_{01}$ frequency ratios, and $\Teff$ as constraints
\citep[\eg][]{Buldgen2015b}.  Accordingly, the reported uncertainties
only take into account how the observational uncertainties on the
frequencies propagated through the inversion process onto the mean densities.
DRR used an inversion pipeline to select reference models from a grid using
$\log(L)$, $\Teff$, $\numax$, $\Delta \nu$ as constraints.  For the latter
parameters, the coarse values provided with the data were used, and $\Delta\nu$
was not recalculated.  The constraint on [Fe/H] was discarded as it could lead
to some of the targets having no reference models.  Inversion results from each
reference model were combined after being weighted by the $\chi^2$ value
associated with the constraints used to select the reference models. 
Accordingly, the error bars take into account the observational error propagated
through the inversion procedures and the scatter between the results from the
different models.

\section{The results}
\label{sect:results}

In this section, we compare the results obtained by the various hounds with the
actual properties of the artificial stars.  We start by introducing various
average error and bias measurements which will help assess the quality of the
results and reported uncertainties.  This is followed by a comparison of the
results obtained for global properties, before we focus on the properties
related to the base of the convection zone and the \HeII\ ionisation zone.

\subsection{Average errors and biases}

In order to summarise the quality of the results it is helpful to introduce the
following average error measurements: 
\begin{eqnarray}
\label{eq:error_relative}
\erel  &=& \sqrt{\frac{1}{N} \sum_{i=1}^N \left( \frac{p^{\mathrm{fit}}_i -
                 p^{\mathrm{exact}}}{p^{\mathrm{exact}}} \right)^2}, \\
\label{eq:error_normalised}
\enorm &=& \sqrt{\frac{1}{N} \sum_{i=1}^N \left( \frac{p^{\mathrm{fit}}_i -
                 p^{\mathrm{exact}}}{\sigma^{\mathrm{fit}}_i} \right)^2},
\end{eqnarray}
where $p$ is a given property, the superscripts ``exact'' and ``fit'' refer to
the exact and fitted values, $\sigma$ the estimated error bar (or the average
if the error bar is not symmetric), $N$ the number of relevant cases, and $i$
each particular case.  In what follows, these errors will be averaged
over:
\begin{itemize}
\item particular stars
\item particular hounds
\item over all the stars and hounds.
\end{itemize}
The error from Eq.~(\ref{eq:error_relative}), which we will call the ``average
relative error'', gives a measure of the relative accuracy with which a
particular parameter is determined, whereas the error from
Eq.~(\ref{eq:error_normalised}), the ``average normalised error'', is used to
see how realistic the reported error bars are.  Large values indicate that the
error bars are underestimated, small values mean the error bars are
overestimated, and values close to unity correspond to well-estimated error
bars. In cases, where a particular value is not provided, it is excluded from
the average relative error.  If an error bar is not provided, the associated
value is excluded from the average normalised error.

In addition, we also define biases, which come in the same two flavours as
above:
\begin{eqnarray}
\brel  &=& \frac{1}{N} \sum_{i=1}^N \left( \frac{p^{\mathrm{fit}}_i -
               p^{\mathrm{exact}}}{p^{\mathrm{exact}}} \right), \\
\bnorm &=& \frac{1}{N} \sum_{i=1}^N \left( \frac{p^{\mathrm{fit}}_i -
               p^{\mathrm{exact}}}{\sigma^{\mathrm{fit}}_i} \right).
\end{eqnarray}
These are useful for detecting a systematic offset between fitted results and
the true values.  We note in passing that the scatter, $\sigma$, of the results
around the bias is given by the formula:
\begin{equation}
\sigma_{j}^2 = \varepsilon_j^2 - b_j^2,
\end{equation}
where ``$j$'' could stand for ``$\mathrm{rel.}$'' or ``$\mathrm{norm.}$''.

\subsection{Global properties}

The most important global properties are radius, $R$, mass, $M$, and age, $t$.
Indeed, these are key properties in stellar evolution and have a direct impact
on the study of exoplanetary systems as well as that of galactic stellar
populations.  We also decided to include two other properties, namely the mean
density, $\bar{\rho}$, and $\log(g)$, $g$ being the surface gravity. Although it is
straightforward to derive these properties from $M$ and $R$, their error bars
cannot straightforwardly be deduced from the error bars on $M$ and $R$ alone,
given the correlations between these two quantities.

\begin{figure*}[htbp]
\includegraphics[width=\textwidth]{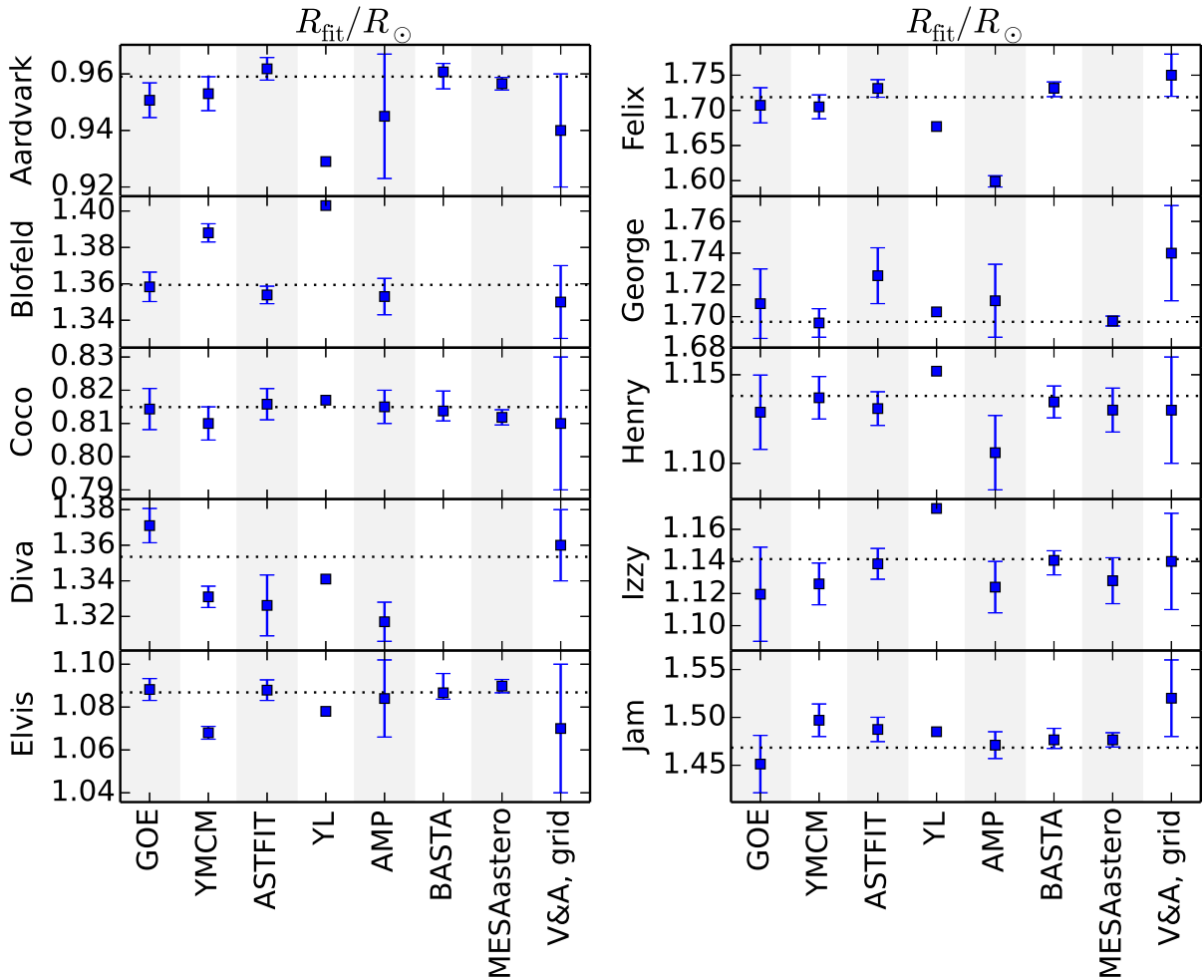}
\caption{Fitted results for the radius.  Each panel corresponds to a stellar
target, the ``columns'' in each plot to particular hounds, and the horizontal
dotted line to the true value. \label{fig:radius}}
\end{figure*}

\begin{figure*}[htbp]
\includegraphics[width=\textwidth]{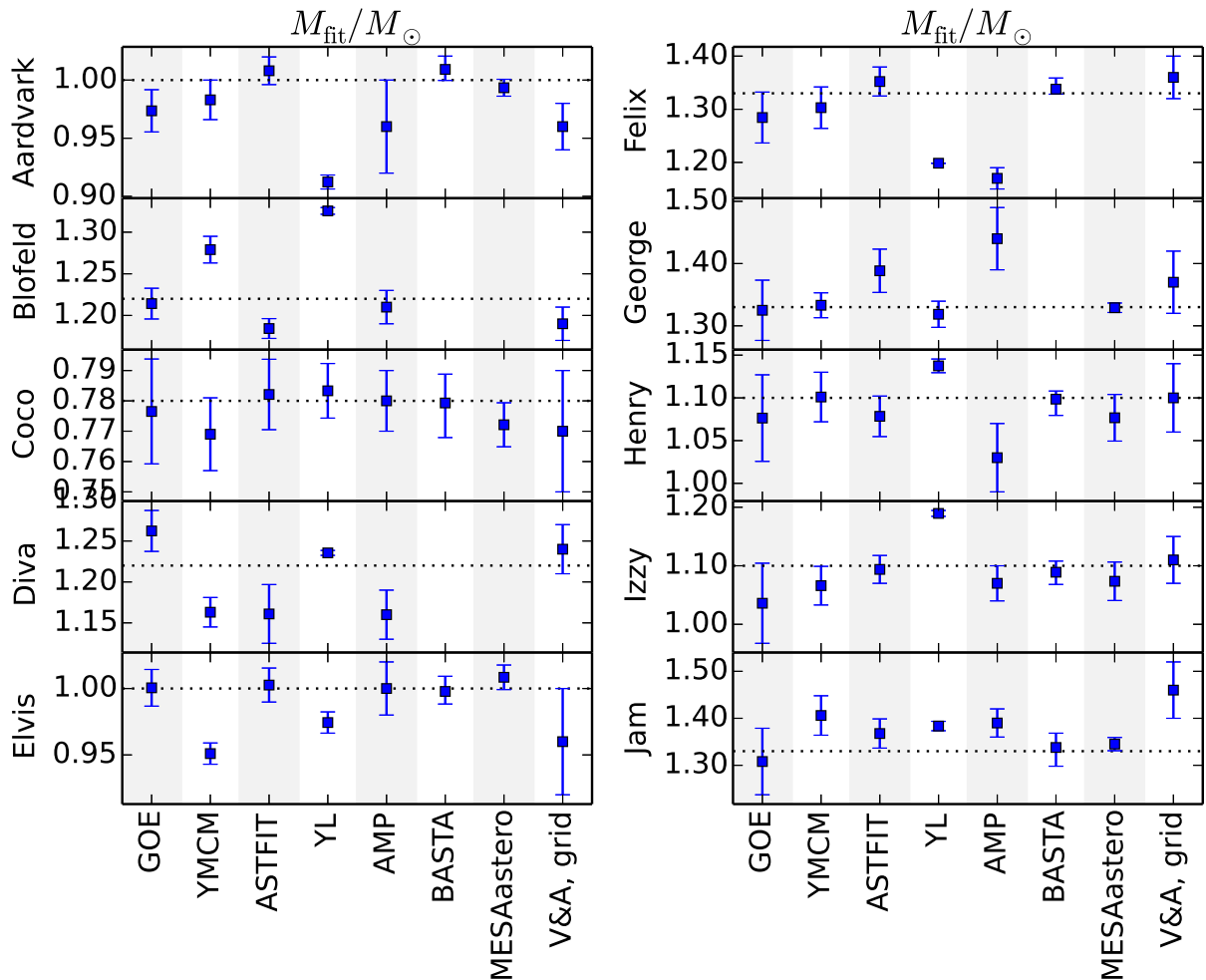}
\caption{Fitted results for the mass. \label{fig:mass}}
\end{figure*}

\begin{figure*}[htbp]
\includegraphics[width=\textwidth]{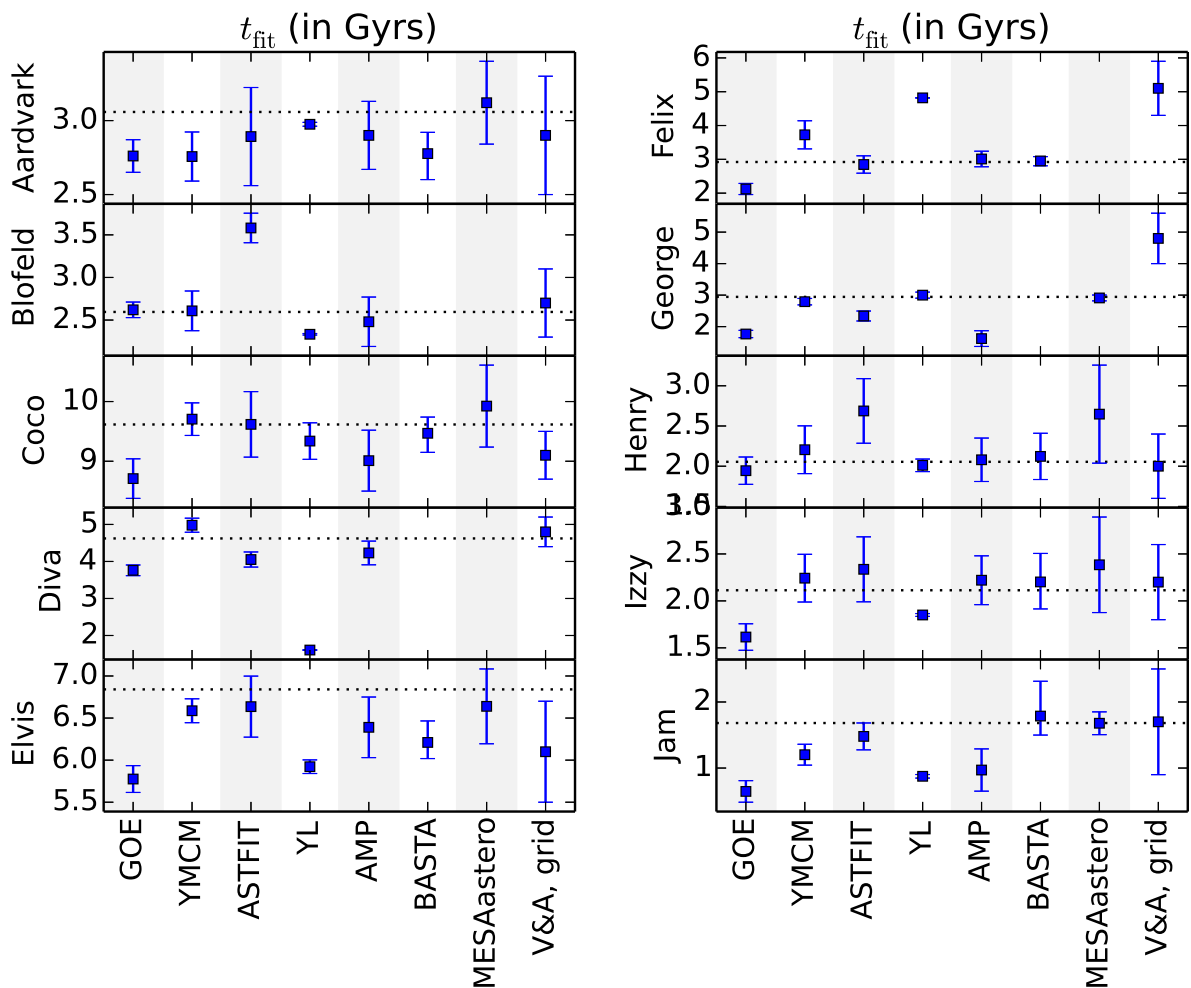}
\caption{Fitted results for the age. \label{fig:age}}
\end{figure*}

\begin{figure*}[htbp]
\includegraphics[width=\textwidth]{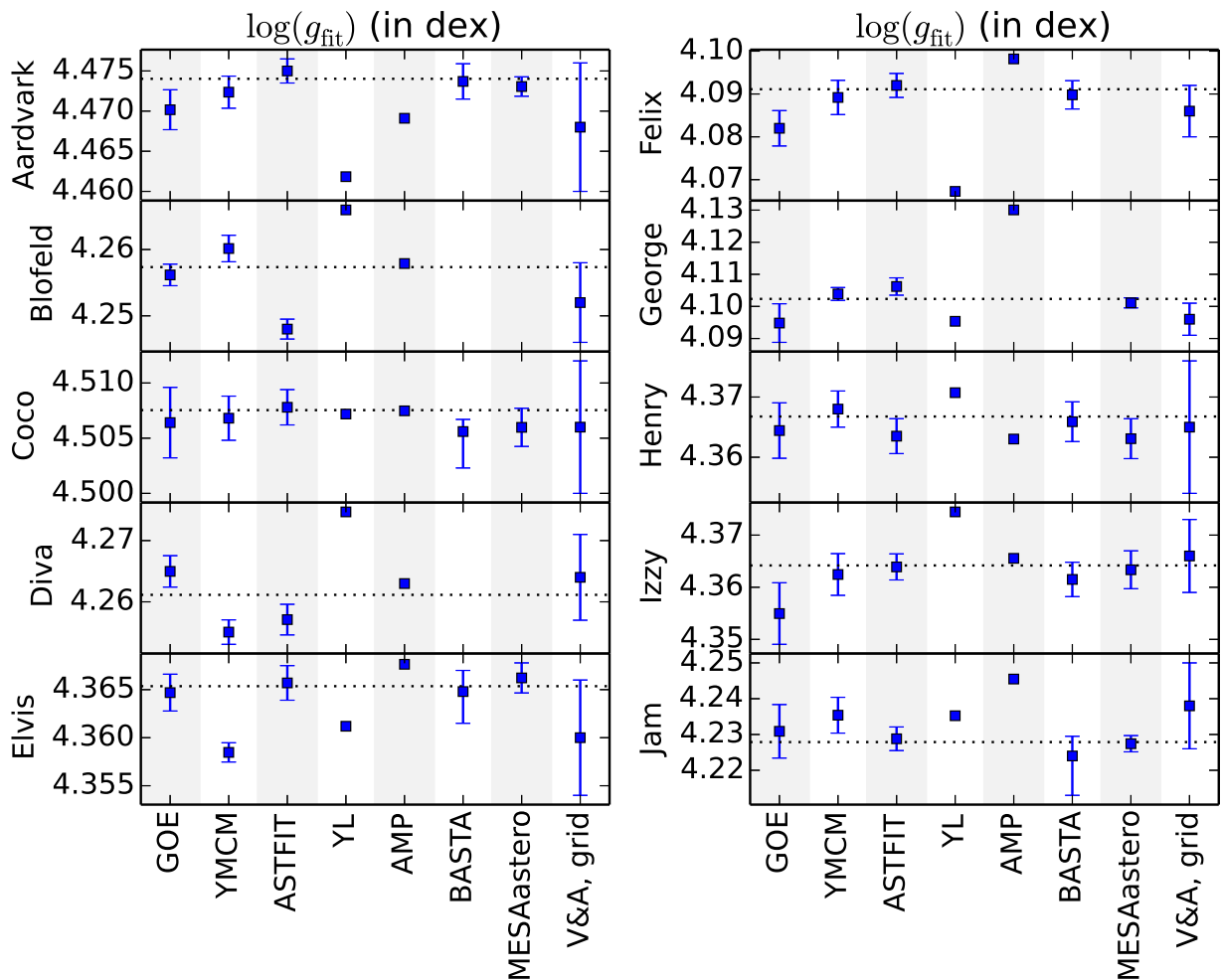}
\caption{Fitted results for $\log(g)$.  Different symbols and colours
represent different techniques for obtaining the result.  \label{fig:logg}}
\end{figure*}

\begin{figure*}[htbp]
\includegraphics[width=\textwidth]{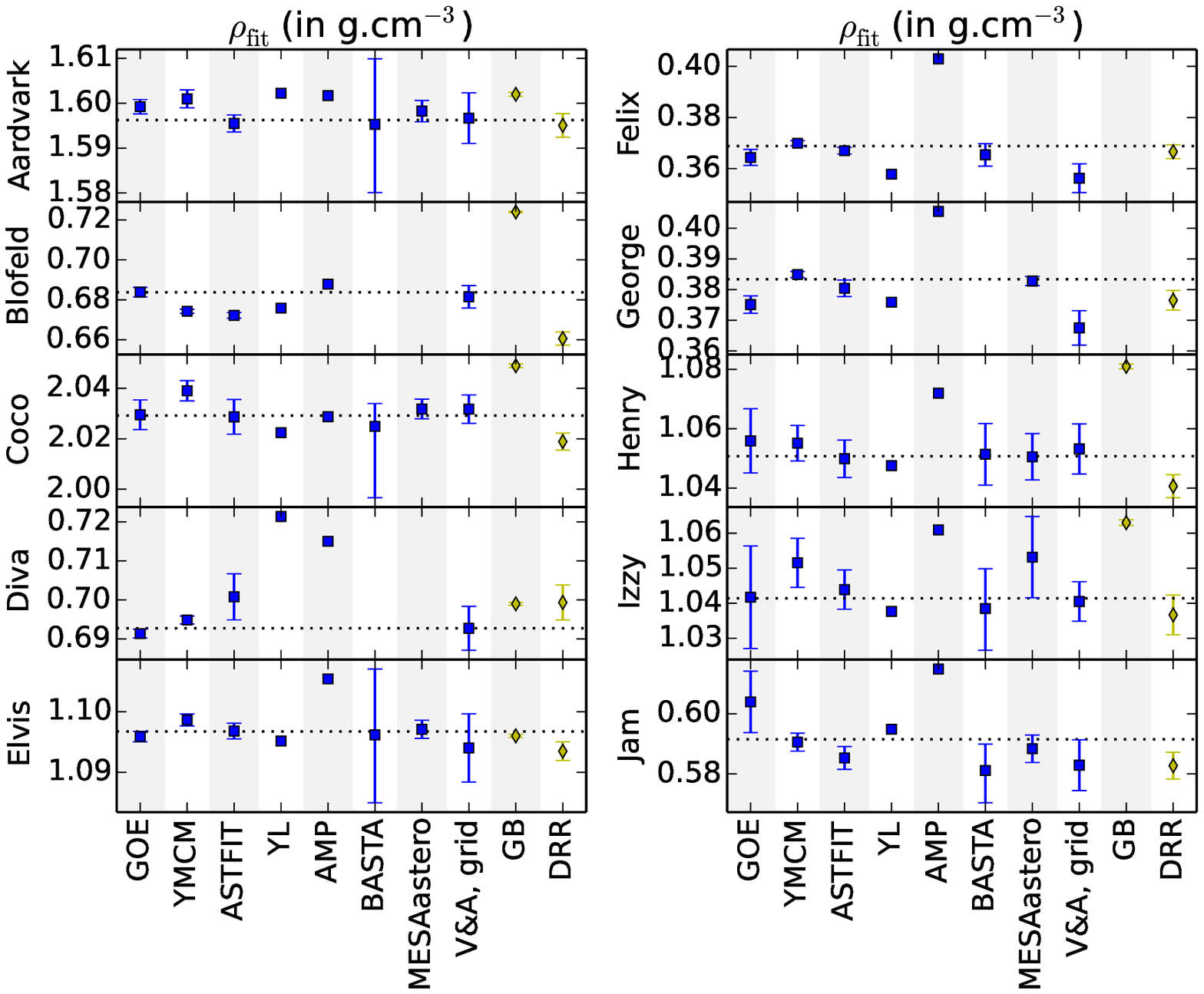}
\caption{Fitted results for $\bar{\rho}$.\label{fig:rho}}
\end{figure*}

Tables~\ref{tab:radius} to~\ref{tab:rho} list the results from the various hounds
as well as the associated average errors and biases.  Figures~\ref{fig:radius}
to~\ref{fig:rho} illustrate these results.

\begin{table*}[htbp]
\caption{Fitted values for the radius in solar units, $R$, and
associated average errors and biases. \label{tab:radius}}
\begin{tabular}{lcccccccc}
\hhline{=========}
\multicolumn{9}{c}{\textbf{Radius}} \\
\hhline{------~--}
\textbf{Hounds} & \textbf{Aardvark} & \textbf{Blofeld} & \textbf{Coco} & \textbf{Diva} & \textbf{Elvis} & & $\erel$ & $\brel$ \\
\hhline{------~--}
\SolutionCell{Solution} & $0.959$ & $1.359$ & $0.815$ & $1.353$ & $1.087$  & & -- & -- \\
\GridCell{GOE}& $0.951 \pm 0.006$ & $1.358 \pm 0.008$ & $0.814 \pm 0.006$ & $1.371 \pm 0.010$ & $1.088 \pm 0.005$  & & $0.95\,\%$  & $-0.35\,\%$  \\
\GridCell{YMCM}& $0.953 \pm 0.006$ & $1.388 \pm 0.005$ & $0.810 \pm 0.005$ & $1.331 \pm 0.006$ & $1.068 \pm 0.003$  & & $1.31\,\%$  & $-0.29\,\%$  \\
\GridCell{ASTFIT}& $0.962 \pm 0.004$ & $1.354 \pm 0.005$ & $0.816 \pm 0.005$ & $1.326 \pm 0.017$ & $1.088 \pm 0.005$  & & $1.00\,\%$  & $0.09\,\%$  \\
\GridCell{YL}& $0.929$ & $1.403$ & $0.817$ & $1.341$ & $1.078$  & & $1.95\,\%$  & $0.17\,\%$  \\
\GridCell{AMP}& $0.945 \pm 0.022$ & $1.353 \pm 0.010$ & $0.815 \pm 0.005$ & $1.317 \pm 0.011$ & $1.084 \pm 0.018$  & & $2.63\,\%$  & $-1.52\,\%$  \\
\GridCell{BASTA}& $0.961_{-0.006}^{+0.003}$  & -- & $0.814_{-0.003}^{+0.006}$  & -- & $1.087_{-0.003}^{+0.009}$  & & $0.38\,\%$  & $0.13\,\%$  \\
\GridCell{MESAastero}& $0.957 \pm 0.002$  & -- & $0.812 \pm 0.002$  & -- & $1.090 \pm 0.003$  & & $0.60\,\%$  & $-0.24\,\%$  \\
\GridCell{V\&A, grid}& $0.940 \pm 0.020$ & $1.350 \pm 0.020$ & $0.810 \pm 0.020$ & $1.360 \pm 0.020$ & $1.070 \pm 0.030$  & & $1.74\,\%$  & $0.27\,\%$  \\
$\erel$ & $1.47\,\%$ & $1.61\,\%$ & $0.35\,\%$ & $1.68\,\%$ & $0.88\,\%$ & & $1.55\,\%$ & -- \\
$\brel$ & $-0.99\,\%$ & $0.61\,\%$ & $-0.18\,\%$ & $-0.92\,\%$ & $-0.49\,\%$ & & -- & $-0.23\,\%$ \\
$\enorm$ & $0.93$ & $2.63$ & $0.65$ & $2.49$ & $2.42$ & & -- & -- \\
$\bnorm$ & $-0.57$ & $0.66$ & $-0.39$ & $-1.30$ & $-0.80$ & & -- & -- \\
\hhline{------~--}
\textbf{Hounds} & \textbf{Felix} & \textbf{George} & \textbf{Henry} & \textbf{Izzy} & \textbf{Jam} & & $\enorm$ & $\bnorm$ \\
\hhline{------~--}
\SolutionCell{Solution} & $1.719$ & $1.697$ & $1.138$ & $1.141$ & $1.468$  & & -- & -- \\
\GridCell{GOE}& $1.707 \pm 0.025$ & $1.708 \pm 0.022$ & $1.129 \pm 0.021$ & $1.120 \pm 0.029$ & $1.451 \pm 0.030$  & & $0.83$ & $-0.12$ \\
\GridCell{YMCM}& $1.705 \pm 0.017$ & $1.696 \pm 0.009$ & $1.137 \pm 0.012$ & $1.126 \pm 0.013$ & $1.497 \pm 0.017$  & & $3.05$ & $-0.68$ \\
\GridCell{ASTFIT}& $1.731 \pm 0.013$ & $1.726 \pm 0.018$ & $1.131 \pm 0.009$ & $1.139 \pm 0.010$ & $1.487 \pm 0.013$  & & $1.05$ & $0.14$ \\
\GridCell{YL}& $1.677$ & $1.703$ & $1.152$ & $1.173$ & $1.485$  & & --  & --  \\
\GridCell{AMP}& $1.599 \pm 0.008$ & $1.710 \pm 0.023$ & $1.106 \pm 0.021$ & $1.124 \pm 0.016$ & $1.471 \pm 0.014$  & & $4.90$ & $-2.16$ \\
\GridCell{BASTA}& $1.731_{-0.012}^{+0.009}$  & -- & $1.135 \pm 0.009$ & $1.141_{-0.009}^{+0.006}$ & $1.477_{-0.009}^{+0.012}$  & & $0.59$ & $0.23$ \\
\GridCell{MESAastero} & -- & $1.697 \pm 0.003$ & $1.130 \pm 0.012$ & $1.128 \pm 0.014$ & $1.477 \pm 0.007$  & & $0.96$ & $-0.27$ \\
\GridCell{V\&A, grid}& $1.750 \pm 0.030$ & $1.740 \pm 0.030$ & $1.130 \pm 0.030$ & $1.140 \pm 0.030$ & $1.520 \pm 0.040$  & & $0.81$ & $0.16$ \\
$\erel$ & $2.93\,\%$ & $1.23\,\%$ & $1.20\,\%$ & $1.46\,\%$ & $1.63\,\%$ & & -- & -- \\
$\brel$ & $-1.08\,\%$ & $0.87\,\%$ & $-0.60\,\%$ & $-0.46\,\%$ & $1.00\,\%$ & & -- & -- \\
$\enorm$ & $6.17$ & $0.95$ & $0.73$ & $0.77$ & $1.13$ & & $2.39$ & -- \\
$\bnorm$ & $-2.17$ & $0.71$ & $-0.59$ & $-0.63$ & $0.85$ & & -- & $-0.42$ \\
\hhline{------~--}
\end{tabular}

\tablefoot{The last two columns contain errors and biases for
individual hounds, averaged over the 10 stellar targets (and not over 5
as the layout of the table may suggest).  These have been slightly offset from
the table to make this point clearer.  The last four rows in each half of the
table contain errors and biases for individuals stars, averaged
over the relevant hounds.  At the intersection between the two, overall
averages have been included in logical places.}
\end{table*}

\begin{table*}[htbp]
\caption{Fitted values for the mass in solar units,
$M$, and associated average errors and biases. \label{tab:mass}}
\begin{tabular}{lcccccccc}
\hhline{=========}
\multicolumn{9}{c}{\textbf{Mass}} \\
\hhline{------~--}
\textbf{Hounds} & \textbf{Aardvark} & \textbf{Blofeld} & \textbf{Coco} & \textbf{Diva} & \textbf{Elvis} & & $\erel$ & $\brel$ \\
\hhline{------~--}
\SolutionCell{Solution} & $1.000$ & $1.220$ & $0.780$ & $1.220$ & $1.000$  & & -- & -- \\
\GridCell{GOE}& $0.974 \pm 0.018$ & $1.214 \pm 0.018$ & $0.777 \pm 0.017$ & $1.262 \pm 0.025$ & $1.001 \pm 0.014$  & & $2.69\,\%$  & $-1.35\,\%$  \\
\GridCell{YMCM}& $0.983 \pm 0.017$ & $1.279 \pm 0.016$ & $0.769 \pm 0.012$ & $1.163 \pm 0.018$ & $0.951 \pm 0.008$  & & $3.47\,\%$  & $-0.69\,\%$  \\
\GridCell{ASTFIT}& $1.008 \pm 0.012$ & $1.184 \pm 0.012$ & $0.782 \pm 0.012$ & $1.161 \pm 0.036$ & $1.003 \pm 0.013$  & & $2.59\,\%$  & $-0.01\,\%$  \\
\GridCell{YL}& $0.912 \pm 0.006$ & $1.326 \pm 0.004$ & $0.783 \pm 0.009$ & $1.235 \pm 0.003$ & $0.974 \pm 0.008$  & & $5.94\,\%$  & $0.38\,\%$  \\
\GridCell{AMP}& $0.960 \pm 0.040$ & $1.210 \pm 0.020$ & $0.780 \pm 0.010$ & $1.160 \pm 0.030$ & $1.000 \pm 0.020$  & & $5.68\,\%$  & $-1.81\,\%$  \\
\GridCell{BASTA}& $1.009_{-0.009}^{+0.011}$  & -- & $0.779_{-0.011}^{+0.009}$  & -- & $0.998_{-0.009}^{+0.011}$  & & $0.61\,\%$  & $0.09\,\%$  \\
\GridCell{MESAastero}& $0.993 \pm 0.007$  & -- & $0.772 \pm 0.007$  & -- & $1.008 \pm 0.009$  & & $1.40\,\%$  & $-0.61\,\%$  \\
\GridCell{V\&A, grid}& $0.960 \pm 0.020$ & $1.190 \pm 0.020$ & $0.770 \pm 0.020$ & $1.240 \pm 0.030$ & $0.960 \pm 0.040$  & & $3.91\,\%$  & $0.58\,\%$  \\
$\erel$ & $3.88\,\%$ & $4.35\,\%$ & $0.80\,\%$ & $3.78\,\%$ & $2.43\,\%$ & & $3.86\,\%$ & -- \\
$\brel$ & $-2.51\,\%$ & $1.13\,\%$ & $-0.44\,\%$ & $-1.34\,\%$ & $-1.32\,\%$ & & -- & $-0.44\,\%$ \\
$\enorm$ & $5.28$ & $10.96$ & $0.56$ & $2.79$ & $2.49$ & & -- & -- \\
$\bnorm$ & $-2.43$ & $4.12$ & $-0.28$ & $0.12$ & $-1.17$ & & -- & -- \\
\hhline{------~--}
\textbf{Hounds} & \textbf{Felix} & \textbf{George} & \textbf{Henry} & \textbf{Izzy} & \textbf{Jam} & & $\enorm$ & $\bnorm$ \\
\hhline{------~--}
\SolutionCell{Solution} & $1.330$ & $1.330$ & $1.100$ & $1.100$ & $1.330$  & & -- & -- \\
\GridCell{GOE}& $1.284 \pm 0.048$ & $1.325 \pm 0.048$ & $1.076 \pm 0.051$ & $1.036 \pm 0.068$ & $1.308 \pm 0.071$  & & $0.85$ & $-0.30$ \\
\GridCell{YMCM}& $1.303 \pm 0.039$ & $1.333 \pm 0.020$ & $1.101 \pm 0.029$ & $1.066 \pm 0.033$ & $1.406 \pm 0.042$  & & $2.60$ & $-0.72$ \\
\GridCell{ASTFIT}& $1.352 \pm 0.027$ & $1.388 \pm 0.035$ & $1.078 \pm 0.024$ & $1.094 \pm 0.024$ & $1.368 \pm 0.031$  & & $1.34$ & $-0.11$ \\
\GridCell{YL}& $1.198 \pm 0.000$ & $1.319 \pm 0.021$ & $1.137 \pm 0.008$ & $1.189 \pm 0.005$ & $1.384 \pm 0.010$  & & $12.09$ & $4.61$ \\
\GridCell{AMP}& $1.170 \pm 0.020$ & $1.440 \pm 0.050$ & $1.030 \pm 0.040$ & $1.070 \pm 0.030$ & $1.390 \pm 0.030$  & & $2.87$ & $-1.01$ \\
\GridCell{BASTA}& $1.338_{-0.009}^{+0.021}$  & -- & $1.099_{-0.019}^{+0.009}$ & $1.089_{-0.021}^{+0.019}$ & $1.338_{-0.040}^{+0.030}$  & & $0.46$ & $0.10$ \\
\GridCell{MESAastero} & -- & $1.329 \pm 0.008$ & $1.077 \pm 0.027$ & $1.074 \pm 0.033$ & $1.345 \pm 0.014$  & & $0.88$ & $-0.26$ \\
\GridCell{V\&A, grid}& $1.360 \pm 0.040$ & $1.370 \pm 0.050$ & $1.100 \pm 0.040$ & $1.110 \pm 0.040$ & $1.460 \pm 0.060$  & & $1.18$ & $-0.04$ \\
$\erel$ & $6.17\,\%$ & $3.74\,\%$ & $2.85\,\%$ & $3.95\,\%$ & $4.71\,\%$ & & -- & -- \\
$\brel$ & $-3.27\,\%$ & $2.08\,\%$ & $-1.15\,\%$ & $-0.82\,\%$ & $3.37\,\%$ & & -- & -- \\
$\enorm$ & $3.34$ & $1.11$ & $1.83$ & $6.36$ & $2.33$ & & $4.55$ & -- \\
$\bnorm$ & $-1.26$ & $0.58$ & $0.08$ & $1.69$ & $1.69$ & & -- & $0.26$ \\
\hhline{------~--}
\end{tabular}

\end{table*}

\begin{table*}[htbp]
\caption{Fitted values for the age in Gyrs, $t$, and associated
average errors and biases. \label{tab:age}}
\begin{tabular}{lcccccccc}
\hhline{=========}
\multicolumn{9}{c}{\textbf{Age}} \\
\hhline{------~--}
\textbf{Hounds} & \textbf{Aardvark} & \textbf{Blofeld} & \textbf{Coco} & \textbf{Diva} & \textbf{Elvis} & & $\erel$ & $\brel$ \\
\hhline{------~--}
\SolutionCell{Solution} & $3.058$ & $2.595$ & $9.616$ & $4.622$ & $6.841$  & & -- & -- \\
\GridCell{GOE}& $2.761 \pm 0.110$ & $2.620 \pm 0.091$ & $8.709 \pm 0.331$ & $3.761 \pm 0.144$ & $5.775 \pm 0.159$  & & $27.39\,\%$  & $-21.03\,\%$  \\
\GridCell{YMCM}& $2.757 \pm 0.166$ & $2.608 \pm 0.233$ & $9.705 \pm 0.273$ & $4.979 \pm 0.190$ & $6.587 \pm 0.142$  & & $13.63\,\%$  & $0.28\,\%$  \\
\GridCell{ASTFIT}& $2.892 \pm 0.332$ & $3.581 \pm 0.173$ & $9.618 \pm 0.549$ & $4.052 \pm 0.204$ & $6.636 \pm 0.363$  & & $18.08\,\%$  & $2.34\,\%$  \\
\GridCell{YL}& $2.975 \pm 0.013$ & $2.333 \pm 0.009$ & $9.338 \pm 0.306$ & $1.607 \pm 0.007$ & $5.922 \pm 0.081$  & & $33.51\,\%$  & $-9.01\,\%$  \\
\GridCell{AMP}& $2.900 \pm 0.230$ & $2.480 \pm 0.290$ & $9.010 \pm 0.510$ & $4.230 \pm 0.320$ & $6.390 \pm 0.360$  & & $20.13\,\%$  & $-10.89\,\%$  \\
\GridCell{BASTA}& $2.777_{-0.176}^{+0.144}$  & -- & $9.468_{-0.319}^{+0.271}$  & -- & $6.210_{-0.192}^{+0.256}$  & & $5.86\,\%$  & $-0.73\,\%$  \\
\GridCell{MESAastero}& $3.121 \pm 0.280$  & -- & $9.924 \pm 0.686$  & -- & $6.639 \pm 0.444$  & & $12.08\,\%$  & $6.09\,\%$  \\
\GridCell{V\&A, grid}& $2.900 \pm 0.400$ & $2.700 \pm 0.400$ & $9.100 \pm 0.400$ & $4.800 \pm 0.400$ & $6.100 \pm 0.600$  & & $31.25\,\%$  & $12.67\,\%$  \\
$\erel$ & $6.81\,\%$ & $16.24\,\%$ & $4.73\,\%$ & $28.58\,\%$ & $9.35\,\%$ & & $22.94\,\%$ & -- \\
$\brel$ & $-5.66\,\%$ & $4.82\,\%$ & $-2.67\,\%$ & $-15.51\,\%$ & $-8.16\,\%$ & & -- & $-2.96\,\%$ \\
$\enorm$ & $2.64$ & $12.12$ & $1.22$ & $175.84$ & $4.85$ & & -- & -- \\
$\bnorm$ & $-1.76$ & $-3.87$ & $-0.73$ & $-73.06$ & $-3.27$ & & -- & -- \\
\hhline{------~--}
\textbf{Hounds} & \textbf{Felix} & \textbf{George} & \textbf{Henry} & \textbf{Izzy} & \textbf{Jam} & & $\enorm$ & $\bnorm$ \\
\hhline{------~--}
\SolutionCell{Solution} & $2.921$ & $2.944$ & $2.055$ & $2.113$ & $1.681$  & & -- & -- \\
\GridCell{GOE}& $2.122 \pm 0.164$ & $1.765 \pm 0.116$ & $1.944 \pm 0.170$ & $1.615 \pm 0.141$ & $0.646 \pm 0.163$  & & $5.25$ & $-4.34$ \\
\GridCell{YMCM}& $3.725 \pm 0.416$ & $2.792 \pm 0.102$ & $2.205 \pm 0.297$ & $2.242 \pm 0.254$ & $1.202 \pm 0.158$  & & $1.61$ & $-0.29$ \\
\GridCell{ASTFIT}& $2.847 \pm 0.257$ & $2.337 \pm 0.158$ & $2.687 \pm 0.402$ & $2.336 \pm 0.346$ & $1.479 \pm 0.203$  & & $2.44$ & $-0.11$ \\
\GridCell{YL}& $4.818 \pm 0.004$ & $2.999 \pm 0.096$ & $2.011 \pm 0.078$ & $1.850 \pm 0.015$ & $0.875 \pm 0.027$  & & $203.11$ & $-5.17$ \\
\GridCell{AMP}& $3.010 \pm 0.230$ & $1.620 \pm 0.250$ & $2.080 \pm 0.270$ & $2.220 \pm 0.260$ & $0.970 \pm 0.320$  & & $1.96$ & $-1.14$ \\
\GridCell{BASTA}& $2.953_{-0.144}^{+0.128}$  & -- & $2.122 \pm 0.287$ & $2.202_{-0.287}^{+0.303}$ & $1.787_{-0.287}^{+0.527}$  & & $1.29$ & $-0.58$ \\
\GridCell{MESAastero} & -- & $2.910 \pm 0.093$ & $2.647 \pm 0.610$ & $2.384 \pm 0.509$ & $1.678 \pm 0.172$  & & $0.51$ & $0.19$ \\
\GridCell{V\&A, grid}& $5.100 \pm 0.800$ & $4.800 \pm 0.800$ & $2.000 \pm 0.400$ & $2.200 \pm 0.400$ & $1.700 \pm 0.800$  & & $1.28$ & $0.29$ \\
$\erel$ & $40.18\,\%$ & $33.93\,\%$ & $15.35\,\%$ & $11.64\,\%$ & $33.32\,\%$ & & -- & -- \\
$\brel$ & $20.18\,\%$ & $-6.72\,\%$ & $7.66\,\%$ & $0.85\,\%$ & $-23.15\,\%$ & & -- & -- \\
$\enorm$ & $179.24$ & $4.70$ & $0.75$ & $6.34$ & $10.88$ & & $74.71$ & -- \\
$\bnorm$ & $67.76$ & $-2.61$ & $0.25$ & $-2.31$ & $-5.27$ & & -- & $-1.49$ \\
\hhline{------~--}
\end{tabular}

\end{table*}

\begin{table*}[htbp]
\caption{Fitted values for $\log(g)$ in dex and
average errors and biases for $g$ (in cm.s$^{-2}$). \label{tab:logg}}
\begin{tabular}{lcccccccc}
\hhline{=========}
\multicolumn{9}{c}{\textbf{Surface gravity}} \\
\hhline{------~--}
\textbf{Hounds} & \textbf{Aardvark} & \textbf{Blofeld} & \textbf{Coco} & \textbf{Diva} & \textbf{Elvis} & & $\erel$ & $\brel$ \\
\hhline{------~--}
\SolutionCell{Solution} & $4.474$ & $4.257$ & $4.508$ & $4.261$ & $4.365$  & & -- & -- \\
\GridCell{GOE}& $4.470 \pm 0.002$ & $4.256 \pm 0.002$ & $4.506 \pm 0.003$ & $4.265 \pm 0.003$ & $4.365 \pm 0.002$  & & $1.19\,\%$  & $-0.64\,\%$  \\
\GridCell{YMCM}& $4.472 \pm 0.002$ & $4.260 \pm 0.002$ & $4.507 \pm 0.002$ & $4.255 \pm 0.002$ & $4.358 \pm 0.001$  & & $0.93\,\%$  & $-0.13\,\%$  \\
\GridCell{ASTFIT}& $4.475 \pm 0.002$ & $4.248 \pm 0.002$ & $4.508 \pm 0.002$ & $4.257 \pm 0.003$ & $4.366 \pm 0.002$  & & $0.83\,\%$  & $-0.22\,\%$  \\
\GridCell{YL}& $4.462$ & $4.266$ & $4.507$ & $4.275$ & $4.361$  & & $2.51\,\%$  & $-0.06\,\%$  \\
\GridCell{AMP}& $4.469$ & $4.258$ & $4.507$ & $4.263$ & $4.368$  & & $2.57\,\%$  & $1.18\,\%$  \\
\GridCell{BASTA}& $4.474 \pm 0.002$  & -- & $4.506_{-0.003}^{+0.001}$  & -- & $4.365_{-0.003}^{+0.002}$  & & $0.47\,\%$  & $-0.38\,\%$  \\
\GridCell{MESAastero}& $4.473 \pm 0.001$  & -- & $4.506 \pm 0.002$  & -- & $4.366 \pm 0.002$  & & $0.39\,\%$  & $-0.26\,\%$  \\
\GridCell{V\&A, grid}& $4.468 \pm 0.008$ & $4.252 \pm 0.006$ & $4.506 \pm 0.006$ & $4.264 \pm 0.007$ & $4.360 \pm 0.006$  & & $1.22\,\%$  & $-0.38\,\%$  \\
$\erel$ & $1.22\,\%$ & $1.33\,\%$ & $0.26\,\%$ & $1.54\,\%$ & $0.81\,\%$ & & $1.54\,\%$ & -- \\
$\brel$ & $-0.83\,\%$ & $-0.14\,\%$ & $-0.20\,\%$ & $0.47\,\%$ & $-0.40\,\%$ & & -- & $-0.09\,\%$ \\
$\enorm$ & $0.89$ & $3.28$ & $0.57$ & $1.90$ & $2.87$ & & -- & -- \\
$\bnorm$ & $-0.57$ & $-1.63$ & $-0.43$ & $-0.70$ & $-1.28$ & & -- & -- \\
\hhline{------~--}
\textbf{Hounds} & \textbf{Felix} & \textbf{George} & \textbf{Henry} & \textbf{Izzy} & \textbf{Jam} & & $\enorm$ & $\bnorm$ \\
\hhline{------~--}
\SolutionCell{Solution} & $4.091$ & $4.102$ & $4.367$ & $4.364$ & $4.228$  & & -- & -- \\
\GridCell{GOE}& $4.082 \pm 0.004$ & $4.095 \pm 0.006$ & $4.364 \pm 0.005$ & $4.355 \pm 0.006$ & $4.231 \pm 0.007$  & & $1.22$ & $-0.67$ \\
\GridCell{YMCM}& $4.089 \pm 0.004$ & $4.104 \pm 0.002$ & $4.368 \pm 0.003$ & $4.362 \pm 0.004$ & $4.235 \pm 0.005$  & & $2.53$ & $-0.81$ \\
\GridCell{ASTFIT}& $4.092 \pm 0.003$ & $4.106 \pm 0.003$ & $4.364 \pm 0.003$ & $4.364 \pm 0.003$ & $4.229 \pm 0.003$  & & $2.15$ & $-0.61$ \\
\GridCell{YL}& $4.067$ & $4.095$ & $4.371$ & $4.374$ & $4.235$  & & --  & --  \\
\GridCell{AMP}& $4.098$ & $4.130$ & $4.363$ & $4.366$ & $4.245$  & & --  & --  \\
\GridCell{BASTA}& $4.090 \pm 0.003$  & -- & $4.366 \pm 0.003$ & $4.362 \pm 0.003$ & $4.224_{-0.011}^{+0.005}$  & & $0.53$ & $-0.46$ \\
\GridCell{MESAastero} & -- & $4.101 \pm 0.002$ & $4.363 \pm 0.003$ & $4.363 \pm 0.004$ & $4.227 \pm 0.002$  & & $0.73$ & $-0.50$ \\
\GridCell{V\&A, grid}& $4.086 \pm 0.006$ & $4.096 \pm 0.005$ & $4.365 \pm 0.011$ & $4.366 \pm 0.007$ & $4.238 \pm 0.012$  & & $0.75$ & $-0.36$ \\
$\erel$ & $2.31\,\%$ & $2.73\,\%$ & $0.65\,\%$ & $1.17\,\%$ & $1.94\,\%$ & & -- & -- \\
$\brel$ & $-1.07\,\%$ & $0.40\,\%$ & $-0.30\,\%$ & $-0.03\,\%$ & $1.23\,\%$ & & -- & -- \\
$\enorm$ & $1.12$ & $1.14$ & $0.71$ & $0.76$ & $0.75$ & & $1.59$ & -- \\
$\bnorm$ & $-0.73$ & $-0.22$ & $-0.46$ & $-0.49$ & $0.39$ & & -- & $-0.58$ \\
\hhline{------~--}
\end{tabular}

\tablefoot{As stated in the caption, the various average errors and biases have
been calculated for $g$ rather than $\log(g)$.}
\end{table*}

\begin{table*}[htbp]
\caption{Fitted values for $\bar{\rho}$ in g.cm$^{-3}$,
and associated average errors and biases. \label{tab:rho}}
\begin{tabular}{lcccccccc}
\hhline{=========}
\multicolumn{9}{c}{\textbf{Mean density}} \\
\hhline{------~--}
\textbf{Hounds} & \textbf{Aardvark} & \textbf{Blofeld} & \textbf{Coco} & \textbf{Diva} & \textbf{Elvis} & & $\erel$ & $\brel$ \\
\hhline{------~--}
\SolutionCell{Solution} & $1.5962$ & $0.6838$ & $2.0292$ & $0.6928$ & $1.0967$  & & -- & -- \\
\GridCell{GOE}& $1.5992 \pm 0.0016$ & $0.6838 \pm 0.0024$ & $2.0295 \pm 0.0059$ & $0.6914 \pm 0.0011$ & $1.0959 \pm 0.0009$  & & $1.04\,\%$  & $-0.08\,\%$  \\
\GridCell{YMCM}& $1.6010 \pm 0.0020$ & $0.6742 \pm 0.0010$ & $2.0391 \pm 0.0040$ & $0.6948 \pm 0.0010$ & $1.0986 \pm 0.0010$  & & $0.61\,\%$  & $0.18\,\%$  \\
\GridCell{ASTFIT}& $1.5955 \pm 0.0019$ & $0.6722 \pm 0.0014$ & $2.0287 \pm 0.0069$ & $0.7008 \pm 0.0059$ & $1.0968 \pm 0.0013$  & & $0.79\,\%$  & $-0.28\,\%$  \\
\GridCell{YL}& $1.6022$ & $0.6758$ & $2.0224$ & $0.7214$ & $1.0952$  & & $1.79\,\%$  & $-0.22\,\%$  \\
\GridCell{AMP}& $1.6017$ & $0.6879$ & $2.0287$ & $0.7150$ & $1.1054$  & & $3.92\,\%$  & $2.78\,\%$  \\
\GridCell{BASTA}& $1.5953_{-0.0152}^{+0.0146}$  & -- & $2.0249_{-0.0283}^{+0.0091}$  & -- & $1.0962_{-0.0112}^{+0.0109}$  & & $0.76\,\%$  & $-0.46\,\%$  \\
\GridCell{MESAastero}& $1.5982 \pm 0.0024$  & -- & $2.0318 \pm 0.0039$  & -- & $1.0971 \pm 0.0015$  & & $0.48\,\%$  & $0.10\,\%$  \\
\GridCell{V\&A, grid}& $1.5967 \pm 0.0056$ & $0.6815 \pm 0.0056$ & $2.0318 \pm 0.0056$ & $0.6927 \pm 0.0056$ & $1.0940 \pm 0.0056$  & & $1.77\,\%$  & $-0.93\,\%$  \\
\InversionCell{GB}& $1.6020 \pm 0.0004$ & $0.7240 \pm 0.0003$ & $2.0490 \pm 0.0007$ & $0.6990 \pm 0.0004$ & $1.0960 \pm 0.0003$  & & $2.65\,\%$  & $1.86\,\%$  \\
\InversionCell{DRR}& $1.5951 \pm 0.0026$ & $0.6606 \pm 0.0033$ & $2.0189 \pm 0.0034$ & $0.6994 \pm 0.0045$ & $1.0935 \pm 0.0015$  & & $1.40\,\%$  & $-0.86\,\%$  \\
$\erel$ & $0.23\,\%$ & $2.57\,\%$ & $0.40\,\%$ & $1.95\,\%$ & $0.29\,\%$ & & $1.85\,\%$ & -- \\
$\brel$ & $0.15\,\%$ & $-0.18\,\%$ & $0.06\,\%$ & $1.30\,\%$ & $0.01\,\%$ & & -- & $0.18\,\%$ \\
$\enorm$ & $4.67$ & $55.08$ & $9.70$ & $6.99$ & $1.44$ & & -- & -- \\
$\bnorm$ & $2.13$ & $18.17$ & $3.43$ & $3.42$ & $-0.51$ & & -- & -- \\
\hhline{------~--}
\textbf{Hounds} & \textbf{Felix} & \textbf{George} & \textbf{Henry} & \textbf{Izzy} & \textbf{Jam} & & $\enorm$ & $\bnorm$ \\
\hhline{------~--}
\SolutionCell{Solution} & $0.3688$ & $0.3834$ & $1.0508$ & $1.0414$ & $0.5915$  & & -- & -- \\
\GridCell{GOE}& $0.3643 \pm 0.0031$ & $0.3751 \pm 0.0028$ & $1.0559 \pm 0.0108$ & $1.0417 \pm 0.0146$ & $0.6039 \pm 0.0102$  & & $1.35$ & $-0.29$ \\
\GridCell{YMCM}& $0.3699 \pm 0.0010$ & $0.3849 \pm 0.0010$ & $1.0551 \pm 0.0060$ & $1.0516 \pm 0.0070$ & $0.5906 \pm 0.0030$  & & $3.41$ & $0.38$ \\
\GridCell{ASTFIT}& $0.3670 \pm 0.0013$ & $0.3804 \pm 0.0027$ & $1.0499 \pm 0.0063$ & $1.0439 \pm 0.0056$ & $0.5853 \pm 0.0038$  & & $2.76$ & $-1.11$ \\
\GridCell{YL}& $0.3578$ & $0.3759$ & $1.0476$ & $1.0377$ & $0.5949$  & & --  & --  \\
\GridCell{AMP}& $0.4029$ & $0.4055$ & $1.0720$ & $1.0609$ & $0.6149$  & & --  & --  \\
\GridCell{BASTA}& $0.3654 \pm 0.0044$  & -- & $1.0514 \pm 0.0104$ & $1.0385_{-0.0119}^{+0.0113}$ & $0.5812_{-0.0108}^{+0.0088}$  & & $0.51$ & $-0.34$ \\
\GridCell{MESAastero} & -- & $0.3828 \pm 0.0015$ & $1.0505 \pm 0.0078$ & $1.0532 \pm 0.0116$ & $0.5884 \pm 0.0046$  & & $0.64$ & $0.23$ \\
\GridCell{V\&A, grid}& $0.3562 \pm 0.0056$ & $0.3675 \pm 0.0056$ & $1.0532 \pm 0.0084$ & $1.0405 \pm 0.0056$ & $0.5829 \pm 0.0084$  & & $1.21$ & $-0.63$ \\
\InversionCell{GB} & --  & -- & $1.0810 \pm 0.0008$ & $1.0630 \pm 0.0008$  & --  & & $55.01$ & $35.80$ \\
\InversionCell{DRR}& $0.3666 \pm 0.0027$ & $0.3765 \pm 0.0032$ & $1.0406 \pm 0.0039$ & $1.0367 \pm 0.0057$ & $0.5827 \pm 0.0045$  & & $2.86$ & $-1.95$ \\
$\erel$ & $3.70\,\%$ & $2.81\,\%$ & $1.18\,\%$ & $1.02\,\%$ & $1.80\,\%$ & & -- & -- \\
$\brel$ & $-0.01\,\%$ & $-0.61\,\%$ & $0.47\,\%$ & $0.51\,\%$ & $0.02\,\%$ & & -- & -- \\
$\enorm$ & $1.38$ & $2.04$ & $12.91$ & $9.21$ & $1.24$ & & $17.40$ & -- \\
$\bnorm$ & $-0.92$ & $-1.32$ & $4.40$ & $3.45$ & $-0.78$ & & -- & $3.01$ \\
\hhline{------~--}
\end{tabular}

\tablefoot{The hounds using inversion techniques (GB and DRR) have been
highlighted in black since their methodology is different from that of the other
hounds.  The overall average errors and biases have been calculated using the
results from all of the hounds in the table.  If the last two hounds (GB, DRR)
are excluded due to their different methodologies, these averages become: $\erel
= 1.81\,\%$, $\brel=0.16\,\%$, $\enorm=2.07$, $\bnorm=-0.32$.}
\end{table*}

The relative error bars on the radius, mass, and age, averaged over all
of the stars and relevant hounds, are $1.5\,\%$, $3.9\,\%$, and $23\,\%$,
respectively.  The first two are well within the requirements for PLATO 2.0 and
are comparable to the results recently obtained in the KAGES project
\citep{SilvaAguirre2015}.  The age, on the other hand, is determined  with a
higher uncertainty than what was achieved in \citet{SilvaAguirre2015}. However,
we note that in the present work, the uncertainties are calculated with respect
to the exact solutions rather than as a dispersion between different results. 
If, however, the exact solutions are replaced by the average of the results
obtained by the hounds, then the overall relative error bar (averaged over all
of the stars and hounds) becomes $20\,\%$ which is closer to the result
obtained by \citet{SilvaAguirre2015} who found $14\,\%$. Also, the
proportion of massive (and problematic) stars seems to be slightly higher in the
present sample. Nonetheless, it is important to note that the age estimates for
Aardvark and Elvis, the two stars that match quite well the PLATO 2.0 reference
case, are on average accurate to within $10\,\%$, thereby satisfying the
requirements for PLATO 2.0.

It is also interesting to look at how well the error bars were estimated.  On
the whole, the error estimates are quite reasonable, with only a few outlying
cases.  In some cases, the error bars  were underestimated, for instance on the
age by YL, and on the mean density by GB.  In YL's case, the reason for this may
be related to the fact that they are using a Levenberg-Marquardt approach, which
is more prone to getting stuck in local minima.  In GB's case, inversions are
applied to a single reference model.  Hence, his error bars only take into
account the errors on the frequencies as they propagate through the inversion. 
However, such errors do not allow for the fact that the reference model may be
sub-optimal (thus requiring non-linear corrections).  DRR also applied
inversions, but to a set of reference models selected according to classical
constraints.  Hence, his error bars include the scatter between the results from
the different reference models and are thus more realistic.  Nonetheless, in
both cases, the error bars do not account for mismatches between averaging
kernels and relevant target functions.  We also note that GB included a surface
correction term in his inversions whereas DRR did not.  This, in fact, leads to
worse results for all of the stars except Aardvark, due to the reduced quality
of the averaging kernels, as indicated by further tests by GB.  Inversions
naturally mitigate surface effects, so including a surface correction
term yields little improvement while degrading the quality of the
averaging kernel.

\subsubsection{Comparisons between similar stars}

\paragraph{Aardvark and Elvis}

Aardvark and Elvis differ according to age (Aardvark being approximately half
the age of Elvis) and the use of an isothermal boundary condition on Elvis. 
Both stars were well characterised by the hounds with slightly better results
for Aardvark.  Interestingly, the average relative error on the age was slightly
smaller on Aardvark, even though the star is younger, thereby also implying an
absolute error on the age more than twice as small.  This is somewhat
surprising because the central hydrogen abundance decreases roughly linearly in
time over the main sequence, thereby leading to the expectation that the
absolute age error should be similar at all ages on the main sequence.  Hence,
one would expect the relative error to be larger for younger stars.  As
mentioned above, these stars are the closest to PLATO's reference case, and the
results for these stars satisfy all of the requirements for PLATO 2.0.

\paragraph{Felix, George, and Jam}

It is also interesting to see which stars were the most problematic.  In this
particular case, the star Felix proved to be challenging for a number of hounds
and was even excluded by one of them. We also note that George and Jam also
yielded poor results and not all of the hounds proposed results for these
either.  Their average relative errors show substantial scatter between the
different hounds, and the average relative biases were also
significant.  In the case of the age, the error bars were often highly
underestimated.

The common factor between these stars is their high mass, $1.33\,\Msun$.
High mass leads to various phenomena, which make these stars more difficult to
model and their pulsations more difficult to interpret.  For instance, these
stars are hotter thus leading to shorter mode lifetimes and, hence, larger error
bars on the frequencies.  Furthermore, these stars contain convective cores,
which may result in sharp density gradients.  Different stellar evolution codes
use different criteria for defining the boundary of the convective core
\citep[\eg][]{Gabriel2014}, different core overshoot prescriptions, and
different numerical approaches, all of which affect the size of the convective
core and its transition to the radiative region above.  Accordingly, there can
be large discrepancies in the sizes of convective cores in models from different
evolution codes.

It is interesting to compare these stars.  The main difference between Felix and
George is the higher overshoot parameter in the latter star.  Although the two
have the same mass, it is interesting to note that, on the whole, the mass of
Felix was underestimated whereas the mass of George tended to be overestimated. 
Even more dramatic are the age differences between the two obtained by the
various hounds.  Hence, on average, Felix was found to be $0.79$ Gyrs older and
$0.082\,\Msun$ lighter than George (if we limit ourselves to hounds which gave
results for both stars), even though both have the same mass and nearly the same
age.  Of course, it is normal that the mass and age are anti-correlated,
because lighter stars evolve more slowly and will therefore be older
for a given evolutionary stage.  Jam had a similar overshoot parameter as Felix
but was substantially younger.  On the whole, Jam seemed to yield better results
(with substantially less scatter on the mass), but it still proved to be
difficult to obtain accurate ages for this star.

\paragraph{Blofeld and Diva}

The stars Blofeld and Diva also proved to be problematic.  Although
these stars share the same mass ($1.22\, M_{\odot}$), they are different in a
number of ways.  Indeed, the distinguishing features of Blofeld include a
different heavy abundance mixture (AGS05), diffusion, a truncated atmosphere, and
the use of an isothermal boundary condition on the pulsations to reduce the
effects of the truncated atmosphere.  In contrast, Diva, is much more
like the other stellar targets.

It is interesting to look at each of the distinguishing features of Blofeld and
discuss how likely they are to affect the results from the hounds.  The
truncated atmosphere can lead to important surface effects.   However, as
pointed out above, the isothermal boundary condition reduces these effects --
accordingly, some of the hounds reported rather small surface effects on the
frequencies.  We note in passing that Elvis did not seem to be much affected by
the isothermal boundary condition (but unlike Blofeld, its atmosphere was not
truncated). The composition can also be a source of error.  SB reran the YMCM
pipeline on Blofeld using the correct mixture (AGS05 as opposed to
\citealt{Grevesse1998} as used in the previous run).  The results are shown in
Table~\ref{tab:Blofeld_Basu}.  Apart from the age (which is likely to be
affected by a fortuitous agreement with the original result judging from its
extreme accuracy), all of the stellar properties are determined with increased
accuracy, thereby highlighting the importance of using the correct composition. 

Finally, atomic diffusion, in which radiative accelerations are
neglected, is expected to be problematic given the relatively high mass of
this star.  Indeed, this tends to over-deplete surface heavy abundances
compared to observations in higher mass stars.  Radiative accelerations
counteract this effect by levitating heavy elements to specific regions
of the star which in turn affects local opacities and may lead to supplementary convection zones
\citep[\eg][]{Richard2001}.  In addition, element accumulation can
lead to double-diffusive convection, another means by which elements can be
redistributed within stars \citep[\eg][]{Deal2016}.
Given that radiative accelerations were not taken into account by the
various diffusion prescriptions used in the present exercise,  several of the hounds
(VSA, KV \& HMA) only included atomic diffusion up to a certain mass
threshold which is close to Blofeld's mass, thereby potentially leading to
greater discrepancies in the results.

\begin{table*}[htbp]
\caption{Original versus new YMCM results (with the correct mixture,
\ie\ AGS05) for Blofeld. \label{tab:Blofeld_Basu}}
\begin{tabular}{lccccc}
\hline
\hline
\textbf{Blofeld}
& $R$ (in R$_{\odot}$) 
& $M$ (in M$_{\odot}$)
& age (in Gyrs)
& $\log(g)$ (in dex)
& $\bar{\rho}$ (in g.cm$^{-3}$) \\
\hline
\SolutionCell{Solution} & $1.359$ & $1.220$ & $2.595$ & $4.257$ & $0.684$ \\
\GridCell{Original} & $1.388$ & $1.279$ & $2.608$ & $4.260$ & $0.674$ \\
\GridCell{$\brel$(original)} & $2.10\,\%$ & $4.84\,\%$ & $0.49\,\%$ & $0.65\,\%$ & $-1.39\,\%$ \\
\GridCell{Correct mixture} & $1.373$ & $1.251$ & $2.423$ & $4.259$ & $0.680$ \\
\GridCell{$\brel$(correct mixture)} & $1.00\,\%$ & $2.54\,\%$ & $-6.65\,\%$ & $0.47\,\%$ & $-0.53\,\%$ \\
\hline
\end{tabular}
\end{table*}

A comparison of the results for both stars showed that Blofeld had
better age and surface gravity estimates, whereas Diva had better mass and mean
density estimates.  The radius estimates were very similar in quality between
the two stars.  Interpreting these differences is not straightforward.  One
would expect surface effects to affect ``structural'' properties (namely, $M$,
$R$, $\bar{\rho}$, $\log(g)$), but the results are not clear cut.  However, as
argued above, surface effects may not be the dominant factor in Blofeld.  One
would also expect diffusion to make it more difficult to estimate the age of
Blofeld, but the opposite is true in the present case.  It is not clear why this
is so, although we do note that the older star has a bigger relative error
as was the case for Aardvark and Elvis.

\paragraph{Henry and Izzy}

The main difference between Henry and Izzy is the fact that the former includes
diffusion.  However, in terms of results, both were fairly similar, with results
for Henry being slightly better on structural properties (for the most part) and
results for Izzy being better on the age.  Hence, diffusion seems to have a
small impact on the results as expected for stars of this mass.

\subsection{Properties related to the base of the convection zone}
\label{sect:tau_results}

We now turn our attention to properties related to the base of the convection
zone.  As stated earlier, the hounds were asked to provide the fractional radius
of the base of the convection zone (for those carrying out grid modelling) as
well as the acoustic radius or depth of the base of the convection zone along
with the acoustic radius of the star.

Tables~\ref{tab:r_bcz} and~\ref{tab:AR_bcz} list the results obtained for the
fractional radius of the base of the convection zone, $r_{\mathrm{BCZ}}/R$, and
the acoustic radius of the base of the convection zone, $\tau_{\mathrm{BCZ}}$.
The fractional radius can only be obtained from a forward modelling
approach since it is not an acoustic variable.  In contrast, the acoustic radius
can be found through both forward modelling and glitch fitting.  It is
also interesting to note that one of the hounds has applied both approaches. 
These results have been kept as separate entries in Table~\ref{tab:AR_bcz} and
shown separately in Fig.~\ref{fig:AR_bcz} under the headers ``V\&A, grid'' for
the grid modelling, and ``V\&A, glitch'' for the glitch fitting.

\begin{table*}[htbp]
\caption{Fitted values for $r_{\mathrm{BCZ}}/R$ and associated average errors
and biases. \label{tab:r_bcz}}
\begin{tabular}{lcccccccc}
\hhline{=========}
\multicolumn{9}{c}{\textbf{$r_{\mathrm{BCZ}}/R$}} \\
\hhline{------~--}
\textbf{Hounds} & \textbf{Aardvark} & \textbf{Blofeld} & \textbf{Coco} & \textbf{Diva} & \textbf{Elvis} & & $\erel$ & $\brel$ \\
\hhline{------~--}
\SolutionCell{Solution} & $0.731$ & $0.838$ & $0.746$ & $0.767$ & $0.727$  & & -- & -- \\
\GridCell{GOE}& $0.725 \pm 0.003$ & $0.839 \pm 0.005$ & $0.736 \pm 0.009$ & $0.750 \pm 0.004$ & $0.710 \pm 0.004$  & & $2.98\,\%$  & $-1.94\,\%$  \\
\GridCell{YMCM}& $0.739$ & $0.825$ & $0.759$ & $0.774$ & $0.728$  & & $1.12\,\%$  & $0.64\,\%$  \\
\GridCell{ASTFIT}& $0.729 \pm 0.004$ & $0.811 \pm 0.005$ & $0.743 \pm 0.010$ & $0.789 \pm 0.019$ & $0.728 \pm 0.005$  & & $1.53\,\%$  & $-0.41\,\%$  \\
\GridCell{YL}& $0.724$ & $0.827$ & $0.724$ & $0.854$ & $0.709$  & & $4.73\,\%$  & $-1.01\,\%$  \\
\GridCell{AMP}& $0.723$ & $0.828$ & $0.727$ & $0.763$ & $0.708$  & & $3.64\,\%$  & $0.52\,\%$  \\
\GridCell{BASTA}& $0.722 \pm 0.006$  & -- & $0.724_{-0.006}^{+0.009}$  & -- & $0.713_{-0.006}^{+0.003}$  & & $2.06\,\%$  & $-1.90\,\%$  \\
\GridCell{V\&A, grid}& $0.734$ & $0.843$ & $0.742$ & $0.777$ & $0.709$  & & $3.94\,\%$  & $-1.95\,\%$  \\
$\erel$ & $0.91\,\%$ & $1.67\,\%$ & $2.02\,\%$ & $4.89\,\%$ & $2.03\,\%$ & & $3.15\,\%$ & -- \\
$\brel$ & $-0.43\,\%$ & $-1.12\,\%$ & $-1.25\,\%$ & $2.29\,\%$ & $-1.67\,\%$ & & -- & $-0.82\,\%$ \\
$\enorm$ & $1.50$ & $3.82$ & $1.80$ & $3.37$ & $3.12$ & & -- & -- \\
$\bnorm$ & $-1.38$ & $-2.59$ & $-1.41$ & $-1.72$ & $-2.49$ & & -- & -- \\
\hhline{------~--}
\textbf{Hounds} & \textbf{Felix} & \textbf{George} & \textbf{Henry} & \textbf{Izzy} & \textbf{Jam} & & $\enorm$ & $\bnorm$ \\
\hhline{------~--}
\SolutionCell{Solution} & $0.842$ & $0.875$ & $0.839$ & $0.849$ & $0.905$  & & -- & -- \\
\GridCell{GOE}& $0.794 \pm 0.016$ & $0.819 \pm 0.013$ & $0.842 \pm 0.013$ & $0.834 \pm 0.022$ & $0.910 \pm 0.014$  & & $2.74$ & $-1.94$ \\
\GridCell{YMCM}& $0.853$ & $0.879$ & $0.848$ & $0.860$ & $0.904$  & & --  & --  \\
\GridCell{ASTFIT}& $0.833 \pm 0.007$ & $0.865 \pm 0.011$ & $0.841 \pm 0.010$ & $0.851 \pm 0.008$ & $0.893 \pm 0.006$  & & $1.92$ & $-0.85$ \\
\GridCell{YL}& $0.780$ & $0.832$ & $0.835$ & $0.842$ & $0.904$  & & --  & --  \\
\GridCell{AMP}& $0.917$ & $0.923$ & $0.830$ & $0.834$ & $0.919$  & & --  & --  \\
\GridCell{BASTA}& $0.820 \pm 0.012$  & -- & $0.835_{-0.013}^{+0.012}$ & $0.836_{-0.007}^{+0.013}$ & $0.882_{-0.021}^{+0.013}$  & & $1.99$ & $-1.77$ \\
\GridCell{V\&A, grid}& $0.778$ & $0.794$ & $0.842$ & $0.846$ & $0.888$  & & --  & --  \\
$\erel$ & $5.76\,\%$ & $5.49\,\%$ & $0.67\,\%$ & $1.29\,\%$ & $1.44\,\%$ & & -- & -- \\
$\brel$ & $-2.00\,\%$ & $-2.61\,\%$ & $0.01\,\%$ & $-0.69\,\%$ & $-0.56\,\%$ & & -- & -- \\
$\enorm$ & $2.16$ & $3.18$ & $0.26$ & $0.87$ & $1.36$ & & $2.27$ & -- \\
$\bnorm$ & $-2.02$ & $-2.67$ & $0.04$ & $-0.58$ & $-0.96$ & & -- & $-1.50$ \\
\hhline{------~--}
\end{tabular}

\end{table*}

\begin{table*}[htbp]
\caption{Fitted values for $\tau_{\mathrm{BCZ}}$ (in s) and associated average
errors and biases. \label{tab:AR_bcz}}
\begin{tabular}{lcccccccc}
\hhline{=========}
\multicolumn{9}{c}{\textbf{$\tau_{\mathrm{BCZ}}$}} \\
\hhline{------~--}
\textbf{Hounds} & \textbf{Aardvark} & \textbf{Blofeld} & \textbf{Coco} & \textbf{Diva} & \textbf{Elvis} & & $\erel$ & $\brel$ \\
\hhline{------~--}
\SolutionCell{Solution} & $1405$ & $2811$ & $1293$ & $2288$ & $1656$  & & -- & -- \\
\GridCell{GOE}& $1378 \pm 10$ & $2806 \pm 37$ & $1256 \pm 28$ & $2193 \pm 19$ & $1583 \pm 16$  & & $6.31\,\%$  & $-4.10\,\%$  \\
\GridCell{YMCM}& $1431$ & $2724$ & $1338$ & $2335$ & $1661$  & & $2.37\,\%$  & $1.41\,\%$  \\
\GridCell{YL}& $1377$ & $2738$ & $1227$ & $2869$ & $1583$  & & $10.30\,\%$  & $-1.78\,\%$  \\
\GridCell{BASTA}& $1377$  & -- & $1234$  & -- & $1612$  & & $4.75\,\%$  & $-4.21\,\%$  \\
\GridCell{MESAastero}& $1423$  & -- & $1301$  & -- & $1674$  & & $1.95\,\%$  & $0.70\,\%$  \\
\GridCell{V\&A, grid}& $1399$ & $2849$ & $1277$ & $2368$ & $1590$  & & $8.69\,\%$  & $-4.42\,\%$  \\
\GlitchCell{V\&A, glitch}& $1447 \pm 104$ & $2776 \pm 77$ & $1231 \pm 77$ & $2713 \pm 92$ & $1711 \pm 98$  & & $13.21\,\%$  & $-4.60\,\%$  \\
\GlitchCell{HRC}& $1446 \pm 156$ & $2783 \pm 81$ & $552 \pm 186$ & $2761 \pm 125$ & $1748 \pm 94$  & & $33.56\,\%$  & $-17.87\,\%$  \\
\GlitchCell{GH}& $1451_{-437}^{+156}$ & $2791 \pm 77$ & $1284_{-517}^{+157}$ & $2748_{-243}^{+155}$  & --  & & $14.90\,\%$  & $-4.77\,\%$  \\
\GlitchCell{AM}& $1407_{-134}^{+144}$ & $2782_{-77}^{+93}$ & $595 \pm 144$ & $2712_{-109}^{+112}$ & $1701_{-161}^{+124}$  & & $32.07\,\%$  & $-19.57\,\%$  \\
$\erel$ & $2.13\,\%$ & $1.67\,\%$ & $25.08\,\%$ & $16.58\,\%$ & $3.53\,\%$ & & $16.78\,\%$ & -- \\
$\brel$ & $0.61\,\%$ & $-1.06\,\%$ & $-12.66\,\%$ & $13.09\,\%$ & $-0.29\,\%$ & & -- & $-5.97\,\%$ \\
$\enorm$ & $1.21$ & $0.32$ & $2.89$ & $3.97$ & $2.34$ & & -- & -- \\
$\bnorm$ & $-0.36$ & $-0.30$ & $-2.20$ & $1.93$ & $-0.68$ & & -- & -- \\
\hhline{------~--}
\textbf{Hounds} & \textbf{Felix} & \textbf{George} & \textbf{Henry} & \textbf{Izzy} & \textbf{Jam} & & $\enorm$ & $\bnorm$ \\
\hhline{------~--}
\SolutionCell{Solution} & $3830$ & $4156$ & $2280$ & $2353$ & $3718$  & & -- & -- \\
\GridCell{GOE}& $3379 \pm 138$ & $3583 \pm 125$ & $2293 \pm 79$ & $2254 \pm 123$ & $3784 \pm 154$  & & $2.93$ & $-2.16$ \\
\GridCell{YMCM}& $3947$ & $4209$ & $2333$ & $2428$ & $3705$  & & --  & --  \\
\GridCell{YL}& $3247$ & $3694$ & $2255$ & $2314$ & $3717$  & & --  & --  \\
\GridCell{BASTA}& $3528$  & -- & $2241$ & $2262$ & $3467$  & & --  & --  \\
\GridCell{MESAastero} & -- & $4110$ & $2300$ & $2451$ & $3644$  & & --  & --  \\
\GridCell{V\&A, grid}& $3225$ & $3281$ & $2293$ & $2308$ & $3527$  & & --  & --  \\
\GlitchCell{V\&A, glitch}& $3308 \pm 132$ & $2992 \pm 173$ & $2247 \pm 126$ & $1896 \pm 161$ & $3630 \pm 224$  & & $3.03$ & $-0.99$ \\
\GlitchCell{HRC}& $1505 \pm 217$  & -- & $1467 \pm 164$ & $1943 \pm 120$  & --  & & $4.78$ & $-2.30$ \\
\GlitchCell{GH}& $3346_{-187}^{+169}$ & $3033_{-1284}^{+507}$  & -- & $1982_{-197}^{+177}$  & --  & & $1.62$ & $-0.54$ \\
\GlitchCell{AM}& $3363_{-125}^{+145}$ & $2955_{-226}^{+208}$ & $1434_{-132}^{+156}$ & $1927_{-145}^{+272}$ & $1271_{-238}^{+225}$  & & $4.81$ & $-2.85$ \\
$\erel$ & $23.26\,\%$ & $19.72\,\%$ & $17.20\,\%$ & $11.50\,\%$ & $23.49\,\%$ & & -- & -- \\
$\brel$ & $-16.30\,\%$ & $-16.21\,\%$ & $-8.08\,\%$ & $-7.49\,\%$ & $-10.09\,\%$ & & -- & -- \\
$\enorm$ & $5.68$ & $4.96$ & $3.85$ & $2.39$ & $6.11$ & & $3.68$ & -- \\
$\bnorm$ & $-4.83$ & $-4.52$ & $-2.74$ & $-2.22$ & $-3.51$ & & -- & $-1.83$ \\
\hhline{------~--}
\end{tabular}

\tablefoot{The hounds using glitch analysis have been highlighted in
grey, since their methodology is different from that of the other hounds (who
used forward modelling instead).}
\end{table*}

\begin{figure*}[htbp]
\includegraphics[width=\textwidth]{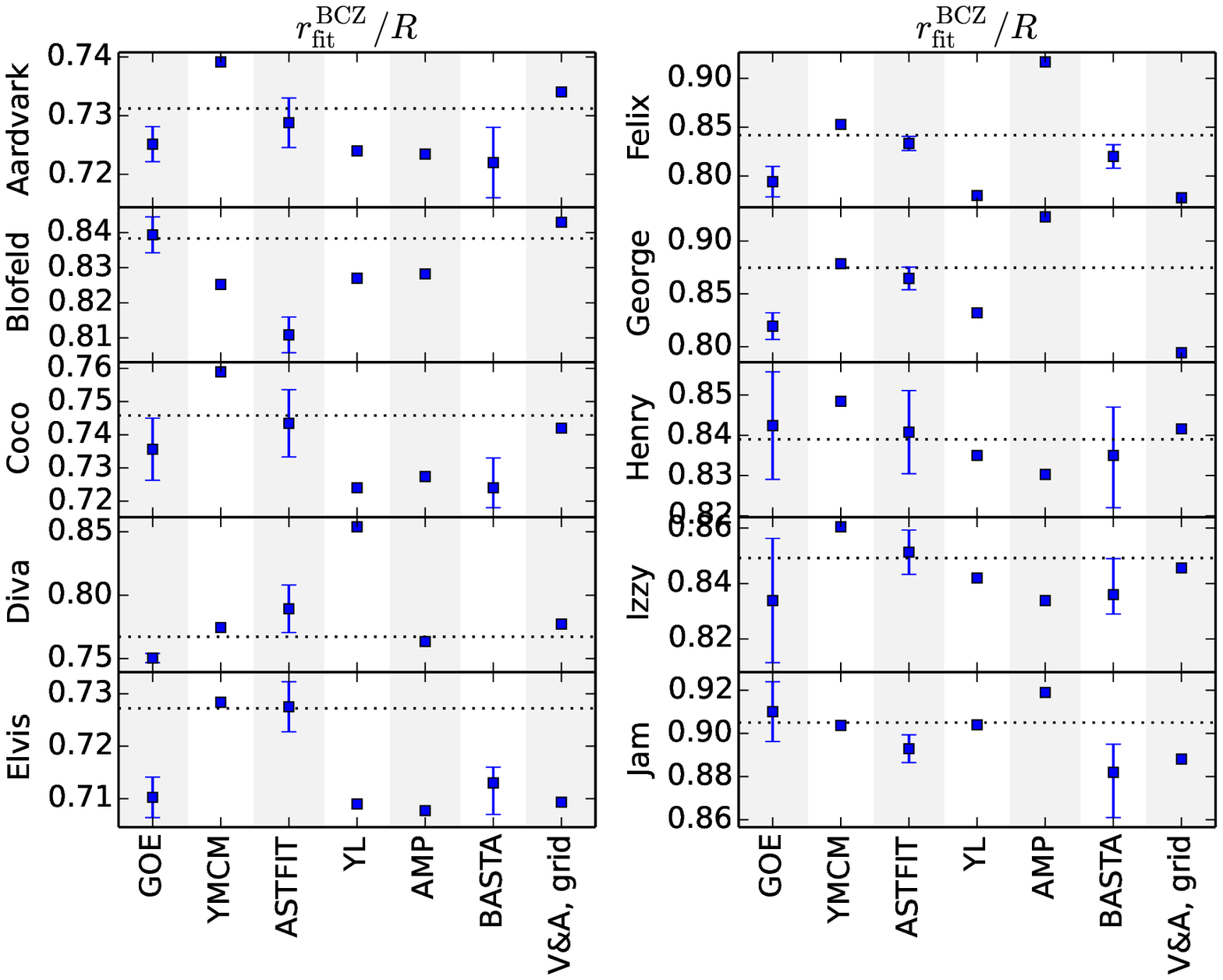}
\caption{Fitted results for $r_{\mathrm{BCZ}}/R$. \label{fig:r_bcz}}
\end{figure*}

\begin{figure*}[htbp]
\includegraphics[width=\textwidth]{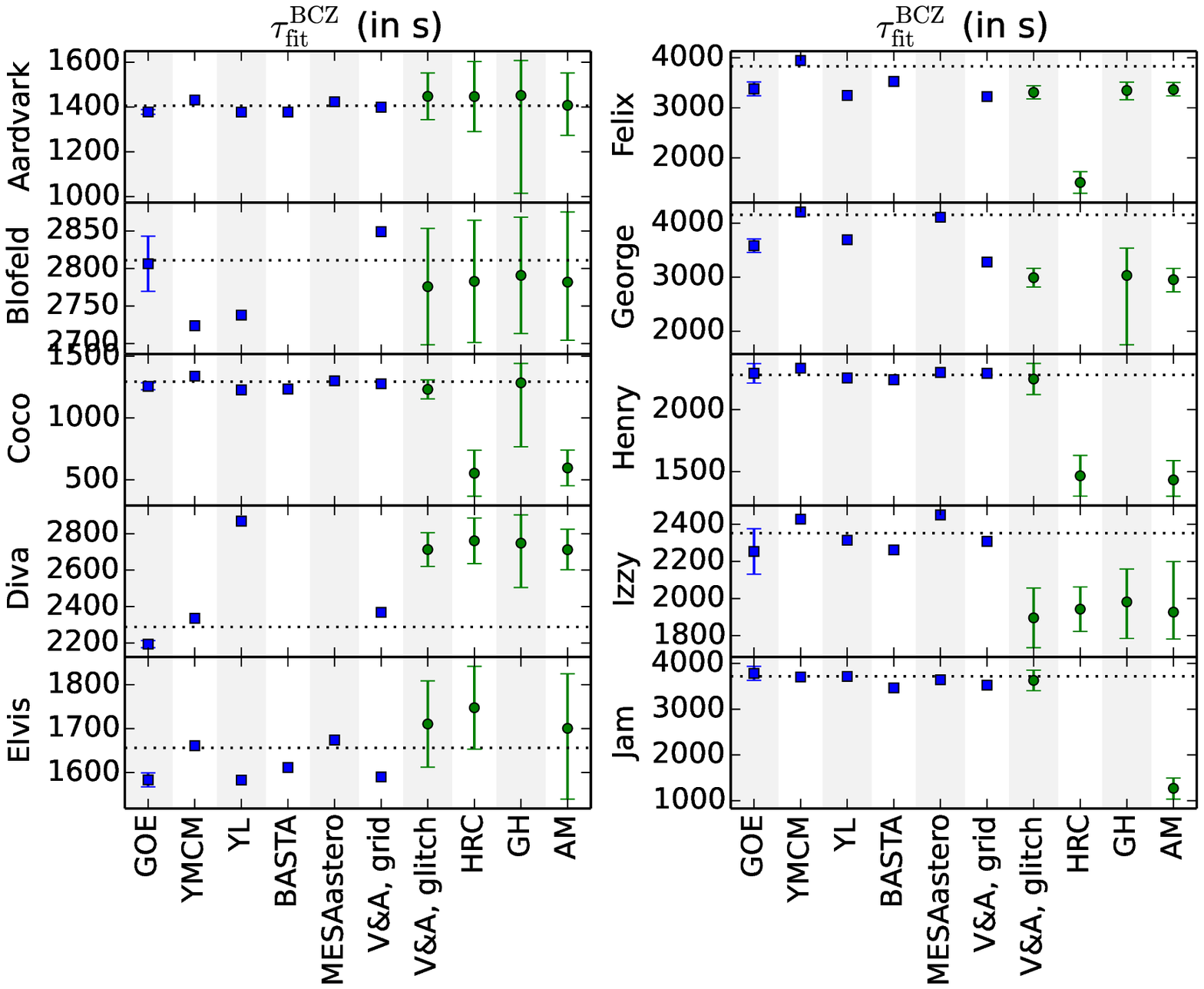}
\caption{Fitted results for $\tau_{\mathrm{BCZ}}$ (in s). \label{fig:AR_bcz}}
\end{figure*}

Overall, the fractional radius at the base of the convection zone is fairly well
fitted, with a global relative average error at $3.2\,\%$.  This is, however,
approximately twice as large as the errors on the stellar radius.  One may be
tempted to think that this larger error on $r_{\mathrm{BCZ}}/R$ is an accumulation
of both the errors on the radius and those on the internal structure.  However,
this simple reasoning does not account for the fact that for some stars, the
error on $R$ is \emph{larger} than on $r_{\mathrm{BCZ}}/R$.  What really emerges
from a detailed comparison, is the fact that the radius is more consistently
determined whereas $r_{\mathrm{BCZ}}/R$ seems to be less consistent, with some
results being very accurate and others very inaccurate.

It is then interesting to have a detailed look at the results for
$\tau_{\mathrm{BCZ}}$. Here, the errors are much larger with a very clear
dichotomy between the two approaches.  Apart from some outliers, forward
modelling produces consistent and reliable results.  In contrast, glitch fitting
seems to be more prone to finding spurious solutions (see \eg\ Diva and Izzy). 
Nonetheless, one should not forget that in the present case, forward modelling
benefits from the fact that it relies on models with similar physical
ingredients to what was used by the hares when constructing the target stars. 
If real stars were used, the errors in the forward modelling would likely
increase due to supplementary physical phenomena that are not currently included
in stellar evolution codes.  In contrast, the glitch-fitting approach is
model-independent and would not be affected in the same way.  We also note that
the stars that were problematic for one method were not always
problematic for the other.  Indeed, Blofeld turned out to be one of the
most well fitted stars by glitch analysis, in spite of diffusion and the
different mixture which seemed to affect forward modelling.
In contrast, Coco proved to be difficult to
model for some of the glitch-fitting hounds, even though it was very
straightforward to model using forward modelling.  It is also interesting to
point out that in some cases, namely for Coco and Henry, ``V\&A, glitch'' did
not make the same mistakes as the other hounds who applied glitch fitting. 
This can be explained by the fact that there are multiple solutions to
the glitch fitting problem and that ``V\&A, glitch'' is helped by ``V\&A, grid''
in selecting the correct solution.  Nonetheless, for Diva and Izzy, ``V\&A,
glitch'' and ``V\&A, grid'' found different solutions.  We also note that GH
managed to find the correct solution for Coco, and gave much larger and, hence,
more realistic error bars for George as a result of finding a bi-modal solution
(although the second mode seems to go the wrong way, judging from the error
bars).  It is not entirely clear why GH obtained better results than HRC and AM
for Coco given that all three used second frequency differences.

\begin{figure}[htbp]
\includegraphics[width=\columnwidth]{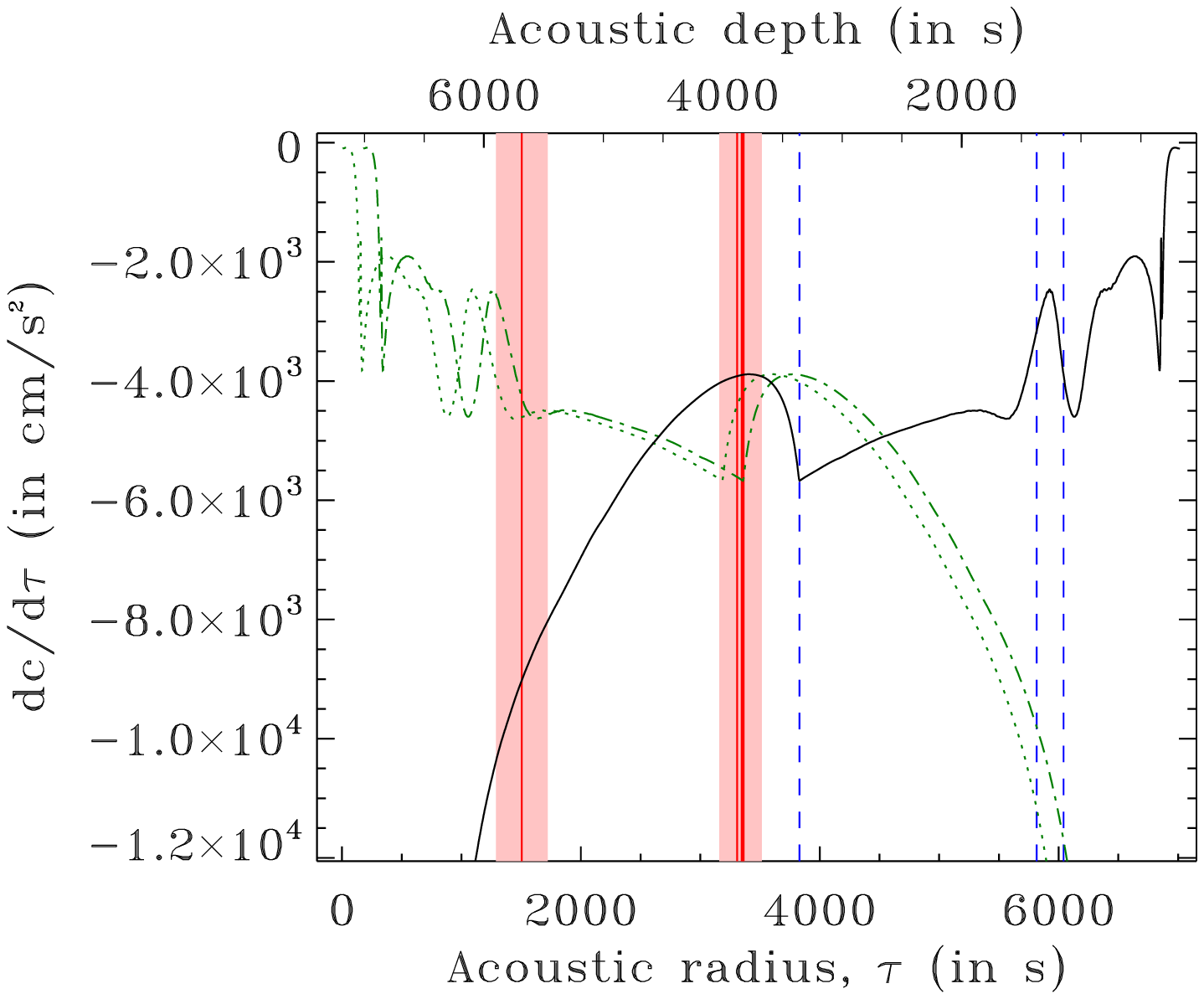}
\caption{(Colour online)  Comparison between glitch-based solutions for the
acoustic radius of the base of the convection zone in Felix (solid vertical red
lines and shaded pink area for the error bars), and the $dc/d\tau$ profile
(solid black curve).  The true solution, the \HeII\ ionisation zone, and the
peak in the $\Gamma_1$ profile, located between the \HeI\ and \HeII\ ionisation
zones, are indicated by the vertical dashed blue lines at $3830$ s, $5816$ s,
and $6041$ s, respectively.  The dotted and the dot-dashed green
curves show $dc/d\tau$ as a function of acoustic depth.  Given the uncertainties
on the determination of the total acoustic radius, the dotted curve uses
$\tau_{\mathrm{Tot.}}$ whereas the dot-dashed curve uses $1/2\Delta\nu$ (see
Table~\ref{tab:tau}).  The upper x-axis also uses $1/2\Delta\nu$
as the total acoustic radius. \label{fig:Felix_tau_BCZ}}
\end{figure}

In order to understand these spurious solutions found by the glitch fitting
approach, we looked at the sound-speed profile as a function of
acoustic radius to see if there are other features in the model which could
produce a glitch signature.  A first analysis showed that none of the spurious
solutions corresponded to a sharp acoustic feature located elsewhere in the
star.  However, a number of spurious solutions appeared to be complements (i.e.
acoustic depths rather than radii) of either the actual solution or of
approximately the \HeII\ ionisation zone, thereby implying an aliasing problem. 
For instance, for Diva, Felix, George, and Izzy, the complement to the solution
was found by some of the hounds.  Some of the solutions for Coco and Felix were
complements to the \HeII\ ionisation zone. Both of these situations are
illustrated in Fig.~\ref{fig:Felix_tau_BCZ} where the
$\mathrm{d}c/\mathrm{d}\tau$ profile and its complements are plotted for Felix.
Strictly speaking, aliasing such as illustrated in Fig.~6 of
\citet{Mazumdar2001} is only applicable for an analysis based on a single $\l$
value, but in practice unless the errors on the frequencies are very small, it
will manifest itself even when multiple $\l$ values are used. In some cases, even with
the knowledge of the ``correct'' value, it may not be possible to find a
corresponding peak in the distribution of glitch values. It was also
noted in \citet{Verma2014b} that fits to acoustic glitches tend to deteriorate
as the mode frequencies approach the Brunt-V{\"a}is{\"a}l{\"a} frequency; a
typical situation in the more massive stars due to sharp gradient that forms
above their convective core.  A possible explanation for this is the fact that
the asymptotic relation used to describe the frequencies is no longer valid,
thereby leading to deviations from the form of the fitting function used in the
glitch analysis.

\begin{figure}[htbp]
\includegraphics[width=\columnwidth]{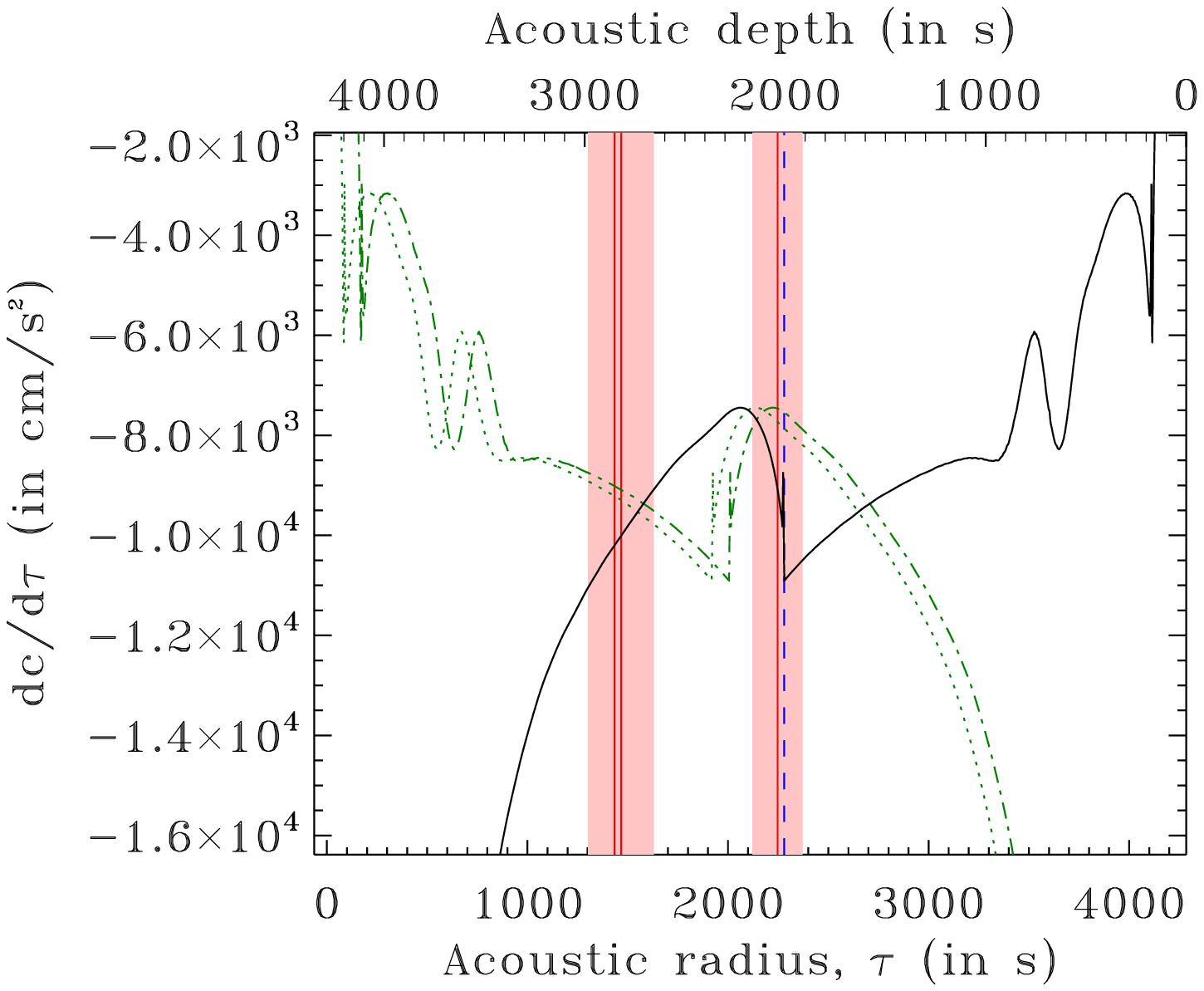}
\caption{(Colour online)  Same as Fig.~\ref{fig:Felix_tau_BCZ} but for Henry.
\label{fig:Henry_tau_BCZ}}
\end{figure}

Finally, no feature was found to explain the spurious solutions in Henry and Jam
(although we do note that the solutions found for Jam may marginally correspond
to the complement of the \HeII\ ionisation zone). Figure~\ref{fig:Henry_tau_BCZ}
compares the solutions for Henry to the $\mathrm{d}c/\mathrm{d}\tau$ profile and
its complements.  A possible explanation for Henry is that the signal to noise
ratio for the glitch signature is too low to allow us to obtain anything meaningful.  
Nonetheless, it is surprising that nearly the same erroneous solution is found
by more than one hound.

In this context, it is important to mention the role of atomic
diffusion.  Indeed, as has been shown in previous studies \citep[\eg][]
{Theado2005, Castro2006}, atomic diffusion reduces the helium content in the
convective envelope thereby leading to a helium gradient near its base and
slightly modifying its location.  This will then alter the amplitude of the
corresponding glitch signature and may facilitate its detection in some cases
(for instance in more massive stars).  This will remain true even when radiative
accelerations are included, as the radiative flux is unable to support this
amount of helium \citep[\eg][]{Vauclair1974}.   The fact that some of the
hounds included atomic diffusion (without radiative accelerations) whereas
others did not leads to inconsistencies.  The resultant increased dispersion
in the results can, however, be used to give us a first qualitative
idea of the effects of neglecting physical phenomena which occur in observed
stars.

\subsection{The \HeII\ ionisation zone and the $\Gamma_1$ peak}

The \HeII\  ionisation zone generally leads to a stronger glitch signature than the
base of the convection zone.  Accordingly, all of the hounds who applied a
glitch analysis returned estimates of the acoustic depth of this zone, which was
subsequently converted to acoustic radii using the $1/2\Delta\nu$ values
provided in Table~\ref{tab:tau}.   However, it is important to bear in mind that
the glitch signature corresponds to a region which tends to be near the
peak in the $\Gamma_1$ profile between the \HeI\ and {\small II} ionisation
zones, rather than the minimum in the $\Gamma_1$ curve resulting from the \HeII\
ionisation zone, as was recently pointed out by \citet{Broomhall2014} and
\citet{Verma2014b}.  Accordingly, in what follows we will compare the results
from V\&A, glitch, HRC, and AM with the acoustic radius of this peak,
which we will denote $\tau_{\mathrm{peak}}$.  The analysis by GH is
somewhat different because he fits both the \HeI\ and \HeII\  ionisation zones. 
Accordingly, his results will be compared with $\tau_{\mathrm{\HeII}}$.

\begin{figure*}[htbp]
\includegraphics[width=\textwidth]{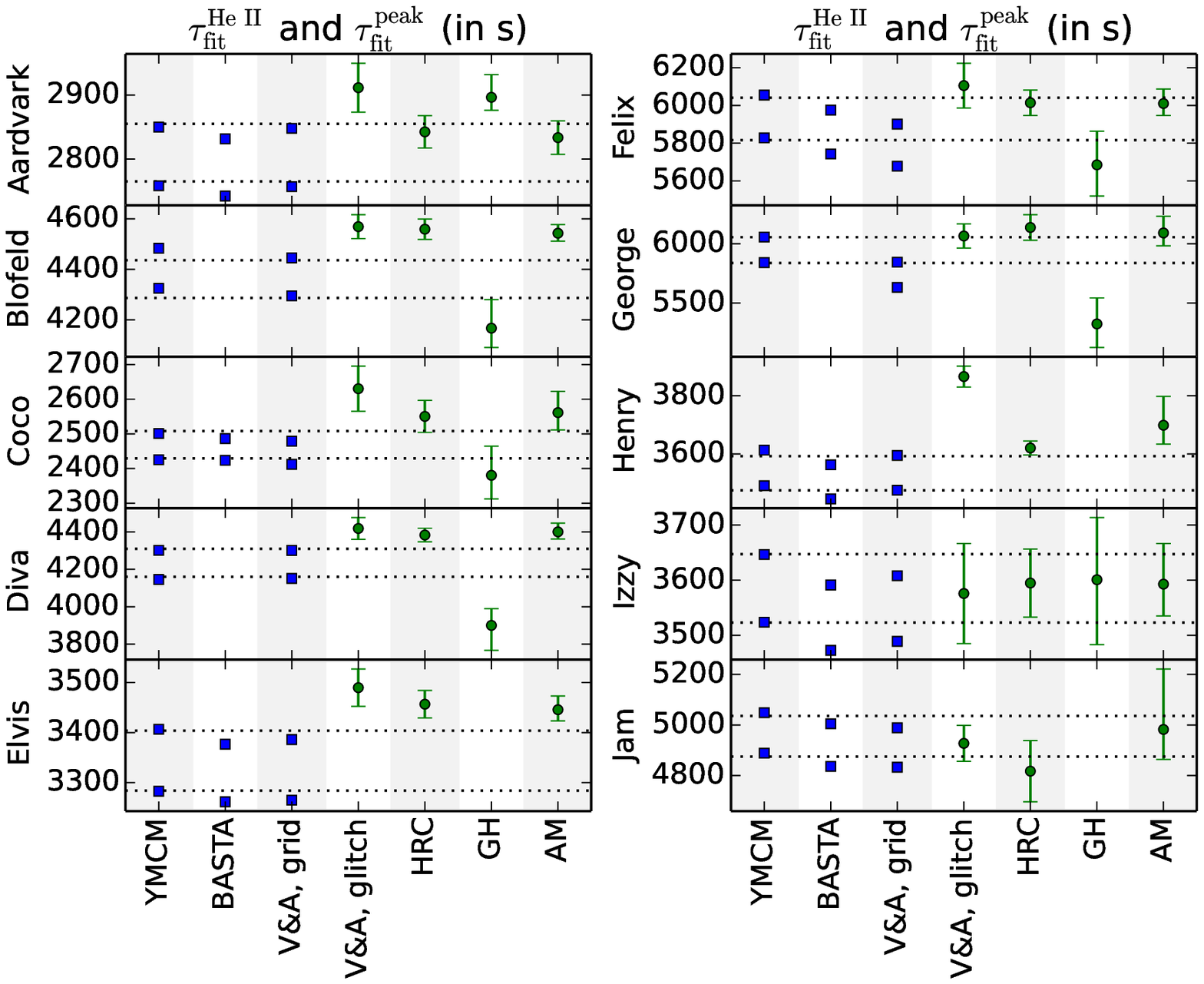}
\caption{Fitted results for $\tau_{\mathrm{\HeII}}$  and $\tau_{\mathrm{peak}}$
(in s). The two horizontal dotted lines correspond to the solutions
($\tau_{\mathrm{\HeII}}$ is always smaller than $\tau_{\mathrm{peak}}$), and the
various symbols correspond to the results from the different hounds.  The type
of symbol corresponds to the type method (forward modelling or glitch analysis).
\label{fig:AR_hiz}}
\end{figure*}

Beside the results from the glitch analysis, KV \& HMA also sent in
both $\tau_{\mathrm{\HeII}}$ and $\tau_{\mathrm{peak}}$ from their best-fitting
models.  The same values were also extracted from the actual solutions, and from
the best-fitting models produced by YMCM and BASTA.  All of these results are
displayed in Fig.~\ref{fig:AR_hiz}.  Tables~\ref{tab:AR_HeII}
and~\ref{tab:AR_peak} list the results and solutions.  Except for the results
from GH, all of the results from the glitch analysis are in
Table~\ref{tab:AR_peak}, which contains the results for $\tau_{\mathrm{peak}}$.

\begin{table*}[htbp]
\caption{Fitted values for $\tau_{\mathrm{\HeII}}$ (in s) and associated average
errors and biases. \label{tab:AR_HeII}}
\begin{tabular}{lcccccccc}
\hhline{=========}
\multicolumn{9}{c}{\textbf{$\tau_{\mathrm{\HeII}}$}} \\
\hhline{------~--}
\textbf{Hounds} & \textbf{Aardvark} & \textbf{Blofeld} & \textbf{Coco} & \textbf{Diva} & \textbf{Elvis} & & $\erel$ & $\brel$ \\
\hhline{------~--}
\SolutionCell{Solution} & $2765$ & $4287$ & $2429$ & $4160$ & $3284$  & & -- & -- \\
\GridCell{YMCM}& $2758$ & $4325$ & $2425$ & $4146$ & $3283$  & & $0.37\,\%$  & $0.11\,\%$  \\
\GridCell{BASTA}& $2742$  & -- & $2424$  & -- & $3262$  & & $0.94\,\%$  & $-0.87\,\%$  \\
\GridCell{V\&A, grid}& $2757$ & $4295$ & $2412$ & $4151$ & $3265$  & & $1.44\,\%$  & $-0.93\,\%$  \\
\GlitchCell{GH}& $2896_{-20}^{+35}$ & $4167_{-76}^{+113}$ & $2380_{-68}^{+84}$ & $3900_{-133}^{+89}$  & --  & & $4.80\,\%$  & $-2.17\,\%$  \\
$\erel$ & $2.42\,\%$ & $1.70\,\%$ & $1.08\,\%$ & $3.61\,\%$ & $0.52\,\%$ & & $2.36\,\%$ & -- \\
$\brel$ & $0.84\,\%$ & $-0.57\,\%$ & $-0.78\,\%$ & $-2.27\,\%$ & $-0.44\,\%$ & & -- & $-0.87\,\%$ \\
$\enorm$ & $4.74$ & $1.26$ & $0.64$ & $2.33$ &  --  & & -- & -- \\
$\bnorm$ & $4.74$ & $-1.26$ & $-0.64$ & $-2.33$ &  --  & & -- & -- \\
\hhline{------~--}
\textbf{Hounds} & \textbf{Felix} & \textbf{George} & \textbf{Henry} & \textbf{Izzy} & \textbf{Jam} & & $\enorm$ & $\bnorm$ \\
\hhline{------~--}
\SolutionCell{Solution} & $5816$ & $5838$ & $3475$ & $3523$ & $4875$  & & -- & -- \\
\GridCell{YMCM}& $5828$ & $5841$ & $3491$ & $3524$ & $4889$  & & --  & --  \\
\GridCell{BASTA}& $5743$  & -- & $3446$ & $3473$ & $4836$  & & --  & --  \\
\GridCell{V\&A, grid}& $5678$ & $5632$ & $3476$ & $3489$ & $4833$  & & --  & --  \\
\GlitchCell{GH}& $5685_{-166}^{+179}$ & $5324_{-199}^{+219}$  & -- & $3601_{-118}^{+113}$  & --  & & $2.30$ & $-0.29$ \\
$\erel$ & $1.75\,\%$ & $5.48\,\%$ & $0.56\,\%$ & $1.40\,\%$ & $0.69\,\%$ & & -- & -- \\
$\brel$ & $-1.42\,\%$ & $-4.10\,\%$ & $-0.12\,\%$ & $-0.04\,\%$ & $-0.46\,\%$ & & -- & -- \\
$\enorm$ & $0.76$ & $2.46$ &  --  & $0.67$ &  --  & & $2.30$ & -- \\
$\bnorm$ & $-0.76$ & $-2.46$ &  --  & $0.67$ &  --  & & -- & $-0.29$ \\
\hhline{------~--}
\end{tabular}

\end{table*}

\begin{table*}[htbp]
\caption{Fitted values for $\tau_{\mathrm{peak}}$ (in s) and associated average
errors and biases. \label{tab:AR_peak}}
\begin{tabular}{lcccccccc}
\hhline{=========}
\multicolumn{9}{c}{\textbf{$\tau_{\mathrm{peak}}$}} \\
\hhline{------~--}
\textbf{Hounds} & \textbf{Aardvark} & \textbf{Blofeld} & \textbf{Coco} & \textbf{Diva} & \textbf{Elvis} & & $\erel$ & $\brel$ \\
\hhline{------~--}
\SolutionCell{Solution} & $2855$ & $4436$ & $2508$ & $4310$ & $3404$  & & -- & -- \\
\GridCell{YMCM}& $2850$ & $4483$ & $2501$ & $4301$ & $3407$  & & $0.42\,\%$  & $0.16\,\%$  \\
\GridCell{BASTA}& $2832$  & -- & $2486$  & -- & $3377$  & & $0.97\,\%$  & $-0.93\,\%$  \\
\GridCell{V\&A, grid}& $2848$ & $4445$ & $2479$ & $4301$ & $3386$  & & $1.46\,\%$  & $-0.97\,\%$  \\
\GlitchCell{V\&A, glitch}& $2911 \pm 38$ & $4569 \pm 47$ & $2630 \pm 65$ & $4418 \pm 58$ & $3490 \pm 37$  & & $3.41\,\%$  & $1.96\,\%$  \\
\GlitchCell{HRC}& $2842 \pm 25$ & $4559 \pm 40$ & $2550 \pm 46$ & $4383 \pm 36$ & $3457 \pm 27$  & & $1.99\,\%$  & $0.32\,\%$  \\
\GlitchCell{AM}& $2833 \pm 26$ & $4543_{-31}^{+34}$ & $2561_{-50}^{+61}$ & $4400_{-38}^{+46}$ & $3446_{-22}^{+27}$  & & $1.71\,\%$  & $0.76\,\%$  \\
$\erel$ & $0.95\,\%$ & $2.17\,\%$ & $2.36\,\%$ & $1.65\,\%$ & $1.37\,\%$ & & $1.94\,\%$ & -- \\
$\brel$ & $-0.08\,\%$ & $1.89\,\%$ & $1.07\,\%$ & $1.18\,\%$ & $0.68\,\%$ & & -- & $0.27\,\%$ \\
$\enorm$ & $1.02$ & $3.03$ & $1.33$ & $2.00$ & $1.99$ & & -- & -- \\
$\bnorm$ & $0.05$ & $3.03$ & $1.25$ & $1.99$ & $1.97$ & & -- & -- \\
\hhline{------~--}
\textbf{Hounds} & \textbf{Felix} & \textbf{George} & \textbf{Henry} & \textbf{Izzy} & \textbf{Jam} & & $\enorm$ & $\bnorm$ \\
\hhline{------~--}
\SolutionCell{Solution} & $6041$ & $6056$ & $3593$ & $3647$ & $5036$  & & -- & -- \\
\GridCell{YMCM}& $6055$ & $6056$ & $3613$ & $3647$ & $5049$  & & --  & --  \\
\GridCell{BASTA}& $5975$  & -- & $3563$ & $3591$ & $5005$  & & --  & --  \\
\GridCell{V\&A, grid}& $5901$ & $5845$ & $3595$ & $3608$ & $4989$  & & --  & --  \\
\GlitchCell{V\&A, glitch}& $6105 \pm 119$ & $6066 \pm 102$ & $3866 \pm 36$ & $3576 \pm 91$ & $4927 \pm 71$  & & $2.89$ & $1.62$ \\
\GlitchCell{HRC}& $6014 \pm 68$ & $6138 \pm 108$ & $3621 \pm 24$ & $3595 \pm 62$ & $4817 \pm 121$  & & $1.55$ & $0.63$ \\
\GlitchCell{AM}& $6010_{-64}^{+77}$ & $6092_{-108}^{+140}$ & $3699_{-65}^{+99}$ & $3593_{-58}^{+74}$ & $4982_{-118}^{+239}$  & & $1.49$ & $0.72$ \\
$\erel$ & $1.17\,\%$ & $1.69\,\%$ & $3.36\,\%$ & $1.39\,\%$ & $2.08\,\%$ & & -- & -- \\
$\brel$ & $-0.51\,\%$ & $-0.27\,\%$ & $1.85\,\%$ & $-1.25\,\%$ & $-1.47\,\%$ & & -- & -- \\
$\enorm$ & $0.46$ & $0.47$ & $4.51$ & $0.82$ & $1.37$ & & $2.08$ & -- \\
$\bnorm$ & $-0.10$ & $0.38$ & $3.35$ & $-0.82$ & $-1.21$ & & -- & $0.99$ \\
\hhline{------~--}
\end{tabular}

\end{table*}

Overall, the results for $\tau_{\mathrm{\HeII}}$ and $\tau_{\mathrm{peak}}$ are
much more accurate than the results for $\tau_{\mathrm{BCZ}}$. This is expected
given the stronger glitch signature from these features.  Also, the results
based on best-fitting models are more accurate than those from the glitch
analysis, as was the case for $\tau_{\mathrm{BCZ}}$. A more detailed look shows
that, except for Aardvark and Izzy, GH found lower results than all of the other
hounds who applied glitch analysis.  This is expected since he is fitting the
\HeI\  and \HeII\ ionisation zones separately.  In two cases, his results are
too low.  Diva can be explained by the fact that the GH analysis measures
acoustic depths relative to the point where the linearly extrapolated $c^2$
profile vanishes, as pointed out in Section~\ref{sect:glitch_group}, rather than
the point corresponding to $1/2\Delta\nu$, which lies below.  For George, there
seems to be multiple local minima, one of which corresponds to the result given
here, and another which is consistent with the results provided by the other
hounds. This shows once more the limitations of local optimisation methods.
Moreover, it is questionable whether the adopted solar-based parameters for
relating the \HeI\ to \HeII\  glitch are still appropriate at such a high
luminosity.

Once more, atomic diffusion can affect the detection of the helium
ionisation zones.  Indeed, the helium abundance in the convective envelope is
reduced by diffusion, even if radiative accelerations are present, thereby
modifying the amplitude of the corresponding glitch signature
\citep[\eg][]{Theado2005, Castro2006}.  Accordingly, glitch analysis will only
yield the helium abundance in the convective envelope, which will differ from
the helium abundance below as a result of atomic diffusion.

\subsection{Structural profiles}

\begin{figure}[htbp]
\includegraphics[width=\columnwidth]{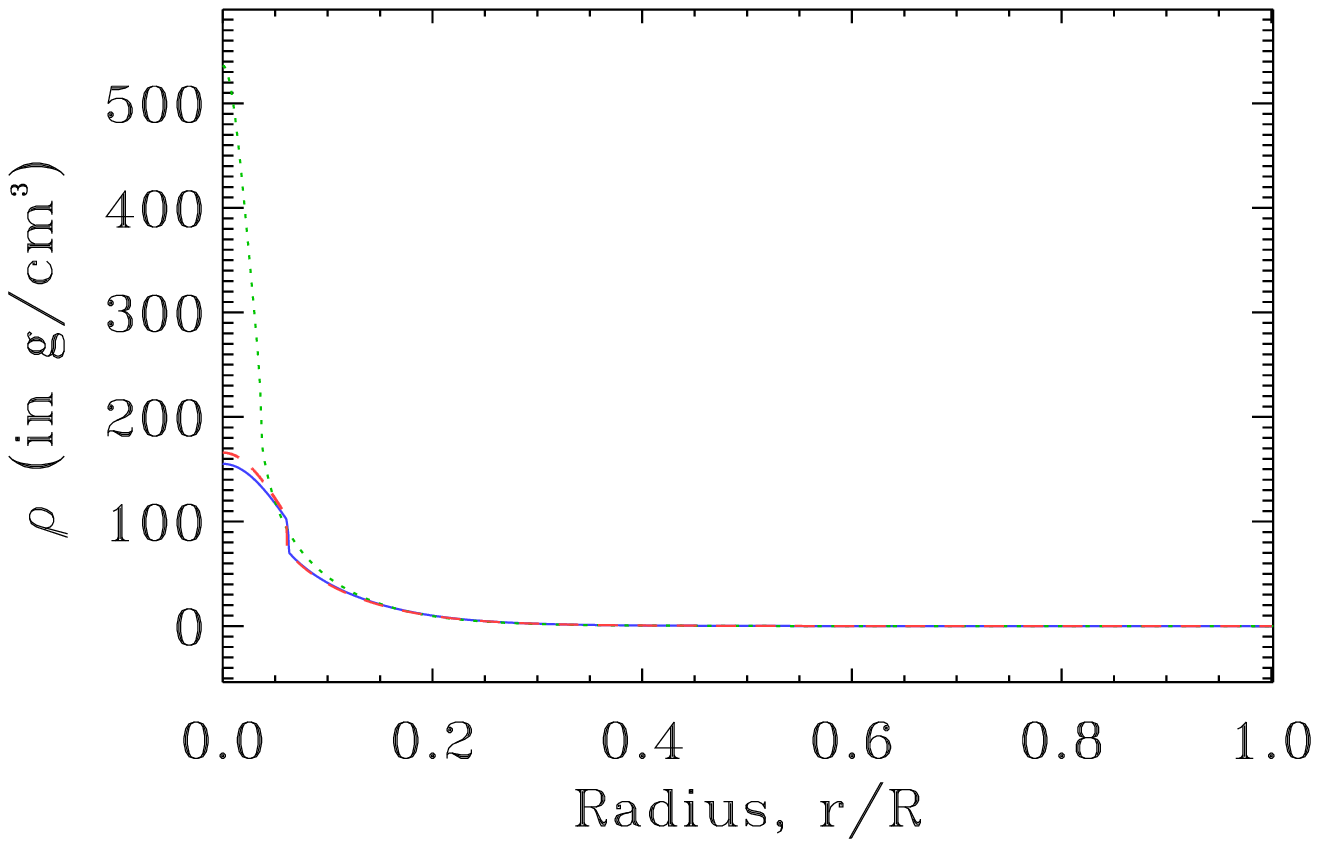} \\
\includegraphics[width=\columnwidth]{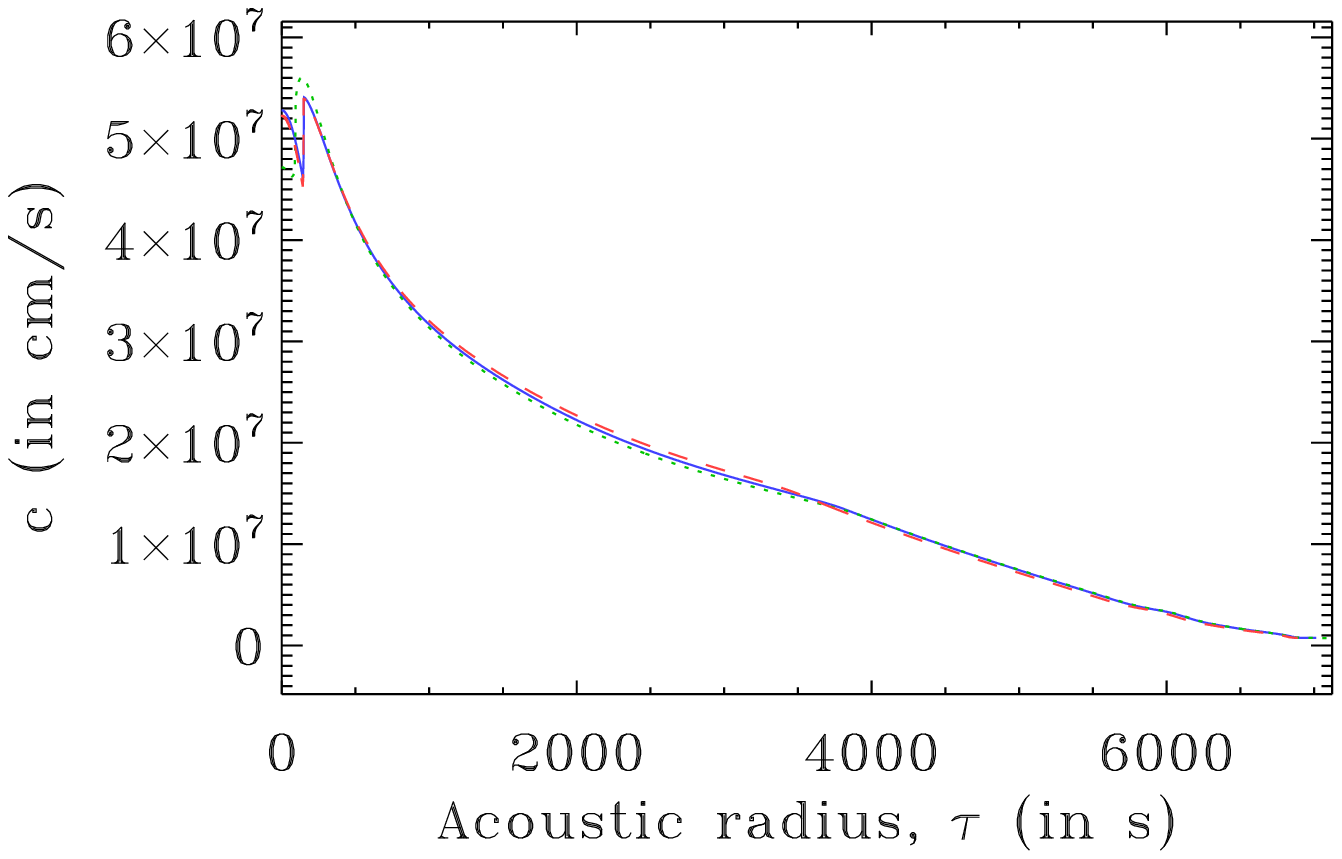} \\
\includegraphics[width=\columnwidth]{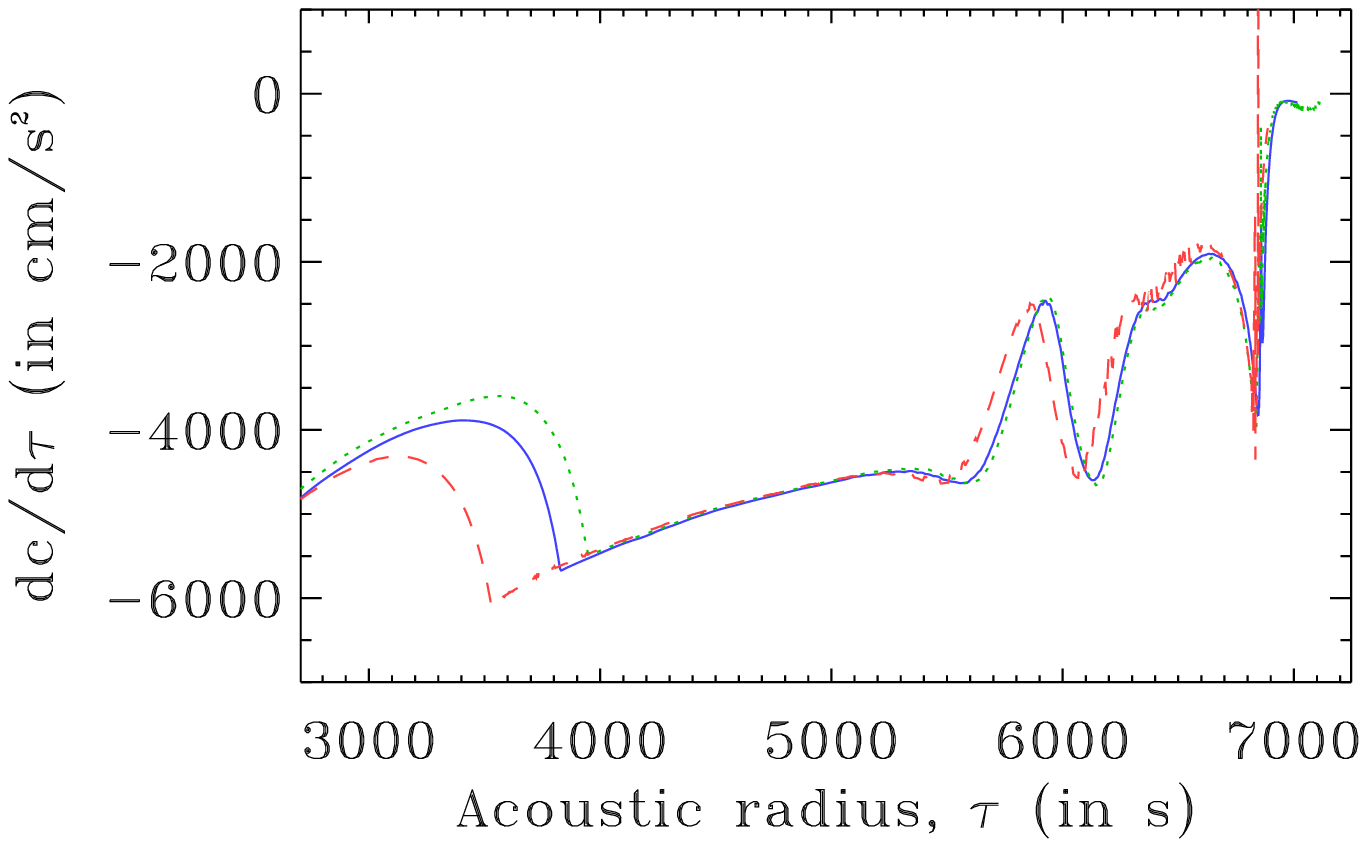}
\caption{(Colour online) Various structural profiles for Felix (solid blue lines)
and for the relevant best-fitting models from YMCM (dotted green lines)
and BASTA (dashed red lines).
\label{fig:profiles_felix}}
\end{figure}

\begin{figure}[htbp]
\includegraphics[width=\columnwidth]{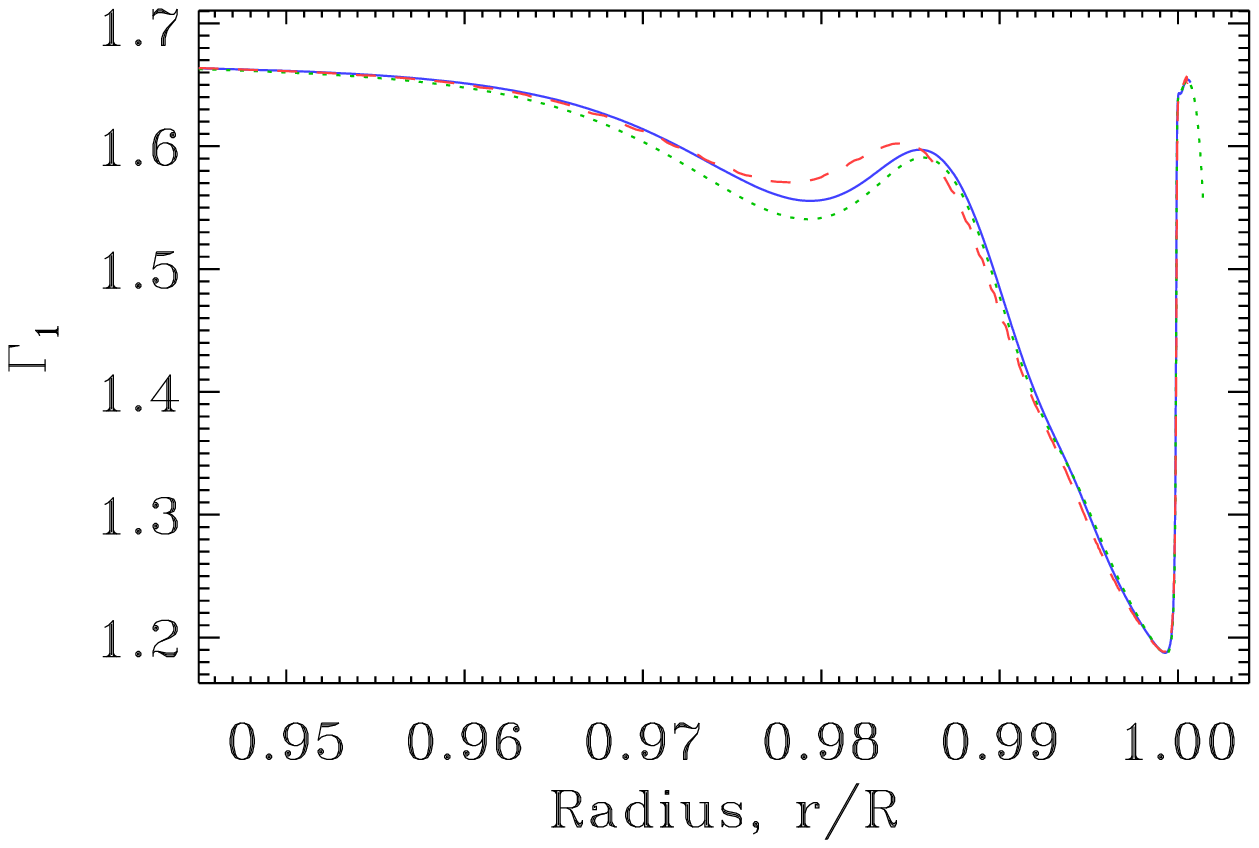}
\caption{(Colour online) $\Gamma_1$ profile for Elvis (solid blue lines)
and for the relevant best-fitting models from YMCM (dotted green lines)
and BASTA (dashed red lines).
\label{fig:gamma1_elvis}}
\end{figure}

Finally, in this section we compare some of the structural profiles from the
solutions and from the best-fitting models from YMCM and BASTA.  A systematic
investigation of all of the targets showed a mixture of results.  For some of
the targets, the results found by the hounds were very similar to the correct
solution, whereas non-negligible differences showed up for other targets.  Sharp
density gradients near the core tended to be problematic and could lead
to incorrect profiles in the entire core, especially for YMCM, as illustrated
in Fig.~\ref{fig:profiles_felix} (upper panel).  However, it is not too
surprising that these features are difficult to reproduce since they only take
up a small portion of the star in terms of acoustic radius and only lead to
small differences in the sound-speed profile, as shown in
Fig.~\ref{fig:profiles_felix} (middle panel).  Also, the extent of the
stellar atmospheres in the models used by BASTA were more limited than those of
the stellar targets, which in turn were more limited than those of the YMCM
models.  This led to differences in the total acoustic radii of the various
models and stellar targets (when integrating $dr/c$ to the last mesh point)
and meant that the hounds were unable to fit both the
acoustic radius and the acoustic depth of features such as the \HeII\ ionisation
zones or the base of the convection zone.  A systematic look at the results
revealed that BASTA did a better job at reproducing acoustic depths of the
\HeII\ ionisation zones to the detriment of their acoustic radii, as illustrated
in Fig.~\ref{fig:profiles_felix} (lower panel), whereas YMCM reproduced acoustic
radii more accurately.  The reason for this difference in behaviour between
the two methods is not entirely clear, but it does highlight the
impact of the extent of the atmosphere.  Sometimes, the \HeII\ ionisation zone
was not well reproduced, both in terms of physical location and depth
in the $\Gamma_1$ profile, as illustrated in Fig.~\ref{fig:gamma1_elvis}. 
Nonetheless, in spite of these differences, the sound-speed profile remained
similar between the targets and the best fitting models in all of the cases. 
This is probably because of the high sensitivity of acoustic modes to the sound
speed.

\section{Conclusion}

This article describes the results of a hare-and-hounds exercise conducted 
within the context of the SpaceInn network.  In this exercise, observational
data, including detailed frequency spectra as well as classic observables
($\Teff$, [Fe/H], $L$), for a set of $10$ artificial stars were provided by a
group of ``hares''.  Other participants, the ``hounds'', applied various
methodologies in order to deduce the properties of these stars as well as
realistic error bars.  The hounds were subdivided into two main groups: the
first applied forward modelling and the second relied on acoustic glitch
signatures.  In addition to these groups, two other hounds used inverse
techniques.

The overall accuracies on radius, mass, age, surface gravity, and mean density,
as based on forward modelling, were  $1.5\,\%$, $3.9\,\%$, $23\,\%$, $1.5\,\%$,
and $1.8\,\%$, respectively.  Furthermore, these accuracies become $1.2\,\%$,
$3.2\,\%$, and $8.2\,\%$ on radius, mass, and age, respectively, for the two $1$
M$_{\odot}$ stars, thereby easily satisfying the requirements for the PLATO 2.0
mission.  The stars that proved to be the most challenging were Felix, George,
and Jam, due to their high mass, and Blofeld, probably because of diffusion at a
relatively high mass and/or the different abundance mixture.  Indeed, high-mass
stars are hotter, thereby leading to shorter mode lifetimes and larger error
bars on their frequencies, and contain convective cores, the sizes of which
strongly depend on the various prescriptions used in stellar evolution codes. 
Atomic diffusion, in which radiative accelerations are
neglected as is the case here, leads to depletion of heavy elements at the
surface of higher mass stars (in contradiction with current
observations) and is therefore usually only included in lower mass models.

Taking into account results from both forward modelling and glitch analysis, the
average errors on the acoustic radii $\tau_{\mathrm{BCZ}}$,
$\tau_{\mathrm{\HeII}}$, and $\tau_{\mathrm{peak}}$ were $17\,\%$,
$2.4\,\%$, and $1.9\,\%$, respectively.  Furthermore, forward modelling
results tended to be more accurate than those from glitch analysis, which seemed
to be affected by aliasing problems in a number of cases.  One possible
explanation is that glitch analysis finds multiple local minima and needs prior
information before being able to select the correct one.

Overall, forward modelling seems to be the most promising way of carrying out
detailed asteroseismology in solar-type stars.  Nonetheless, it is, by
construction, very model-dependent and will benefit greatly from the results of
methods that are less model-dependent like seismic inversions, or
model-independent like glitch analysis.  Indeed, the present exercise only tests
the ability of various asteroseismic methods at reproducing the properties of
artificial stars.  As such, it is unable to test the effects of hitherto unknown
or poorly modelled physical phenomena present in real stars. 
More realistic models should include the effects of radiative
accelerations, a more realistic description of convection \citep[for instance based
on 3D simulations, see \eg][]{Trampedach2014,Magic2015}, rotation and the mixing
it induces \citep[\eg][]{Eggenberger2010}, magnetic activity cycles, a
more realistic atmosphere etc.  In addition to being less model-dependent,
glitch analysis is not always prone to the same difficulties as the forward
modelling approach, thereby making the two methods complementary. Finally,
results from one of the hounds suggest that global optimisation algorithms
should be used instead of local ones in order to obtain robust error bars, given
that the latter are more prone to being ``trapped'' in local minima.

\begin{acknowledgements}

We thank the referee, S. Vauclair, for clarifications concerning the
role of atomic diffusion, and for other comments and suggestions. DRR was
funded by the European Community's Seventh Framework Programme (FP7/2007-2013)
under grant agreement no. 312844 (SPACEINN), which is gratefully acknowledged.
This article made use of InversionKit and InversionPipeline, inversion software
developed in the context of the HELAS and SPACEINN networks, funded by the
European Commission's Sixth and Seventh Framework Programmes. WHB acknowledges
research funding by Deutsche Forschungsgemeinschaft (DFG) under grant SFB 963/1
"Astrophysical flow instabilities and turbulence", Project A18. SB acknowledges
NASA grant NNX13AE70G and NSF grant AST-1514676. GB is supported by the FNRS
(``Fonds National de la Recherche Scientifique'') through a FRIA (``Fonds pour
la Formation à la Recherche dans l'Industrie et l'Agriculture'') doctoral
fellowship. Funding for the Stellar Astrophysics Centre is provided by The
Danish National Research Foundation (Grant DNRF106). The research is supported
by the ASTERISK project (ASTERoseismic Investigations with SONG and Kepler)
funded by the European Research Council (Grant agreement no.: 267864). The
research leading to the presented results has received funding from the European
Research Council under the European Community's Seventh Framework Programme
(FP7/2007-2013) / ERC grant agreement no 338251 (StellarAges). AM was supported
by the NIUS programme of HBCSE (TIFR). VSA acknowledges support from VILLUM
FONDEN (research grant 10118).

\end{acknowledgements}

\bibliographystyle{aa}
\bibliography{biblio}

\appendix

\section{True and ``observed'' hare frequencies}
\label{sect:pulsation_frequencies}

Tables~\ref{tab:freq_hares1} to~\ref{tab:freq_hares3} give the ``observational''
frequencies as well as the genuine model frequencies.

\begin{table*}[p]
\rotatebox[origin=c]{90}{
\parbox[][][t]{24cm}{
\caption{``Observed'' and exact frequencies for Aardvark, Blofeld, Coco, and
Diva. \label{tab:freq_hares1}}
\begin{tabular}{cccc}
\begin{tabular}{cccc}
\hline
\hline
\multicolumn{4}{c}{\textbf{Aardvark}} \\
$\l$ & $n$ & 
$\nu_{\mathrm{obs}}$ & 
$\nu_{\mathrm{exact}}$ \\
\hline
0 & 16 & $2513.93 \pm 0.21$ & 2514.04 \\
0 & 17 & $2656.81 \pm 0.13$ & 2656.78 \\
0 & 18 & $2799.64 \pm 0.10$ & 2799.55 \\
0 & 19 & $2942.98 \pm 0.08$ & 2942.89 \\
0 & 20 & $3087.25 \pm 0.06$ & 3087.15 \\
0 & 21 & $3232.06 \pm 0.05$ & 3232.13 \\
0 & 22 & $3377.04 \pm 0.06$ & 3377.04 \\
0 & 23 & $3522.31 \pm 0.09$ & 3522.22 \\
0 & 24 & $3667.38 \pm 0.14$ & 3667.55 \\
0 & 25 & $3813.47 \pm 0.27$ & 3813.28 \\
0 & 26 & $3960.05 \pm 0.61$ & 3959.52 \\
0 & 27 & $4104.57 \pm 1.74$ & 4105.94 \\
1 & 15 & $2436.54 \pm 0.37$ & 2437.36 \\
1 & 16 & $2581.10 \pm 0.19$ & 2581.06 \\
1 & 17 & $2723.88 \pm 0.12$ & 2723.89 \\
1 & 18 & $2866.68 \pm 0.08$ & 2866.81 \\
1 & 19 & $3010.96 \pm 0.06$ & 3010.94 \\
1 & 20 & $3155.64 \pm 0.05$ & 3155.62 \\
1 & 21 & $3300.81 \pm 0.05$ & 3300.76 \\
1 & 22 & $3446.02 \pm 0.06$ & 3445.98 \\
1 & 23 & $3591.19 \pm 0.09$ & 3591.21 \\
1 & 24 & $3736.86 \pm 0.14$ & 3736.95 \\
1 & 25 & $3883.11 \pm 0.27$ & 3882.94 \\
1 & 26 & $4029.39 \pm 0.60$ & 4029.41 \\
1 & 27 & $4178.11 \pm 1.69$ & 4176.15 \\
2 & 17 & $2787.17 \pm 0.22$ & 2787.17 \\
2 & 18 & $2930.57 \pm 0.15$ & 2930.70 \\
2 & 19 & $3075.21 \pm 0.11$ & 3075.28 \\
2 & 20 & $3220.54 \pm 0.09$ & 3220.54 \\
2 & 21 & $3365.94 \pm 0.09$ & 3365.77 \\
2 & 22 & $3510.96 \pm 0.13$ & 3511.27 \\
2 & 23 & $3656.53 \pm 0.19$ & 3656.85 \\
2 & 24 & $3802.63 \pm 0.31$ & 3802.87 \\
2 & 25 & $3949.44 \pm 0.57$ & 3949.35 \\
2 & 26 & $4096.32 \pm 1.28$ & 4096.03 \\
\hline
~
\end{tabular} &
\begin{tabular}{cccc}
\hline
\hline
\multicolumn{4}{c}{\textbf{Blofeld}} \\
$\l$ & $n$ & 
$\nu_{\mathrm{obs}}$ & 
$\nu_{\mathrm{exact}}$ \\
\hline
0 & 15 & $1517.74 \pm 0.19$ & 1517.59 \\
0 & 16 & $1610.98 \pm 0.12$ & 1610.96 \\
0 & 17 & $1703.06 \pm 0.09$ & 1703.30 \\
0 & 18 & $1796.19 \pm 0.07$ & 1796.09 \\
0 & 19 & $1889.74 \pm 0.06$ & 1889.80 \\
0 & 20 & $1984.57 \pm 0.06$ & 1984.51 \\
0 & 21 & $2079.41 \pm 0.06$ & 2079.42 \\
0 & 22 & $2174.07 \pm 0.09$ & 2174.16 \\
0 & 23 & $2269.01 \pm 0.13$ & 2268.82 \\
0 & 24 & $2363.41 \pm 0.22$ & 2363.43 \\
0 & 25 & $2458.42 \pm 0.47$ & 2458.39 \\
0 & 26 & $2555.02 \pm 1.25$ & 2553.34 \\
1 & 14 & $1467.72 \pm 0.32$ & 1467.36 \\
1 & 15 & $1561.03 \pm 0.16$ & 1560.88 \\
1 & 16 & $1653.93 \pm 0.10$ & 1653.89 \\
1 & 17 & $1746.23 \pm 0.07$ & 1746.19 \\
1 & 18 & $1839.73 \pm 0.06$ & 1839.61 \\
1 & 19 & $1933.82 \pm 0.05$ & 1933.80 \\
1 & 20 & $2028.97 \pm 0.05$ & 2028.94 \\
1 & 21 & $2123.71 \pm 0.06$ & 2123.76 \\
1 & 22 & $2218.68 \pm 0.08$ & 2218.69 \\
1 & 23 & $2313.29 \pm 0.13$ & 2313.30 \\
1 & 24 & $2408.61 \pm 0.21$ & 2408.32 \\
1 & 25 & $2502.62 \pm 0.45$ & 2503.32 \\
1 & 26 & $2598.65 \pm 1.18$ & 2598.60 \\
2 & 15 & $1601.86 \pm 0.31$ & 1602.81 \\
2 & 16 & $1695.17 \pm 0.19$ & 1695.24 \\
2 & 17 & $1787.80 \pm 0.13$ & 1787.96 \\
2 & 18 & $1881.75 \pm 0.10$ & 1881.70 \\
2 & 19 & $1976.38 \pm 0.09$ & 1976.41 \\
2 & 20 & $2071.52 \pm 0.10$ & 2071.48 \\
2 & 21 & $2166.50 \pm 0.13$ & 2166.33 \\
2 & 22 & $2261.31 \pm 0.18$ & 2261.17 \\
2 & 23 & $2355.97 \pm 0.27$ & 2355.89 \\
2 & 24 & $2450.19 \pm 0.45$ & 2451.03 \\
2 & 25 & $2546.46 \pm 0.93$ & 2546.13 \\
\hline
\end{tabular} &
\begin{tabular}{cccc}
\hline
\hline
\multicolumn{4}{c}{\textbf{Coco}} \\
$\l$ & $n$ & 
$\nu_{\mathrm{obs}}$ & 
$\nu_{\mathrm{exact}}$ \\
\hline
0 & 15 & $2659.48 \pm 0.42$ & 2659.41 \\
0 & 16 & $2819.69 \pm 0.23$ & 2819.60 \\
0 & 17 & $2979.64 \pm 0.15$ & 2979.55 \\
0 & 18 & $3140.44 \pm 0.12$ & 3140.41 \\
0 & 19 & $3301.93 \pm 0.09$ & 3301.98 \\
0 & 20 & $3464.39 \pm 0.07$ & 3464.46 \\
0 & 21 & $3626.91 \pm 0.07$ & 3626.93 \\
0 & 22 & $3789.81 \pm 0.10$ & 3789.74 \\
0 & 23 & $3952.92 \pm 0.17$ & 3952.97 \\
0 & 24 & $4116.51 \pm 0.32$ & 4116.59 \\
0 & 25 & $4280.24 \pm 0.70$ & 4280.88 \\
0 & 26 & $4442.54 \pm 1.94$ & 4445.36 \\
1 & 15 & $2735.43 \pm 0.38$ & 2735.35 \\
1 & 16 & $2896.29 \pm 0.21$ & 2895.98 \\
1 & 17 & $3056.18 \pm 0.13$ & 3056.43 \\
1 & 18 & $3218.06 \pm 0.10$ & 3218.11 \\
1 & 19 & $3380.58 \pm 0.07$ & 3380.50 \\
1 & 20 & $3543.23 \pm 0.05$ & 3543.17 \\
1 & 21 & $3706.09 \pm 0.06$ & 3706.17 \\
1 & 22 & $3869.20 \pm 0.10$ & 3869.21 \\
1 & 23 & $4032.80 \pm 0.17$ & 4032.91 \\
1 & 24 & $4196.62 \pm 0.31$ & 4196.94 \\
1 & 25 & $4360.32 \pm 0.67$ & 4361.43 \\
1 & 26 & $4528.35 \pm 1.83$ & 4526.28 \\
2 & 16 & $2969.51 \pm 0.41$ & 2970.07 \\
2 & 17 & $3131.65 \pm 0.25$ & 3131.45 \\
2 & 18 & $3293.38 \pm 0.17$ & 3293.55 \\
2 & 19 & $3456.36 \pm 0.12$ & 3456.54 \\
2 & 20 & $3619.53 \pm 0.10$ & 3619.46 \\
2 & 21 & $3782.88 \pm 0.15$ & 3782.73 \\
2 & 22 & $3946.85 \pm 0.23$ & 3946.32 \\
2 & 23 & $4109.72 \pm 0.37$ & 4110.31 \\
2 & 24 & $4274.87 \pm 0.65$ & 4274.91 \\
2 & 25 & $4439.61 \pm 1.39$ & 4439.66 \\
\hline
~ \\ ~
\end{tabular} &
\begin{tabular}{cccc}
\hline
\hline
\multicolumn{4}{c}{\textbf{Diva}} \\
$\l$ & $n$ & 
$\nu_{\mathrm{obs}}$ & 
$\nu_{\mathrm{exact}}$ \\
\hline
0 & 14 & $1459.23 \pm 0.30$ & 1459.12 \\
0 & 15 & $1554.45 \pm 0.17$ & 1554.63 \\
0 & 16 & $1649.27 \pm 0.11$ & 1649.07 \\
0 & 17 & $1743.25 \pm 0.08$ & 1743.23 \\
0 & 18 & $1837.95 \pm 0.07$ & 1837.83 \\
0 & 19 & $1933.59 \pm 0.06$ & 1933.69 \\
0 & 20 & $2030.18 \pm 0.06$ & 2030.09 \\
0 & 21 & $2126.64 \pm 0.07$ & 2126.64 \\
0 & 22 & $2223.08 \pm 0.10$ & 2223.11 \\
0 & 23 & $2319.52 \pm 0.16$ & 2319.56 \\
0 & 24 & $2416.20 \pm 0.30$ & 2416.37 \\
0 & 25 & $2513.97 \pm 0.69$ & 2513.41 \\
1 & 14 & $1502.60 \pm 0.27$ & 1502.84 \\
1 & 15 & $1598.28 \pm 0.14$ & 1598.23 \\
1 & 16 & $1692.62 \pm 0.09$ & 1692.53 \\
1 & 17 & $1787.00 \pm 0.07$ & 1787.00 \\
1 & 18 & $1882.52 \pm 0.06$ & 1882.48 \\
1 & 19 & $1978.82 \pm 0.05$ & 1978.84 \\
1 & 20 & $2075.65 \pm 0.05$ & 2075.73 \\
1 & 21 & $2172.26 \pm 0.07$ & 2172.43 \\
1 & 22 & $2269.35 \pm 0.10$ & 2269.23 \\
1 & 23 & $2366.03 \pm 0.15$ & 2366.07 \\
1 & 24 & $2463.26 \pm 0.29$ & 2463.32 \\
1 & 25 & $2559.90 \pm 0.66$ & 2560.86 \\
1 & 26 & $2659.59 \pm 1.99$ & 2658.60 \\
2 & 15 & $1641.81 \pm 0.28$ & 1642.13 \\
2 & 16 & $1736.43 \pm 0.17$ & 1736.46 \\
2 & 17 & $1831.09 \pm 0.13$ & 1831.19 \\
2 & 18 & $1927.13 \pm 0.10$ & 1927.23 \\
2 & 19 & $2023.75 \pm 0.09$ & 2023.84 \\
2 & 20 & $2120.58 \pm 0.11$ & 2120.68 \\
2 & 21 & $2217.56 \pm 0.15$ & 2217.42 \\
2 & 22 & $2313.95 \pm 0.21$ & 2314.16 \\
2 & 23 & $2411.21 \pm 0.33$ & 2411.24 \\
2 & 24 & $2508.22 \pm 0.60$ & 2508.56 \\
2 & 25 & $2602.63 \pm 1.38$ & 2606.27 \\
\hline
\end{tabular}
\end{tabular}}}
\end{table*}

\begin{table*}[htbp]
\rotatebox[origin=c]{90}{
\parbox[][][t]{24cm}{
\caption{``Observed'' and exact frequencies for Elvis, Felix, George, and Henry.
\label{tab:freq_hares2}}
\begin{tabular}{cccc}
\begin{tabular}{cccc}
\hline
\hline
\multicolumn{4}{c}{\textbf{Elvis}} \\
$\l$ & $n$ & 
$\nu_{\mathrm{obs}}$ & 
$\nu_{\mathrm{exact}}$ \\
\hline
0 & 14 & $1844.12 \pm 0.29$ & 1843.96 \\
0 & 15 & $1963.29 \pm 0.16$ & 1963.38 \\
0 & 16 & $2082.00 \pm 0.11$ & 2081.97 \\
0 & 17 & $2200.05 \pm 0.08$ & 2200.08 \\
0 & 18 & $2319.04 \pm 0.06$ & 2319.10 \\
0 & 19 & $2439.21 \pm 0.05$ & 2439.27 \\
0 & 20 & $2559.80 \pm 0.05$ & 2559.75 \\
0 & 21 & $2680.52 \pm 0.07$ & 2680.45 \\
0 & 22 & $2801.07 \pm 0.10$ & 2801.09 \\
0 & 23 & $2921.66 \pm 0.18$ & 2921.98 \\
0 & 24 & $3042.91 \pm 0.36$ & 3043.37 \\
0 & 25 & $3165.13 \pm 0.95$ & 3164.99 \\
1 & 14 & $1899.39 \pm 0.26$ & 1899.26 \\
1 & 15 & $2018.53 \pm 0.14$ & 2018.59 \\
1 & 16 & $2136.91 \pm 0.09$ & 2136.99 \\
1 & 17 & $2255.91 \pm 0.07$ & 2255.87 \\
1 & 18 & $2375.64 \pm 0.05$ & 2375.62 \\
1 & 19 & $2496.34 \pm 0.05$ & 2496.34 \\
1 & 20 & $2617.23 \pm 0.05$ & 2617.26 \\
1 & 21 & $2738.05 \pm 0.07$ & 2738.05 \\
1 & 22 & $2858.97 \pm 0.10$ & 2859.12 \\
1 & 23 & $2979.95 \pm 0.18$ & 2980.38 \\
1 & 24 & $3102.13 \pm 0.35$ & 3102.11 \\
1 & 25 & $3224.11 \pm 0.92$ & 3224.13 \\
2 & 15 & $2073.40 \pm 0.27$ & 2073.85 \\
2 & 16 & $2192.17 \pm 0.17$ & 2192.25 \\
2 & 17 & $2311.66 \pm 0.12$ & 2311.59 \\
2 & 18 & $2431.91 \pm 0.09$ & 2432.04 \\
2 & 19 & $2552.84 \pm 0.09$ & 2552.87 \\
2 & 20 & $2674.09 \pm 0.11$ & 2673.95 \\
2 & 21 & $2794.84 \pm 0.15$ & 2794.91 \\
2 & 22 & $2916.25 \pm 0.23$ & 2916.13 \\
2 & 23 & $3037.48 \pm 0.38$ & 3037.82 \\
2 & 24 & $3160.07 \pm 0.76$ & 3159.74 \\
\hline
\end{tabular} &
\begin{tabular}{cccc}
\hline
\hline
\multicolumn{4}{c}{\textbf{Felix}} \\
$\l$ & $n$ & 
$\nu_{\mathrm{obs}}$ & 
$\nu_{\mathrm{exact}}$ \\
\hline
0 & 13 & $976.77 \pm 0.46$ & 976.57 \\
0 & 14 & $1046.11 \pm 0.24$ & 1046.17 \\
0 & 15 & $1115.95 \pm 0.16$ & 1115.89 \\
0 & 16 & $1185.07 \pm 0.12$ & 1185.15 \\
0 & 17 & $1253.65 \pm 0.10$ & 1253.57 \\
0 & 18 & $1322.36 \pm 0.08$ & 1322.41 \\
0 & 19 & $1391.98 \pm 0.10$ & 1392.05 \\
0 & 20 & $1462.63 \pm 0.14$ & 1462.56 \\
0 & 21 & $1532.93 \pm 0.21$ & 1533.34 \\
0 & 22 & $1604.12 \pm 0.38$ & 1604.07 \\
0 & 23 & $1674.96 \pm 0.83$ & 1674.63 \\
1 & 13 & $1006.80 \pm 0.40$ & 1007.25 \\
1 & 14 & $1077.20 \pm 0.21$ & 1077.10 \\
1 & 15 & $1146.57 \pm 0.14$ & 1146.64 \\
1 & 16 & $1215.61 \pm 0.10$ & 1215.54 \\
1 & 17 & $1283.88 \pm 0.08$ & 1284.00 \\
1 & 18 & $1353.16 \pm 0.07$ & 1353.32 \\
1 & 19 & $1423.37 \pm 0.09$ & 1423.45 \\
1 & 20 & $1494.53 \pm 0.13$ & 1494.36 \\
1 & 21 & $1565.23 \pm 0.20$ & 1565.26 \\
1 & 22 & $1636.42 \pm 0.34$ & 1636.20 \\
1 & 23 & $1707.50 \pm 0.75$ & 1706.86 \\
1 & 24 & $1776.20 \pm 2.16$ & 1777.62 \\
2 & 14 & $1110.68 \pm 0.40$ & 1110.56 \\
2 & 15 & $1179.57 \pm 0.25$ & 1179.90 \\
2 & 16 & $1248.67 \pm 0.17$ & 1248.42 \\
2 & 17 & $1317.55 \pm 0.14$ & 1317.27 \\
2 & 18 & $1386.85 \pm 0.15$ & 1386.97 \\
2 & 19 & $1457.18 \pm 0.20$ & 1457.56 \\
2 & 20 & $1528.72 \pm 0.27$ & 1528.52 \\
2 & 21 & $1600.11 \pm 0.40$ & 1599.44 \\
2 & 22 & $1668.97 \pm 0.69$ & 1670.25 \\
\hline
~ \\ ~
\end{tabular} &
\begin{tabular}{cccc}
\hline
\hline
\multicolumn{4}{c}{\textbf{George}} \\
$\l$ & $n$ & 
$\nu_{\mathrm{obs}}$ & 
$\nu_{\mathrm{exact}}$ \\
\hline
0 & 14 & $1053.89 \pm 0.41$ & 1053.63 \\
0 & 15 & $1124.14 \pm 0.26$ & 1124.22 \\
0 & 16 & $1195.37 \pm 0.19$ & 1195.33 \\
0 & 17 & $1265.49 \pm 0.15$ & 1265.53 \\
0 & 18 & $1335.17 \pm 0.12$ & 1335.29 \\
0 & 19 & $1404.87 \pm 0.14$ & 1404.95 \\
0 & 20 & $1475.20 \pm 0.19$ & 1475.35 \\
0 & 21 & $1546.61 \pm 0.28$ & 1546.62 \\
0 & 22 & $1618.92 \pm 0.47$ & 1618.24 \\
0 & 23 & $1690.38 \pm 0.98$ & 1690.19 \\
1 & 13 & $1015.82 \pm 0.72$ & 1015.66 \\
1 & 14 & $1085.80 \pm 0.36$ & 1085.62 \\
1 & 15 & $1156.72 \pm 0.22$ & 1156.85 \\
1 & 16 & $1227.55 \pm 0.16$ & 1227.83 \\
1 & 17 & $1297.96 \pm 0.11$ & 1297.97 \\
1 & 18 & $1367.69 \pm 0.10$ & 1367.91 \\
1 & 19 & $1438.09 \pm 0.13$ & 1438.11 \\
1 & 20 & $1509.01 \pm 0.18$ & 1509.35 \\
1 & 21 & $1580.96 \pm 0.25$ & 1581.15 \\
1 & 22 & $1654.18 \pm 0.41$ & 1653.35 \\
1 & 23 & $1726.67 \pm 0.86$ & 1725.63 \\
1 & 24 & $1794.54 \pm 2.34$ & 1797.66 \\
2 & 14 & $1118.06 \pm 0.68$ & 1118.98 \\
2 & 15 & $1190.08 \pm 0.41$ & 1190.17 \\
2 & 16 & $1260.92 \pm 0.28$ & 1260.52 \\
2 & 17 & $1330.15 \pm 0.20$ & 1330.40 \\
2 & 18 & $1400.11 \pm 0.21$ & 1400.18 \\
2 & 19 & $1470.96 \pm 0.28$ & 1470.65 \\
2 & 20 & $1542.76 \pm 0.36$ & 1542.06 \\
2 & 21 & $1612.77 \pm 0.49$ & 1613.86 \\
2 & 22 & $1686.71 \pm 0.80$ & 1686.02 \\
\hline
~ \\ ~ \\ ~
\end{tabular} &
\begin{tabular}{cccc}
\hline
\hline
\multicolumn{4}{c}{\textbf{Henry}} \\
$\l$ & $n$ & 
$\nu_{\mathrm{obs}}$ & 
$\nu_{\mathrm{exact}}$ \\
\hline
0 & 15 & $1885.08 \pm 0.66$ & 1883.29 \\
0 & 16 & $1999.45 \pm 0.36$ & 1999.06 \\
0 & 17 & $2113.49 \pm 0.25$ & 2113.71 \\
0 & 18 & $2228.09 \pm 0.20$ & 2228.32 \\
0 & 19 & $2343.81 \pm 0.16$ & 2343.89 \\
0 & 20 & $2460.42 \pm 0.12$ & 2460.61 \\
0 & 21 & $2578.00 \pm 0.15$ & 2577.98 \\
0 & 22 & $2695.37 \pm 0.22$ & 2695.29 \\
0 & 23 & $2812.83 \pm 0.30$ & 2812.64 \\
0 & 24 & $2930.19 \pm 0.48$ & 2929.91 \\
0 & 25 & $3048.18 \pm 0.96$ & 3047.63 \\
1 & 15 & $1936.02 \pm 0.58$ & 1936.86 \\
1 & 16 & $2052.96 \pm 0.32$ & 2052.34 \\
1 & 17 & $2167.04 \pm 0.21$ & 2166.78 \\
1 & 18 & $2281.96 \pm 0.17$ & 2282.00 \\
1 & 19 & $2398.26 \pm 0.13$ & 2398.13 \\
1 & 20 & $2515.30 \pm 0.10$ & 2515.48 \\
1 & 21 & $2632.73 \pm 0.15$ & 2632.90 \\
1 & 22 & $2750.57 \pm 0.20$ & 2750.44 \\
1 & 23 & $2868.06 \pm 0.27$ & 2867.77 \\
1 & 24 & $2985.41 \pm 0.43$ & 2985.40 \\
1 & 25 & $3103.56 \pm 0.85$ & 3103.29 \\
1 & 26 & $3219.34 \pm 2.17$ & 3221.56 \\
2 & 16 & $2102.59 \pm 0.61$ & 2102.86 \\
2 & 17 & $2217.55 \pm 0.41$ & 2217.54 \\
2 & 18 & $2333.28 \pm 0.31$ & 2333.22 \\
2 & 19 & $2449.61 \pm 0.21$ & 2450.02 \\
2 & 20 & $2567.61 \pm 0.23$ & 2567.60 \\
2 & 21 & $2685.36 \pm 0.31$ & 2685.13 \\
2 & 22 & $2803.43 \pm 0.38$ & 2802.74 \\
2 & 23 & $2920.43 \pm 0.52$ & 2920.22 \\
2 & 24 & $3037.71 \pm 0.83$ & 3038.15 \\
\hline
~ \\ ~
\end{tabular}
\end{tabular}}}
\end{table*}

\begin{table*}[htbp]
\caption{``Observed'' and exact frequencies for Izzy and Jam.
\label{tab:freq_hares3}}
\begin{tabular}{cc}
\begin{tabular}{cccc}
\hline
\hline
\multicolumn{4}{c}{\textbf{Izzy}} \\
$\l$ & $n$ & 
$\nu_{\mathrm{obs}}$ & 
$\nu_{\mathrm{exact}}$ \\
\hline
0 & 15 & $1871.72 \pm 0.65$ & 1871.62 \\
0 & 16 & $1986.52 \pm 0.36$ & 1987.43 \\
0 & 17 & $2102.10 \pm 0.24$ & 2102.51 \\
0 & 18 & $2216.60 \pm 0.19$ & 2216.46 \\
0 & 19 & $2330.75 \pm 0.16$ & 2330.97 \\
0 & 20 & $2446.45 \pm 0.12$ & 2446.44 \\
0 & 21 & $2563.08 \pm 0.15$ & 2563.27 \\
0 & 22 & $2680.25 \pm 0.21$ & 2680.39 \\
0 & 23 & $2797.61 \pm 0.30$ & 2797.70 \\
0 & 24 & $2914.51 \pm 0.47$ & 2914.70 \\
0 & 25 & $3033.72 \pm 0.93$ & 3031.74 \\
1 & 15 & $1924.83 \pm 0.57$ & 1925.13 \\
1 & 16 & $2041.11 \pm 0.31$ & 2040.92 \\
1 & 17 & $2155.10 \pm 0.21$ & 2155.40 \\
1 & 18 & $2269.87 \pm 0.16$ & 2269.63 \\
1 & 19 & $2384.37 \pm 0.13$ & 2384.61 \\
1 & 20 & $2500.94 \pm 0.10$ & 2500.90 \\
1 & 21 & $2617.94 \pm 0.14$ & 2618.10 \\
1 & 22 & $2735.46 \pm 0.19$ & 2735.44 \\
1 & 23 & $2852.91 \pm 0.26$ & 2852.78 \\
1 & 24 & $2969.72 \pm 0.42$ & 2969.83 \\
1 & 25 & $3087.52 \pm 0.82$ & 3087.10 \\
1 & 26 & $3206.00 \pm 2.09$ & 3204.49 \\
2 & 16 & $2092.21 \pm 0.60$ & 2091.62 \\
2 & 17 & $2205.37 \pm 0.40$ & 2205.74 \\
2 & 18 & $2320.55 \pm 0.30$ & 2320.37 \\
2 & 19 & $2436.10 \pm 0.20$ & 2435.91 \\
2 & 20 & $2553.13 \pm 0.22$ & 2552.87 \\
2 & 21 & $2670.54 \pm 0.30$ & 2670.21 \\
2 & 22 & $2787.88 \pm 0.38$ & 2787.75 \\
2 & 23 & $2905.06 \pm 0.51$ & 2905.02 \\
2 & 24 & $3020.91 \pm 0.81$ & 3022.28 \\
\hline
\end{tabular} &
\begin{tabular}{cccc}
\hline
\hline
\multicolumn{4}{c}{\textbf{Jam}} \\
$\l$ & $n$ & 
$\nu_{\mathrm{obs}}$ & 
$\nu_{\mathrm{exact}}$ \\
\hline
0 & 15 & $1379.65 \pm 0.90$ & 1378.18 \\
0 & 16 & $1464.75 \pm 0.50$ & 1464.53 \\
0 & 17 & $1552.48 \pm 0.34$ & 1551.79 \\
0 & 18 & $1639.61 \pm 0.27$ & 1639.36 \\
0 & 19 & $1725.82 \pm 0.24$ & 1726.14 \\
0 & 20 & $1812.11 \pm 0.22$ & 1812.12 \\
0 & 21 & $1898.10 \pm 0.28$ & 1898.19 \\
0 & 22 & $1984.65 \pm 0.35$ & 1984.62 \\
0 & 23 & $2071.62 \pm 0.51$ & 2071.83 \\
0 & 24 & $2160.89 \pm 0.89$ & 2159.82 \\
0 & 25 & $2249.61 \pm 2.00$ & 2248.00 \\
1 & 15 & $1417.85 \pm 0.78$ & 1417.80 \\
1 & 16 & $1504.29 \pm 0.43$ & 1504.64 \\
1 & 17 & $1592.47 \pm 0.29$ & 1592.18 \\
1 & 18 & $1679.60 \pm 0.23$ & 1679.59 \\
1 & 19 & $1765.97 \pm 0.19$ & 1765.97 \\
1 & 20 & $1851.84 \pm 0.21$ & 1852.02 \\
1 & 21 & $1937.87 \pm 0.24$ & 1938.35 \\
1 & 22 & $2024.94 \pm 0.30$ & 2025.23 \\
1 & 23 & $2113.34 \pm 0.44$ & 2113.01 \\
1 & 24 & $2201.34 \pm 0.77$ & 2201.30 \\
1 & 25 & $2290.63 \pm 1.72$ & 2289.73 \\
2 & 16 & $1544.11 \pm 0.82$ & 1543.85 \\
2 & 17 & $1630.78 \pm 0.56$ & 1631.42 \\
2 & 18 & $1717.91 \pm 0.43$ & 1718.34 \\
2 & 19 & $1805.00 \pm 0.35$ & 1804.41 \\
2 & 20 & $1889.95 \pm 0.41$ & 1890.53 \\
2 & 21 & $1975.94 \pm 0.47$ & 1977.03 \\
2 & 22 & $2063.96 \pm 0.58$ & 2064.28 \\
2 & 23 & $2151.83 \pm 0.84$ & 2152.37 \\
2 & 24 & $2239.81 \pm 1.47$ & 2240.70 \\
\hline
~
\end{tabular}
\end{tabular}
\end{table*}

\end{document}